\documentclass[12pt,a4paper]{article}
\usepackage[margin=29mm]{geometry}
\usepackage{microtype,amsmath,amssymb,amsthm,mathtools}
\usepackage{booktabs,array,enumitem,placeins,xcolor,xurl}
\usepackage[authoryear,round]{natbib}
\usepackage[hidelinks]{hyperref}
\usepackage{aliascnt}
\usepackage[nameinlink,noabbrev]{cleveref}
\hypersetup{pdftitle={Peer Feedback and Forecast Revision},pdfauthor={Georgy Lukyanov},bookmarksdepth=3}
\numberwithin{equation}{section}

\theoremstyle{plain}
\newtheorem{theorem}{Theorem}[section]
\newaliascnt{proposition}{theorem}
\newtheorem{proposition}[proposition]{Proposition}
\aliascntresetthe{proposition}
\newaliascnt{lemma}{theorem}
\newtheorem{lemma}[lemma]{Lemma}
\aliascntresetthe{lemma}
\newaliascnt{corollary}{theorem}
\newtheorem{corollary}[corollary]{Corollary}
\aliascntresetthe{corollary}

\theoremstyle{definition}
\newaliascnt{definition}{theorem}
\newtheorem{definition}[definition]{Definition}
\aliascntresetthe{definition}
\newaliascnt{assumption}{theorem}

\aliascntresetthe{assumption}

\theoremstyle{remark}
\newaliascnt{remark}{theorem}
\newtheorem{remark}[remark]{Remark}
\aliascntresetthe{remark}

\crefname{assumption}{Assumption}{Assumptions}
\Crefname{assumption}{Assumption}{Assumptions}
\crefname{proposition}{Proposition}{Propositions}
\Crefname{proposition}{Proposition}{Propositions}
\crefname{corollary}{Corollary}{Corollaries}
\Crefname{corollary}{Corollary}{Corollaries}

\newcommand{\Prb}{\mathbb{P}}

\newcommand{\ind}{\mathbf{1}}

\crefname{theorem}{Theorem}{Theorems}
\Crefname{theorem}{Theorem}{Theorems}
\crefname{proposition}{Proposition}{Propositions}
\Crefname{proposition}{Proposition}{Propositions}
\crefname{lemma}{Lemma}{Lemmas}
\Crefname{lemma}{Lemma}{Lemmas}
\crefname{corollary}{Corollary}{Corollaries}
\Crefname{corollary}{Corollary}{Corollaries}
\crefname{definition}{Definition}{Definitions}
\Crefname{definition}{Definition}{Definitions}
\crefname{assumption}{Assumption}{Assumptions}
\Crefname{assumption}{Assumption}{Assumptions}
\crefname{remark}{Remark}{Remarks}
\Crefname{remark}{Remark}{Remarks}

\title{Peer Feedback and Forecast Revision}
\author{Georgy Lukyanov\thanks{I thank Darina Cheredina, Olivier Gossner, Vasilii Ivanik, Margarita Kirneva, Yukio Koriyama, Alessandro Riboni, and Anna Vlasova for helpful comments and discussions. All remaining errors are my own.}\\[3pt]{\small Toulouse School of Economics, University of Toulouse Capitole}\\
{\small\href{mailto:georgy.lukyanov@tse-fr.eu}{georgy.lukyanov@tse-fr.eu}}}
\date{November 2026}
\begin{document}
\makeatletter
\let\savedfnsymbol\@fnsymbol
\renewcommand*\@fnsymbol[1]{\ensuremath{\ast}}
\maketitle
\let\@fnsymbol\savedfnsymbol
\makeatother
\begin{abstract}
Should a forecaster see a colleague's forecast before revising her own? If changing her mind casts doubt on her competence, a colleague's agreement may encourage her to disregard new evidence. We study two experts who each report twice. A platform can show one expert the other's first report, and the decision maker eventually sees all reports. We give conditions under which the unique policy that maximises decision accuracy reveals some agreements and withholds every disagreement during revision. Agreements are disclosed just often enough for the forecaster to follow conflicting evidence after silence. The result holds on an open set of parameters within a class of informative equilibria. With two peers, confirming a forecast only when both peers agree can induce revision without a lottery. Both procedures improve on full disclosure with small rewards for accuracy. With unequal error costs, withholding all peer feedback can outperform the rule that maximises accuracy.
\end{abstract}
\begin{center}
\begin{minipage}{0.89\textwidth}
\end{minipage}
\end{center}

\section{Introduction}\label{sec:introduction}

Suppose two forecasters have submitted their opinions about the outcome of an election. Both will have an opportunity to revise their forecasts when more evidence comes in. Before they do so, should one of them be shown the other's opinion? If the two forecasts agree, the answer may seem particularly straightforward. Yet confirmation can have a drawback. A forecaster who subsequently receives evidence against her first opinion can gain a reputation for sound judgement by standing by it, especially if she expects it to turn out to be right. The colleague's agreement can then be enough to keep the new evidence out of her report.

We study this question in a model in which each expert knows the quality of her own information. Both experts receive two private signals about a binary state and submit a report after each signal. Between reports, the platform may show A's first report to B. A receives no feedback. B knows in advance how often the platform will disclose an agreement and how often it will disclose a disagreement. The eventual decision maker sees the initial reports, the feedback message and both final reports. Only after the decision is taken does the state become known. An expert's wage depends on what her reporting history and the realised state imply about her ability. Consequently, changing B's access to information during revision can change what the decision maker learns even though the decision maker has access to the entire record under every policy.

To see how the platform can use this distinction, consider a forecaster whose new evidence conflicts with her first forecast. Under a policy of complete secrecy, hearing nothing from the platform tells her nothing about her colleague's opinion. Now suppose the platform sometimes discloses agreement and always keeps disagreement private during revision. Hearing nothing is then evidence that her colleague may have disagreed. The more agreements the platform discloses, the worse this news is for her first forecast. Eventually even the less able forecaster is willing to report the conflicting evidence. Disclosing further agreements serves little purpose at that point. It moves some forecasters from a situation where they would revise to one where receiving confirmation may persuade them to persist.

Our main result gives sufficient conditions under which this last observation determines the optimal rule. Suppose the decision maker wants to identify the state and makes the same loss from either mistake. Among all policies that specify disclosure probabilities for agreements and disagreements, the unique optimum discloses just enough agreement for B to revise after silence and discloses no disagreement during revision (\cref{thm:analytical-global-optimality}). The conditions hold for an interval of ability shares at given signal precisions and also on an open set on which all five parameters may vary. They ensure that both experts wish to report their first signals truthfully, even allowing an expert who initially lies to make the best possible use of the feedback that follows. At the threshold, the less able expert is indifferent when her second signal contradicts her first after silence. She revises in equilibrium. A slightly higher rate of agreement disclosure makes her strictly prefer to revise, at a loss in accuracy proportional to the increase.

The comparison uses an informative class of sequential equilibria. The first reports convey the experts' first signals, an expert whose signals agree reports their common value, and the more able expert follows her new evidence when her signals differ. Whether the less able expert revises is determined in equilibrium. With purely reputational payoffs there is also an equilibrium in which both experts ignore their information. In that equilibrium the platform has nothing to gain from disclosing a report. This matters for interpretation. The institutional question that our comparison addresses arises when experts already give informative first opinions.

With a single binary peer report a deterministic rule that discloses all agreements and withholds all disagreements tells B what A reported in either event. With two peers there is a simpler way to make silence informative while leaving its meaning uncertain. The platform sends confirmation when both peers agree with B. Otherwise B knows that she lacks unanimous support, but cannot tell whether one or both peers disagree. We show that this deterministic rule induces strict revision after silence and improves on both complete secrecy and full disclosure with exactly the same three experts. Both the one-peer and two-peer rules can also improve decisions when correct final reports earn a small direct reward. An institution that sends a fraction of cases for full review can still use selective feedback on the remaining cases.

The size of the improvement and the use to which the forecasts are put both matter. At one illustrative parameter profile the optimal rule yields about three additional correct decisions per thousand relative to full disclosure. On high-volume questions one could compare that gain with the cost of administering feedback. If the two mistakes carry different costs, even the direction of the comparison can change. On the interval where selective feedback is uniquely optimal under equal losses, complete secrecy is preferable to that same selective policy when missing state one costs nine times as much as acting on a false alarm.

\subsection{Related literature}\label{subsec:related-literature}

Our starting point is the reputational incentive to stand by an opinion.
In \citet{Li2007}, an expert receives two signals and the reporting protocol determines when she must give her opinion. In an informative equilibrium the more able expert follows conflicting new evidence, whereas the less able expert sometimes repeats her first report. Related work studies how reputation shapes professional advice and communication more generally \citep{OttavianiSorensen2006Professional,OttavianiSorensen2006CheapTalk}, and how it can give rise to persistence or an early decision \citep{Tajika2021}. We take the two reporting dates as given. The object of design is the information one expert receives about another between those dates.

The link to \citet{Levy2004} is particularly close. In her model an expert may go against public information because a person with a more precise private signal has more reason to do so. The prospect of being assessed in this way affects the decision to consult an adviser in the first place. Levy allows for concerns about the outcome as well as about reputation, and the outcome is eventually observed. Here an expert's first opinion is already on record when peer feedback arrives. It is the platform that chooses what the expert may see. If the platform shows a disproportionate share of agreements, the absence of news itself makes it easier for the expert to change her mind. Persistence in this comparison means repeating one's own forecast. Following one's second signal can involve moving towards a colleague who disagreed, even though the platform never showed that colleague's report.

A related literature asks who should speak first in a debate \citep{OttavianiSorensen2001}, whether actions and their consequences should be observable \citep{Prat2005}, and how secrecy affects committee deliberation or sequential advice \citep{FehrlerHughes2018,BagSharma2019}. Other work examines an evaluator's choice of information precision \citep{GhoshAnbarciRoy2017}, the effect of a prospective second opinion on the first expert's advice \citep{LiuSanyal2012}, and the choice between independent and collective expertise \citep{CatoniniKurbatovStepanov2024}. In our setting the decision maker obtains all the reports under either disclosure regime. What changes with interim transparency is what the experts are prepared to put into those reports. In the language of information design \citep{KamenicaGentzkow2011,BergemannMorris2019}, the platform commits to a signal about an endogenous report, and its value depends on the reporting behaviour it induces.

Sections~\ref{sec:environment} and~\ref{sec:reputational-response} set out the model and derive the experts' reporting choices. Sections~\ref{sec:policy-design} and~\ref{sec:exact-results} characterise the optimal disclosure rule. Sections~\ref{sec:two-peer-confirmation}--\ref{sec:disagreement-disclosure} study deterministic confirmation, accuracy rewards and disclosure of disagreement. Section~\ref{sec:application} considers a procedure for forecast review. Further proofs appear in the appendix.

\section{Environment and design problem}\label{sec:environment}

An expert submits a forecast and later has a chance to revise it. Between
the two reports she may learn what a peer initially forecast. A platform
chooses when she sees that forecast. It commits to the rule before either
expert acquires information.

\subsection{Experts and private information}\label{subsec:primitives}

There are two experts, A and B. The state \(\theta\) is zero or one with
equal probability. Each expert \(i\) knows her own ability
\(\tau_i\in\{L,H\}\). Abilities are independent across experts and of the
state, and the probability of high ability is \(\omega\in(0,1)\).

At each date \(t\in\{1,2\}\), expert \(i\) privately observes a signal
\(s_{it}\in\{0,1\}\). Given the state and the two abilities, the four
signals are independent. A high-ability expert has the more accurate
signal at each date. For either state \(\vartheta\),

\begin{equation}\label{eq:signal-structure}
\Prb(s_{it}=\vartheta\mid\theta=\vartheta,\tau_i=\tau)
=p_{\tau t},
\qquad
\frac12<p_{Lt}<p_{Ht}<1.
\end{equation}
When an expert receives conflicting signals, her forecast depends on how
much more she trusts one date's evidence than the other's.
\footnote{The symmetric prior gives neither state an advantage. With an
asymmetric prior, equally precise conflicting signals restore the prior
odds, which can themselves favour one report.}

Let the signal odds and relative learning rates be
\begin{equation}\label{eq:learning-rates}
\kappa_{\tau t}=\frac{p_{\tau t}}{1-p_{\tau t}},
\qquad
\eta_\tau=\frac{\kappa_{\tau2}}{\kappa_{\tau1}}.
\end{equation}
For instance, following \((s_{i1},s_{i2})=(1,0)\), type \(\tau\)'s posterior
odds in favour of the state indicated by her first signal are
\begin{equation}\label{eq:private-conflict-odds}
\frac{\Prb(\theta=1\mid s_{i1}=1,s_{i2}=0,\tau_i=\tau)}
 {\Prb(\theta=0\mid s_{i1}=1,s_{i2}=0,\tau_i=\tau)}
=\frac1{\eta_\tau}.
\end{equation}
Thus \(\eta_\tau\) measures how much conflicting evidence favours the new
signal. We consider technologies for which \(\eta_H>\eta_L\). Two experts
who face the same peer evidence and the same wage schedule can then draw
different conclusions about whether to revise.

\subsection{Timing and feasible feedback rules}\label{subsec:timing-policy}

Before the experts receive their signals, the platform announces a policy
\(\pi=(\alpha,\delta)\in[0,1]^2\). It will show A's first report to B with
probability \(\alpha\) if their first reports agree and with probability
\(\delta\) if they disagree. In all other cases it sends B a message of
non-revelation, which we call silence. A receives no feedback.

\begin{enumerate}[label=(\roman*),leftmargin=*,itemsep=2pt]
\item After the policy is announced, nature draws the state and the two
abilities. Each expert observes her ability and first signal.
\item The experts independently submit their initial reports \(r_{A1}\) and
\(r_{B1}\). These reports are locked, meaning that neither expert can
subsequently alter the entry.
\item Each expert receives her second private signal. The platform observes
both first reports and sends B the message prescribed by its policy.
\item The experts simultaneously submit their terminal reports \(r_{A2}\) and
\(r_{B2}\).
\item A decision maker sees all four reports and the platform's message,
then chooses an action. After the decision, the state is verified and a
competitive evaluator pays each expert according to her posterior
reputation for high ability.
\end{enumerate}

Write

\begin{equation}\label{eq:locked-report-labels}
y=r_{A1},
\qquad
b=r_{B1},
\qquad
r_i^2=(r_{i1},r_{i2}).
\end{equation}
The realised platform message is \(m\in\{R_0,R_1,N\}\). If A reported
\(y\), B can receive either \(R_y\) or \(N\). The disclosure probabilities
are
\begin{equation}\label{eq:policy}
\Prb_\pi(R_y\mid y=b)=\alpha,\qquad
\Prb_\pi(R_y\mid y\ne b)=\delta.
\end{equation}
The corresponding probabilities of \(N\) are \(1-\alpha\) and \(1-\delta\).
Disclosure is truthful and treats the two state labels symmetrically. We
write \(\Pi^{\mathrm{RS}}\) for this class of one-way reveal-or-silence
rules. It includes sealing, \(\mathsf S=(0,0)\), and full disclosure,
\(\mathsf F=(1,1)\). The platform's discretion concerns the information
available at revision. Under every rule, the decision maker eventually
observes the entire record.

\subsection{Reputation and the design objective}\label{subsec:payoffs-value}

After state verification, expert \(i\) receives her terminal reputation as a
competitive wage:
\begin{equation}\label{eq:terminal-reputation}
\rho_i
=
\Prb(\tau_i=H\mid r_A^2,r_B^2,m,\theta).
\end{equation}
A competitive market pays this wage if productivity is normalised to zero
for the low type and one for the high type.\footnote{A positive affine
transformation of productivity leaves the reporting incentives unchanged.}
The baseline expert receives no other payment. We allow her to earn a
reward for a correct terminal report in Section~\ref{sec:accuracy-rewards}.

Let
\begin{equation}\label{eq:public-outcome}
Z=(r_A^2,r_B^2,m)
\end{equation}
denote the public outcome before the state is known. A finite downstream
decision problem is \(D=(\mathcal A_D,u_D)\), where \(\mathcal A_D\) is the
action set and \(u_D(a,\theta)\) the payoff. For policy \(\pi\) and
assessment \(\sigma\), let \(P_\theta^{\pi,\sigma}(z)\) be the conditional
probability of outcome \(z\) in state \(\theta\). The ex ante decision value is
\begin{equation}\label{eq:decision-value}
V_D(\pi,\sigma)
=
\sum_z
\max_{a\in\mathcal A_D}
\left\{
\frac12\sum_{\theta\in\{0,1\}}
P_\theta^{\pi,\sigma}(z)u_D(a,\theta)
\right\}.
\end{equation}
The platform chooses a policy to maximise the decision maker's expected
payoff. The reporting strategies enter this objective because feedback can
change which signals the experts convey. Our main criterion is the
probability of choosing the correct state, so \(\mathcal A_D=\{0,1\}\) and
\(u_D(a,\theta)=\ind\{a=\theta\}\). We allow a general finite decision
problem when asking which features of the policy comparison depend on this
criterion.

\subsection{Strategies, beliefs, and sequential equilibrium}
\label{subsec:equilibrium}

An initial-report strategy depends on type and first signal. At the terminal
date, A conditions on her type, signals and initial report; B also observes
the platform's message:
\begin{align}
\sigma_{i1}(\cdot\mid\tau_i,s_{i1})
&\in\Delta(\{0,1\}),\label{eq:locked-strategy}\\
\sigma_{A2}(\cdot\mid\tau_A,s_A^2,r_{A1})
&\in\Delta(\{0,1\}),\label{eq:a-terminal-strategy}\\
\sigma_{B2}(\cdot\mid\tau_B,s_B^2,r_{B1},m)
&\in\Delta(\{0,1\}).\label{eq:b-terminal-strategy}
\end{align}

\begin{definition}\label{def:sequential-equilibrium}
For a fixed policy, an assessment is a sequential equilibrium if reporting is sequentially optimal at every private history, terminal wages equal Bayesian posterior reputations, and beliefs are limits of Bayes-consistent beliefs induced by completely mixed reporting strategies.
\end{definition}

At the initial reporting date an expert can misreport and still make the
best use of her subsequent information. For B this option includes the
change in peer feedback caused by her first report. We check her incentives
against this deviation throughout.

\subsection{The maintained informative response}
\label{subsec:maintained-prediction}

If the evaluator expects every report to be uninformative about ability,
an expert cannot improve her reputation by changing a report. This gives
an equilibrium common to all feedback policies.

\begin{proposition}
\label{prop:policy-independent-babbling}
For every \(\pi\in\Pi^{\mathrm{RS}}\), there is a sequential equilibrium in which each type of each expert chooses each report with probability \(1/2\) at every private history, independently of signals and messages. Every wage equals \(\omega\), the public outcome is independent of the state, and symmetric state-classification accuracy is \(1/2\).
\end{proposition}

\begin{proof}
When initial reports ignore signals, neither those reports nor the platform
message conveys information about the state or types. Uniform terminal
randomisation makes the likelihood of each complete history identical across
types. The posterior reputation is consequently \(\omega\) at every feasible
history, and every report is optimal. Reporting is already completely mixed.
Chance branches that have zero probability after every report history are
absent from the fixed-policy game. They remain impossible under reporting
trembles, so no off-path wage has to be specified there.
\end{proof}

\begin{definition}
\label{def:regular-monotone}
A sequential equilibrium under \(\pi\) is \emph{regular-monotone} if:
\begin{enumerate}[label=(\roman*),leftmargin=*,itemsep=2pt]
\item both experts and both types report their first signals truthfully;
\item after confirming private signals, each expert repeats her locked report;
\item after conflicting private signals, the high type reports the second signal;
\item after conflicting private signals, the low type may mix label-symmetrically between the locked report and the second signal, assigning positive probability to the second signal;
\item B's requirements in (ii)--(iv) apply at every positive-probability message branch, whereas A receives no peer feedback; and
\item behaviour following a locked-report deviation is unrestricted except by sequential rationality and consistency.
\end{enumerate}
\end{definition}

These restrictions select the informative behaviour we compare across
policies. All feasible deviations are available to each expert. In
particular, truthful first reports must be optimal even when a lie changes
what B will learn. Babbling is completely mixed, so restrictions on beliefs
at unreached binary messages alone cannot rule it out.\footnote{A small
correctness reward eliminates uniform babbling but can leave an equilibrium
in which both experts always repeat their first reports. Section~OA.1
also examines multiplicity under small recording errors.}

\subsection{Can a new message select an informative equilibrium?}
\label{subsec:neologisms}

In \citet[pp.~520--522]{Farrell1993}, a new message can announce that its
sender belongs to a specified set of types. The announcement is credible
if each named type strictly benefits when the message is believed and no
other type strictly benefits. In our game the meaning of such an
announcement depends on when it is made. At the terminal date the peer
feedback has already been sent. A sender can then announce a set of private
histories, each specifying her ability and signals, and the evaluator can
condition her assessment on the announcement and the subsequently verified
state.

At the benchmark in Section~OA.11, neither expert can profitably distinguish
herself in this way under babbling, for any feasible feedback rule. In the
equilibrium with full repetition, by contrast, A can credibly announce that
her two signals agree. Under informative play, neither expert has a
credible terminal announcement on any supported branch when
\(\eta_H>\eta_L\geq1\). For a given branch, each possible announcement
offers a pair of state-dependent wages. An expert's willingness to make it
therefore depends on her posterior odds that her initial report was wrong.
Section~OA.11 uses the ordering of these odds across the four private
histories to obtain the result. Both informative play and babbling can
thus survive the terminal test at the benchmark. An
extension to new initial messages would require specifying how the platform
compares such a message with a peer's report and what feedback it sends.

\section{Reputational response to peer feedback}
\label{sec:reputational-response}

Suppose B receives a second signal at odds with her initial forecast. To
decide whether to revise, she asks which state is more likely and how the
market will assess a revision in each state. The first calculation uses the
peer evidence. The second uses the market's beliefs about who revises
after receiving that evidence.

\subsection{Wages and the decision to revise}
\label{subsec:factorization-cutoff}

Once the evaluator knows the state, the reports of the two experts play
different roles in the wage calculation. A saw no peer information, so her
two reports suffice for assessing her ability. For B the evaluator must also
account for the information the platform gave her.

\begin{lemma}
\label{lem:reputational-factorization}
For every committed policy \(\pi\in\Pi^{\mathrm{RS}}\), at histories with
positive probability,
\begin{align}
\Prb(\tau_A=H\mid r_A^2,r_B^2,m,\theta)
&=\Prb(\tau_A=H\mid r_A^2,\theta),\label{eq:a-factorization}\\
\Prb(\tau_B=H\mid r_A^2,r_B^2,m,\theta)
&=\Prb(\tau_B=H\mid r_B^2,y,m,\theta).
\label{eq:b-factorization}
\end{align}
Thus A faces the sealed-feedback problem. The policy affects B through her interim state belief and the message-contingent distribution of her report history.
\end{lemma}

\begin{proof}
Conditional on the state and types, a complete-history likelihood is the
product of A's reporting likelihood, the policy probability and B's reporting
likelihood. In A's type odds the latter two factors cancel. In B's type odds,
A's reporting factor cancels, and the policy probability is type-independent
conditional on \((y,b)\).
The same identities hold in consistent limits of completely mixed reporting
strategies at histories compatible with the policy. Messages assigned zero
probability by the committed rule remain impossible.
\end{proof}

Suppose the initial reports reveal the experts' first signals. A's report
then identifies the state correctly with probability

\begin{equation}\label{eq:peer-accuracy}
q=\omega p_{H1}+(1-\omega)p_{L1}>\frac12,
\qquad
\gamma=\frac{q}{1-q}>1.
\end{equation}
For example, seeing that A reported one multiplies B's odds on state one
by \(\gamma\). After the state is verified, that same report carries no
direct information about B's ability.

Fix a feedback branch that can occur. Let \(h\) be the probability that a
low type repeats her first report when her own signals conflict. The high
type follows her second signal, and either type repeats when her signals
agree. Once the state is known, a pair of reports has one of four
correctness patterns \(\chi\in\{CC,CI,IC,II\}\), in chronological order.
For instance, \(IC\) records an initially incorrect forecast followed by a
correct one. Write \(\ell_\tau^\chi(h)\) for the probability of this
pattern if the expert used the branch's reporting rule on the two private
signals. The likelihood of the feedback itself, conditional on the state
and the initial report, is common to both types and cancels in Bayes' rule.
The wage is therefore

\begin{equation}\label{eq:pattern-wage}
W_\chi(h)
=
\frac{\omega\ell_H^\chi(h)}
{\omega\ell_H^\chi(h)+(1-\omega)\ell_L^\chi(h)}.
\end{equation}
Define the wage gaps
\begin{equation}\label{eq:wage-gaps}
G_C(h)=W_{CC}(h)-W_{CI}(h),
\qquad
G_I(h)=W_{IC}(h)-W_{II}(h),
\qquad
Q(h)=\frac{G_I(h)}{G_C(h)}.
\end{equation}
Thus \(G_C\) is the wage gain from repetition when the first report was
correct, and \(G_I\) the wage gain from revision when it was incorrect.
The weights in the expert's expected payoff are her own beliefs about these
two events. The pattern likelihoods and wages are derived in Section~OA.2.

Let \(\lambda\) be the feedback likelihood ratio in favour of the state named
by B's initial report. After a private conflict, type \(\tau\)'s posterior
odds for that state are \(\lambda/\eta_\tau\). The gain from repetition is
\begin{equation}\label{eq:generic-conflict-gain}
\mathcal G_\tau(h;\lambda)
=
\frac{\lambda G_C(h)-\eta_\tau G_I(h)}
{\lambda+\eta_\tau}.
\end{equation}
Thus, when \(G_C(h)>0\), repetition is optimal if and only if
\(\lambda/\eta_\tau\geq Q(h)\). The left side measures the support for the
initial forecast, while the right side is the relative reputational return
to revision. Since the wage ratio depends on anticipated persistence,
equilibrium requires consistency between \(h\) and this comparison.

\begin{lemma}
\label{lem:message-cutoff}
There is a unique \(h^\dagger\in(0,1)\) with \(G_C(h^\dagger)=0\). On \([0,h^\dagger)\), \(G_C\) and \(G_I\) are positive and
\[
Q:[0,h^\dagger)\longrightarrow[Q_0,\infty),
\qquad
Q_0=Q(0),
\]
is continuous and strictly increasing. Let
\begin{equation}\label{eq:pressure-threshold}
\bar\lambda=\eta_LQ_0.
\end{equation}
At a conflicting private history:
\begin{enumerate}[label=(\roman*),leftmargin=*]
\item if \(\lambda\leq\bar\lambda\), the low type uses the second signal; at equality she is indifferent when \(h=0\);
\item if \(\lambda>\bar\lambda\), there is a unique \(h(\lambda)\in(0,h^\dagger)\) satisfying
\begin{equation}\label{eq:generic-mixing-root}
Q(h(\lambda))=\frac{\lambda}{\eta_L};
\end{equation}
\item if \(\eta_H>\eta_L\), the high type strictly uses the second signal at both the pure and interior low-type solutions.
\end{enumerate}
\end{lemma}

\begin{proof}
An increase in \(h\) moves low-type probability from \(CI\) and \(IC\) to
\(CC\) and \(II\). Accordingly, \(W_{CC}\) and \(W_{II}\) decrease, whereas
\(W_{CI}\) and \(W_{IC}\) increase. Hence \(G_C\) is strictly decreasing and
\(G_I\) strictly increasing. Both gaps are positive at \(h=0\). As
\(h\uparrow1\), \(W_{CI}\) tends to one and \(W_{CC}\) remains below one.
There is therefore a unique zero \(h^\dagger\) of \(G_C\), and \(Q\) is
strictly increasing on \([0,h^\dagger)\), diverging at the upper endpoint.

By \cref{eq:generic-conflict-gain}, the low type is indifferent precisely when
\(Q(h)=\lambda/\eta_L\). Its range gives (i) and (ii). At an interior root,
\(\lambda/\eta_H<\lambda/\eta_L=Q(h)\) under \(\eta_H>\eta_L\), so the
high type strictly revises. The same comparison establishes strict revision
at the pure solution.
\end{proof}

The mechanism behind the cutoff is reputational. When the market expects
more low-ability experts to repeat, a repeated report is less persuasive
evidence of ability. A revision is correspondingly more persuasive.
Stronger peer support for the initial forecast is then needed to keep the
low type willing to repeat.

\subsection{Sealing: the no-feedback benchmark}\label{subsec:sealed-benchmark}

With no peer information, the likelihood ratio is \(\lambda=1\). Sealing
therefore has interior low-type persistence exactly when \(\bar\lambda<1\).
Write \(h^{\mathsf S}\) for the unique solution of
\begin{equation}\label{eq:sealed-root}
Q(h^{\mathsf S})=\frac1{\eta_L}.
\end{equation}

\begin{corollary}
\label{cor:sealed-continuation}
Suppose \(\bar\lambda<1\) and \(\eta_H>\eta_L\). Conditional on truthful locked reports, the sealed branch has a unique regular-monotone continuation: both types repeat after confirmation, the high type uses the second signal after a conflict, and the low type repeats the locked report with probability \(h^{\mathsf S}\) after a conflict.
\end{corollary}

\begin{proof}
Set \(\lambda=1\) in \cref{lem:message-cutoff}. At the interior solution
\(Q(h^{\mathsf S})=1/\eta_L\). After two confirming signals, the low
type's state odds exceed this cutoff since
\(\eta_L\kappa_{L1}\kappa_{L2}=\kappa_{L2}^2>1\). The high type's
confirmation odds are higher still.
\end{proof}

After a conflict, both types face the same wage schedule. The low type
is willing to randomise at that schedule. The high type strictly revises
because she puts greater relative weight on the new signal.

\begin{remark}\label{rem:stationary-precision}
If \(p_{\tau1}=p_{\tau2}\) for both types, then \(\eta_H=\eta_L=1\). Under the symmetric state prior, two conflicting signals cancel for both types. At a common public history and wage schedule, one type cannot strictly retain the old report while the other strictly follows the new signal.
\end{remark}

\subsection{Feedback messages on a common likelihood-ratio scale}
\label{subsec:feedback-continuation}

Revealed agreement has likelihood ratio \(\gamma\), revealed disagreement
\(1/\gamma\), and sealing one. Under \(\pi=(\alpha,\delta)\), silence gives
\begin{equation}\label{eq:silence-likelihood}
\lambda_N(\alpha,\delta)
=
\frac{q(1-\alpha)+(1-q)(1-\delta)}
 {(1-q)(1-\alpha)+q(1-\delta)}.
\end{equation}
Silence favours B's first report when \(\delta>\alpha\), and favours the
opposite state when \(\alpha>\delta\). If the two probabilities are equal
and less than one, silence leaves B's beliefs unchanged. Multiplying both
concealment probabilities by the same positive constant changes how often B
receives silence, while preserving its likelihood ratio.

Following revealed agreement, define \(h^{\mathsf F}\) by
\begin{equation}\label{eq:agreement-root}
Q(h^{\mathsf F})=\frac{\gamma}{\eta_L}.
\end{equation}
Since \(\gamma>1\), we have \(h^{\mathsf F}>h^{\mathsf S}\). Agreement from
the peer makes the low type more willing to retain her earlier forecast.

Two incentive checks matter at the extremes. Even when the peer agrees
with B, the high type should revise after a private conflict. Even when
the peer disagrees with B, both types should confirm two agreeing private
signals. Between the two extremes, the likelihood ratio after silence
determines how often the low type revises.

\begin{proposition}
\label{prop:selective-continuation}
Suppose
\begin{equation}\label{eq:continuation-region}
\bar\lambda<1,
\qquad
\eta_H>\eta_L,
\qquad
\frac1{\eta_LQ_0}
<\gamma
<\frac{\kappa_{L1}\kappa_{L2}}{Q_0}.
\end{equation}
Conditional on truthful locked reports, there is a unique regular-monotone continuation at every positive-probability feedback branch:
\begin{enumerate}[label=(\roman*),leftmargin=*]
\item both types repeat after confirming signals, and the high type uses the second signal after every conflict;
\item after revealed agreement, the low type repeats at a conflict with probability \(h^{\mathsf F}\);
\item after revealed disagreement, the low type uses the second signal;
\item whenever silence has positive probability, the low type uses the second signal if \(\lambda_N\leq\bar\lambda\), and otherwise repeats with the unique \(h_N\in(0,h^\dagger)\) satisfying
\begin{equation}\label{eq:silent-root}
Q(h_N)=\frac{\lambda_N}{\eta_L}.
\end{equation}
At \(\lambda_N=\bar\lambda\), pure use of the second signal is optimal but weak.
\end{enumerate}
\end{proposition}

\begin{proof}
Use \cref{lem:message-cutoff} at \(\gamma,1/\gamma\) and \(\lambda_N\).
The lower bound on \(\gamma\) gives strict revision after revealed
disagreement. To check confirmation, first take a mixing branch. The low
type's odds after confirming signals are
\(\lambda\kappa_{L1}\kappa_{L2}\), exceeding
\(Q(h)=\lambda/\eta_L\) by the factor \(\kappa_{L2}^2>1\).
On a pure branch, the same odds are at least
\(\kappa_{L1}\kappa_{L2}/\gamma>Q_0\) by the upper bound on
\(\gamma\). Both inequalities are strict for the high type as well.
\end{proof}

Writing \(h_N=0\) on a pure-revision silent branch and using
\cref{eq:silent-root} otherwise, the wage for pattern \(\chi\) is

\begin{equation}\label{eq:message-wage}
\widetilde W_\chi(m,y,b)
=
\begin{cases}
W_\chi(h^{\mathsf F}),&m=R_y\text{ and }y=b,\\
W_\chi(0),&m=R_y\text{ and }y\ne b,\\
W_\chi(h_N),&m=N.
\end{cases}
\end{equation}
The eventual release of \(y\) adds no type-dependent factor to this wage:
conditional on \((\theta,b)\), it is independent of B's type, even though it
affects her interim belief about the state.

\subsection{From continuation behaviour to equilibrium feasibility}
\label{subsec:equilibrium-feasibility}

It remains to ask whether the first reports really do reveal the first
signals. A can misstate her first observation to influence how her
eventual report will be judged. B has a second reason to do so. Her
submission determines which of the peer's reports counts as agreement
and can thereby change the feedback she receives.

For each policy, let \(\widehat\sigma^\pi\) consist of truthful initial
reports, A's sealed continuation, B's message-contingent continuation, and an
optimal terminal report at every private history following an initial lie.
The construction covers every message that the fixed policy can generate
after any report history. Impossible chance branches are omitted.

Let \(V_{B\tau}^{T,\pi}\) be B's value from truthful initial reporting and
the candidate continuation. Let \(V_{B\tau}^{D,\pi}\) be her value from
lying initially and then choosing an optimal terminal report at each resulting
history. Define
\begin{equation}\label{eq:b-locked-slack}
\Delta_{B\tau}^\pi
=V_{B\tau}^{T,\pi}-V_{B\tau}^{D,\pi}.
\end{equation}
This deviation value allows the lie to change both the feedback lottery and
the subsequent reporting choice. Label symmetry gives the same constraint
after either first signal. By \cref{lem:reputational-factorization}, A's
constraint is independent of policy; denote her truthful-report slack by
\(\Delta_{A\tau}^{\mathsf S}\).

Let \(\Pi^{\mathrm C}\subseteq\Pi^{\mathrm{RS}}\) consist of policies for
which cutoff behaviour and confirmation are optimal on every supported
branch. The conditions of \cref{prop:selective-continuation} ensure
\(\Pi^{\mathrm C}=\Pi^{\mathrm{RS}}\). A policy with fewer supported branches
may satisfy weaker conditions. Set
\begin{equation}\label{eq:rm-feasible-set}
\Pi^{\mathrm{RM}}
=
\left\{
\pi\in\Pi^{\mathrm C}:
\Delta_{A\tau}^{\mathsf S}\geq0
\ \text{and}\
\Delta_{B\tau}^{\pi}\geq0
\text{ for }\tau\in\{L,H\}
\right\}.
\end{equation}
For \(\pi\in\Pi^{\mathrm{RM}}\), denote the induced equilibrium value by
\(V_D^\pi=V_D(\pi,\widehat\sigma^\pi)\). The finite deviation sums are given
in the Online Appendix.

\begin{proposition}
\label{prop:equilibrium-feasibility}
For every \(\pi\in\Pi^{\mathrm{RM}}\), the candidate \(\widehat\sigma^\pi\), together with Bayesian terminal wages and sequentially optimal continuations following locked-report deviations, is a regular-monotone sequential equilibrium. If all four locked-report slacks are strict, both types of both experts lodge their first signals strictly truthfully.
\end{proposition}

\begin{proof}
Membership in \(\Pi^{\mathrm C}\) gives optimal terminal reporting on every
equilibrium branch. The deviation-value maximisations give optimal terminal
reporting after an initial lie. Nonnegativity of the four slacks then ensures
optimality of each initial report, including deviations followed by an optimal
terminal plan.

For consistency, observe that the high type reports her two signals and their
distribution has full support. Every public history compatible with the
policy consequently has positive probability, so Bayes' rule determines its
terminal wage. This also covers public histories generated by a private
initial lie. Introduce a common type-independent tremble at every reporting
choice. The resulting strategies are completely mixed on the fixed-policy
game and their Bayesian wages converge to the candidate wages. Following a
private lie, the expert retains her actual signals and uses the policy kernel
at her submitted report. The limiting conditional probabilities therefore
coincide with those in the deviation sums. Impossible chance branches remain
absent. This establishes consistency.
\end{proof}

Within the informative class the branch likelihoods pin down the response
to every message that occurs. Thus a feasible policy has a well-defined
decision value even if off-path private choices are nonunique.

\section{How much peer feedback to disclose}\label{sec:policy-design}

If B learns from silence that her peer almost certainly disagreed, she
places less weight on her first report. How often she makes this inference
also matters for the decision maker. We first vary the frequency of
silence while holding its informational content fixed. We then ask how
much agreement must be disclosed before B revises after silence.

\subsection{Why one-peer selective feedback requires randomisation}
\label{subsec:deterministic-collapse}

Partial revelation requires B to remain uncertain about A's initial report
after some message. With binary reports, a deterministic rule that reveals
one category makes the other inferable from silence.

\begin{lemma}
\label{lem:deterministic-collapse}
With binary locked reports, every policy \(\pi=(\alpha,\delta)\in\Pi^{\mathrm{RS}}\) satisfying \(\alpha=1\) or \(\delta=1\) gives B the same interim information about A's locked report as full disclosure. Under the regular-monotone candidate, it consequently induces the same reporting behaviour and the same value for every downstream decision problem as full disclosure.
\end{lemma}

\begin{proof}
If \(\alpha=1\), silence implies disagreement. Knowing \(b\), B therefore
infers \(y=1-b\). If \(\delta=1\), silence implies agreement and hence
\(y=b\). A revealed message identifies \(y\) directly. In either case, B
knows the peer report after every message and candidate behaviour is the same
as under full disclosure. Conditional on \((y,b)\), the remaining policy
lottery is independent of the state and types. It only partitions identical
public outcomes by a state-independent randomisation, which leaves every
posterior decision value unchanged.
\end{proof}

To pool agreement and disagreement under silence, the platform must conceal
each with positive probability. A lottery is therefore necessary for partial
revelation with one peer. It need not alter silent-branch incentives: equal
concealment rates leave silence uninformative and reproduce the sealed
response.\footnote{This necessity result concerns one binary peer report and
the reveal-or-silence technology. Section~\ref{sec:two-peer-confirmation}
constructs a deterministic rule with two peers and verifies its incentives
and benchmark comparisons.}

\subsection{Concealment rays}\label{subsec:concealment-rays}

Let the concealment probabilities after agreement and disagreement be
\begin{equation}\label{eq:concealment-coordinates}
c_A=1-\alpha,
\qquad
c_D=1-\delta.
\end{equation}
When silence has positive probability,
\begin{equation}\label{eq:silence-concealment-ratio}
\lambda_N(c_A,c_D)
=
\frac{qc_A+(1-q)c_D}
 {(1-q)c_A+qc_D}.
\end{equation}
Its likelihood ratio depends only on \(c_A/c_D\). For
\((c_A,c_D)\ne(0,0)\), write
\begin{equation}\label{eq:concealment-ray}
(c_A,c_D)=t(\bar c_A,\bar c_D),
\qquad
t=\max\{c_A,c_D\},
\qquad
\max\{\bar c_A,\bar c_D\}=1.
\end{equation}
The normalised pair \((\bar c_A,\bar c_D)\) determines the composition of
the silent pool, and \(t\in(0,1]\) determines its scale. The normalised
endpoint is on \(\alpha=0\) or \(\delta=0\); the origin corresponds to full
disclosure.

We first compare the candidate assessments from
Section~\ref{subsec:equilibrium-feasibility}, allowing their initial-report
constraints to fail. Let \(\widehat P_\theta^\pi(z)\) be the candidate
probability of public outcome \(z\) conditional on \(\theta\). Define
\(\widehat V_D^\pi\) by substituting \(\widehat P_\theta^\pi\) for
\(P_\theta^{\pi,\sigma}\) in \cref{eq:decision-value}. On
\(\Pi^{\mathrm{RM}}\), the candidate is an equilibrium and
\(\widehat V_D^\pi=V_D^\pi\); elsewhere, it serves only for comparison.
Below, \(\widehat V_D(c_A,c_D)\) denotes the candidate value of policy
\((1-c_A,1-c_D)\). The reduction holds for every finite decision problem.

\begin{theorem}\label{thm:concealment-ray}
Fix primitives for which the candidate continuation is defined on \(\Pi^{\mathrm{RS}}\). For every concealment ray and every finite downstream decision problem \(D\),
\begin{equation}\label{eq:ray-affinity}
\widehat V_D(t\bar c_A,t\bar c_D)
=(1-t)\widehat V_D(0,0)
+t\widehat V_D(\bar c_A,\bar c_D),
\qquad t\in[0,1].
\end{equation}
Moreover:
\begin{enumerate}[label=(\roman*),leftmargin=*]
\item each of B's locked-report incentive slacks is affine in \(t\);
\item each of A's locked-report incentive slacks is constant in \(t\); and
\item a global candidate maximiser over \(\Pi^{\mathrm{RS}}\) can be selected from full disclosure and the two policy boundaries \(\alpha=0\) and
\(\delta=0\).
\end{enumerate}
\end{theorem}

\begin{proof}
Along a ray, \(c_A/c_D\) and therefore \(\lambda_N\) are constant. By
\cref{prop:selective-continuation}, silent-branch reporting and wages remain
unchanged. Every silent public outcome has \(t\) times its probability at
the normalised endpoint. A revealed-agreement outcome has frequency factor
\[
1-t\bar c_A=(1-t)+t(1-\bar c_A),
\]
and revealed disagreement has factor \(1-t\bar c_D\). Each outcome's
two-state likelihood vector is thus multiplied by a nonnegative scalar
affine in \(t\). Its maximised contribution to a finite decision problem
is homogeneous of degree one in this vector. Summation gives
\cref{eq:ray-affinity}.

The same factors scale B's truthful-report payoff. Following an initial lie,
each private-history and message payoff vector has one of these factors.
Its relative state weights remain fixed, so the optimal terminal choice does
not change with \(t\). The deviation value and initial-report slack are
therefore affine as well. A's slack is independent of policy by
\cref{lem:reputational-factorization}. Finally, an affine function attains
its ray maximum at the origin or the normalised endpoint. Taking the union
over rays proves the boundary reduction.
\end{proof}

Feasibility must still be checked. Because initial-report constraints are
affine in \(t\), a ray intersects \(\Pi^{\mathrm{RM}}\) in an interval,
possibly empty. If a constraint binds before the policy boundary, the feasible
endpoint is inside the square. The candidate boundary reduction does not
permit replacing it with an infeasible policy.

\begin{corollary}
\label{cor:candidate-to-equilibrium}
Let \(\pi^*\in\Pi^{\mathrm{RM}}\). If
\[
\widehat V_D^{\pi^*}\geq\widehat V_D^\pi
\qquad\text{for every }\pi\in\Pi^{\mathrm{RS}},
\]
then \(\pi^*\) maximises the regular-monotone equilibrium value over \(\Pi^{\mathrm{RM}}\). If the candidate inequality is strict away from \(\pi^*\), the regular-monotone maximiser is unique.
\end{corollary}

\begin{proof}
On \(\Pi^{\mathrm{RM}}\), candidate values are the equilibrium values.
\end{proof}

\subsection{Disciplinary silence}\label{subsec:disciplinary-silence}

Consider the boundary \(\delta=0\). Increasing \(\alpha\) removes
agreements from the concealed reports and lowers the support that silence
provides for B's initial forecast. The silent likelihood ratio is
\begin{equation}\label{eq:disciplinary-likelihood}
\lambda(\alpha)
=
\frac{1-q\alpha}{1-(1-q)\alpha}.
\end{equation}
It decreases continuously from one at sealing to \(1/\gamma\) at
\(\alpha=1\). Under the continuation conditions,
\(1/\gamma<\bar\lambda<1\). Hence there is a unique interior disclosure
probability at which silence reaches the low type's revision cutoff:
\begin{equation}\label{eq:disciplinary-threshold}
\alpha^{\mathrm{DS}}
=
\frac{1-\bar\lambda}
 {q-(1-q)\bar\lambda}
=
\frac{1-\eta_LQ_0}
 {q-(1-q)\eta_LQ_0}.
\end{equation}

At \(\alpha^{\mathrm{DS}}\), concealed agreements are sufficiently rare
that the low type follows conflicting new information after silence.
Disagreement is never revealed, but B cannot distinguish it from an agreement
that was withheld. Beyond this threshold, further agreement disclosure leaves
the silent response unchanged; its value depends on the downstream objective.

\begin{proposition}
\label{prop:disciplinary-silence}
Suppose the conditions of \cref{prop:selective-continuation} hold and \(\pi^{\mathrm{DS}}=(\alpha^{\mathrm{DS}},0)\in\Pi^{\mathrm{RM}}\). Then:
\begin{enumerate}[label=(\roman*),leftmargin=*]
\item silence satisfies \(\lambda_N=\bar\lambda\);
\item after silence, the low type is indifferent and uses the second signal, while the high type strictly uses the second signal at a conflict;
\item after revealed agreement, the low type retains the locked report with probability \(h^{\mathsf F}\) at a conflict, while the high type uses the second signal;
\item all confirmation actions are strict; and
\item together with truthful locked reports, these strategies form a regular-monotone sequential equilibrium.
\end{enumerate}
\end{proposition}

\begin{proof}
Substitution of \(\alpha^{\mathrm{DS}}\) into
\cref{eq:disciplinary-likelihood} gives \(\lambda_N=\bar\lambda\).
By \cref{lem:message-cutoff}, \(h_N=0\) is weakly optimal for the low type
and the high type strictly revises. Revealed-agreement behaviour and
confirmation follow from \cref{prop:selective-continuation}. Membership in
\(\Pi^{\mathrm{RM}}\) ensures the initial-report constraints.
\end{proof}

At the threshold wages the low type would accept either report after a
silent conflict. Why, then, does she always revise? If she repeated with
positive probability, the market would use a different wage schedule. For
\(0<h<h^\dagger\), the higher ratio \(Q(h)>Q_0\) makes revision strictly
preferable, and for \(h\geq h^\dagger\) repetition loses in both states
(weakly in the correct state at \(h=h^\dagger\)). Zero persistence is
therefore the only rate consistent with the low type's best response.
Section~\ref{subsec:strict-revision-approximation} asks what it costs the
platform to give her a strict incentive at the prevailing wages.

For a reduced-form comparison, set \(r=\bar\lambda\) and
\(D_\alpha=q-(1-q)r\). Holding the other reduced-form object fixed,
\begin{equation}\label{eq:threshold-comparative-statics}
\frac{\partial\alpha^{\mathrm{DS}}}{\partial q}
=-\frac{1-r^2}{D_\alpha^2}<0,
\qquad
\frac{\partial\alpha^{\mathrm{DS}}}{\partial r}
=\frac{1-2q}{D_\alpha^2}<0.
\end{equation}
A more precise peer report or a higher low-type revision cutoff reduces the
agreement disclosure needed for disciplinary silence. Since \(q\) and \(r\)
both depend on the primitives, these derivatives are reduced-form comparisons.

\subsection{An explicit feasible region}
\label{subsec:explicit-feasible-region}

Suppose the low type's second signal is at least as precise as her first.
Along the agreement-disclosure segment ending at disciplinary silence, the
optimal continuation after an initial lie is to follow the second signal.
This gives a lower bound on the gain from initial truth-telling without
solving for the mixing probabilities.

\begin{proposition}
\label{prop:analytical-initial-incentives}
Suppose
\begin{equation}\label{eq:explicit-feasible-region}
\eta_H>\eta_L\geq1,
\qquad
\frac1\gamma<\bar\lambda=\eta_LQ_0<1.
\end{equation}
For every \(\alpha\in[0,\alpha^{\mathrm{DS}}]\), the policy
\((\alpha,0)\) belongs to \(\Pi^{\mathrm{RM}}\). Both types of both experts
report their first signals strictly truthfully. More precisely,
\begin{equation}\label{eq:analytical-initial-lower-bound}
\Delta_{i\tau}^{(\alpha,0)}
\geq(2p_{\tau1}-1)\{W_{CI}(0)-W_{II}(0)\}>0,
\qquad i\in\{A,B\},\quad\tau\in\{L,H\},
\end{equation}
where A's slack is understood to be policy-independent. The lower bound is
uniform over the stated policy segment.
\end{proposition}

\begin{proof}
On this segment, every supported branch satisfies
\(\lambda G_C(h)=\eta_LG_I(h)\). For A this is the sealed branch; for B it
is revealed agreement or silence, including \(h_N=0\) at the disciplinary
endpoint. Conflict choices follow from the cutoff, and confirmation is
strict since \(\eta_L\kappa_{\tau1}\kappa_{\tau2}>1\).

After an initial lie, the odds in favour of the submitted report are
\(\lambda\eta_\tau\) when the second signal agrees with it, and
\(\lambda/(\kappa_{\tau1}\kappa_{\tau2})\) otherwise. Following the second
signal is optimal in both cases because the relevant inequalities reduce to
\(\eta_L\eta_\tau\geq1\) and
\(\eta_L<\kappa_{\tau1}\kappa_{\tau2}\). Under truthful reporting, it also
gives the candidate payoff: the low type mixes only at indifference.

Let \(c_j\) and \(d_j\) be the probabilities of branch \(j\), conditional
on the submitted report being correct and incorrect. Their ratio is
\(\lambda_j\). The truthful-report slack equals
\begin{align*}
\Delta_{i\tau}
=(2p_{\tau1}-1)\bigg[&\sum_j
 \{c_jW_{CI}(h_j)-d_jW_{II}(h_j)\}\\
 &+p_{\tau2}\sum_j\{c_jG_C(h_j)-d_jG_I(h_j)\}\bigg].
\end{align*}
Branch indifference and \(\eta_L\geq1\) make the second sum nonnegative.
The first is bounded below by \(W_{CI}(0)-W_{II}(0)\), since \(W_{CI}\)
increases and \(W_{II}\) decreases with persistence, and both collections
of branch probabilities sum to one. This wage difference is strictly positive
under \(p_{H1}>p_{L1}\). The bound, together with the consistency argument
of \cref{prop:equilibrium-feasibility}, establishes equilibrium.
Section~OA.2 provides the full decomposition.
\end{proof}

The wage difference in \cref{eq:analytical-initial-lower-bound} has a
simple interpretation. Compare two experts whose final forecast was
wrong. The market pays more to the one whose initial forecast was right.
A first signal better than a coin toss makes this premium valuable even
when the expert anticipates choosing the best possible continuation after
a lie. Revealed disagreement has zero probability on this policy segment,
so confirmation after that message need not be checked.

One further condition ensures truthful initial reporting throughout the square.
\begin{corollary}
\label{cor:whole-square-feasibility}
Under \cref{eq:explicit-feasible-region}, if
\begin{equation}\label{eq:whole-square-region}
\gamma Q_0\leq1,
\end{equation}
then \(\Pi^{\mathrm{RM}}=\Pi^{\mathrm{RS}}\). The lower bound in
\cref{eq:analytical-initial-lower-bound} holds for every policy in the square.
\end{corollary}

\begin{proof}
Every message likelihood ratio is at least \(1/\gamma\geq Q_0\).
The continuation conditions in \cref{prop:selective-continuation} follow,
including the upper bound because
\(\gamma Q_0\leq1<\kappa_{L1}\kappa_{L2}\). The previous proof applies
on mixing branches. On a pure-revision branch,
\(Q_0\leq\lambda\leq\eta_LQ_0\); these inequalities make following the
second signal optimal after a lie and imply
\(c_jG_C(0)-d_jG_I(0)=d_jG_C(0)(\lambda-Q_0)\geq0\).
Applying the same payoff decomposition gives the result.
\end{proof}

These conditions use only the primitives and wages at \(h=0\).
No equilibrium mixing root is needed to check them. With \(\eta_L>1\),
\cref{eq:explicit-feasible-region} defines an open region, extending beyond
the benchmark of Section~OA.5.

\begin{corollary}
\label{cor:region-nonempty-share}
Fix a signal technology with \(\eta_H>\eta_L>1\) and
\(p_{L2}<p_{H1}\). There is a nonempty interval of type shares
\(\omega\in(0,1)\) satisfying \cref{eq:explicit-feasible-region}.
Within this region, \(\alpha^{\mathrm{DS}}\) decreases strictly with
\(\omega\).
\end{corollary}

\begin{proof}
For a fixed signal technology, \(Q_0\) increases strictly with \(\omega\),
from \(\kappa_{L1}/\kappa_{H1}\) to \(\kappa_{H1}/\kappa_{L1}\).
Section~OA.2 derives this from the pure-revision likelihoods. Consequently,
\(\bar\lambda=\eta_LQ_0\) increases from
\(\kappa_{L2}/\kappa_{H1}<1\) to a value above one. At its unique crossing
of one, \(\gamma\bar\lambda>1\), so both strict inequalities in
\cref{eq:explicit-feasible-region} hold just below the crossing. Finally,
\(q\) and \(\bar\lambda\) increase with \(\omega\). The two negative
derivatives in \cref{eq:threshold-comparative-statics} give the comparative
static.
\end{proof}

For example, fix the precisions at
\((p_{L1},p_{L2},p_{H1},p_{H2})=(11/20,5/8,33/40,19/20)\).
Every \(\omega\in[1/3,1/2]\) satisfies
\cref{eq:explicit-feasible-region}, and every \(\omega\in[1/3,2/5]\) also
satisfies \cref{eq:whole-square-region}. Monotonicity reduces verification
to three rational endpoint inequalities in Section~OA.2.

\section{Optimal selective feedback}
\label{sec:exact-results}

What agreement probability should the platform choose? If it withholds
every disagreement, revealing additional agreements makes silence less
favourable to the first forecast. This helps the decision maker only so long
as the recipient still repeats after silence. Each newly disclosed
agreement also makes B more inclined to persist when she sees it. The
optimum balances these two effects and must improve on the policies that
sometimes disclose disagreement.

After observing the complete public outcome, a Bayesian decision maker guesses
the state. Candidate accuracy is
\begin{equation}\label{eq:classification-accuracy}
\widehat{\mathcal A}^{\pi}
=
\frac12\sum_z
\max\{\widehat P_0^\pi(z),\widehat P_1^\pi(z)\}.
\end{equation}
On \(\Pi^{\mathrm{RM}}\), denote equilibrium accuracy by \(\mathcal A^\pi\).

\subsection{Analytical global optimality}
\label{subsec:analytical-global-optimality}

Group public histories by initial agreement or disagreement. Write
\[
 A_h=D_{CC}(h),\quad B_h=D_{CI}(h),\quad
 C_h=D_{IC}(h),\quad D_h=D_{II}(h)
\]
for unconditional correctness-pattern probabilities, with
\(D_\chi(h)=\omega\ell_H^\chi(h)+(1-\omega)\ell_L^\chi(h)\).
Let \(A=A_0\), \(B=B_0\), \(C=C_0\), \(D=D_0\), and
\[
 \ell=(1-\omega)p_{L1}(1-p_{L2}),\qquad
 m=(1-\omega)(1-p_{L1})p_{L2}.
\]
Thus \((A_h,B_h,C_h,D_h)=(A+\ell h,B-\ell h,C-mh,D+mh)\).
In this subsection write \(s=h^{\mathsf S}\), \(f=h^{\mathsf F}\), and
\(r(h)=\eta_LQ(h)\), so \(r(s)=1\) and \(r(f)=\gamma\). Define
\begin{align}
 J&=mC_s-\ell B_s,& K&=mA-\ell D,& L&=\ell C-mB,
 \label{eq:classification-costs}\\
 g&=(J+K)s-Jf\alpha^{\mathrm{DS}},\notag\\
 M&=(2-q)g+(1-q)(J-L)(f-s)-Ks.
 \label{eq:classification-comparison-margins}
\end{align}
The signs are \(J>L>0\) and \(K>0\) when \(\eta_H>\eta_L>1\). To
interpret them, count correct decisions separately for histories in which
the first reports agree and histories in which they disagree. In the
decision regions below, raising B's persistence by one unit costs \(J\) on
the agreement histories. On the disagreement histories the cost is \(K\)
up to the sealed persistence rate \(s\). Above \(s\) the decision maker
changes how she treats some disagreements, and the marginal contribution
becomes \(+L\). Thus \((J+K)s\) is what the platform would recover by
eliminating persistence under sealing. It forgoes
\(Jf\alpha^{\mathrm{DS}}\) on agreements it actually discloses. Their
difference \(g\) is the net gain. The last inequality involving \(M\)
below ensures that a policy disclosing disagreements instead cannot provide
a higher value.

\begin{theorem}
\label{thm:analytical-global-optimality}
Suppose the conditions of \cref{cor:whole-square-feasibility} hold, with
\(\eta_L>1\). Assume in addition
\begin{align}
 \gamma&\geq\eta_L,& A_sB_f&>D_sC_f,& r''(0)&\geq0,
 \label{eq:analytical-optimality-regions}\\
 K&\geq\gamma J,& g&>0,& M&\geq0.
 \label{eq:analytical-optimality-comparisons}
\end{align}
Then \(\pi^{\mathrm{DS}}=(\alpha^{\mathrm{DS}},0)\) uniquely maximises
classification accuracy over \(\Pi^{\mathrm{RS}}=\Pi^{\mathrm{RM}}\).
It strictly outperforms sealing and full disclosure, with gains
\begin{equation}\label{eq:analytical-benchmark-gains}
 \mathcal A^{\mathrm{DS}}-\mathcal A^{\mathsf S}=g,
 \qquad
 \mathcal A^{\mathrm{DS}}-\mathcal A^{\mathsf F}
 =Jf(1-\alpha^{\mathrm{DS}})>0.
\end{equation}
All four initial-report incentives satisfy the positive bound in
\cref{eq:analytical-initial-lower-bound}. At the optimum, the low type of B
is indifferent after a silent conflict at the equilibrium wages; the unique
consistent persistence rate in the maintained class is zero.
\end{theorem}

\begin{proof}
Section~OA.4 shows that \(r''(0)\geq0\) implies convexity of \(r\) on
\([0,f]\). Under \cref{eq:analytical-optimality-regions}, agreement histories
contribute \(G(h)=G(0)-Jh\). Disagreement histories contribute
\(H(h)=H(0)-Kh\) for \(h\leq s\) and \(H(h)=H(s)+L(h-s)\) for
\(h\geq s\). Using the complete public history, including the message,
accuracy is therefore
\[
 \alpha G(f)+\delta H(0)+(1-\alpha)G(h_N)+(1-\delta)H(h_N).
\]

On \(\delta=0\), accuracy decreases strictly beyond the disciplinary
threshold. Below the threshold, substitute
\(\alpha=(1-r(h))/(q-(1-q)r(h))\). Convexity implies
\((r(h)-r(0))/h\leq(1-r(0))/s\). Together with \(K\geq\gamma J\), this gives
\[
 \mathcal A^{\mathrm{DS}}-\mathcal A^{(\alpha,0)}
 \geq\frac{h(2q-1)g}{s\{q-(1-q)r(h)\}}>0,
 \qquad 0<h\leq s.
\]

On \(\alpha=0\), set \(t=(h_N-s)/(f-s)\) and \(F=f-s\).
Convexity yields \(\delta\leq t/(1-q+qt)\), and
\[
 \mathcal A^{(0,\delta)}
 =\mathcal A^{\mathsf S}-(J-L)Ft+\delta(Ks-LFt).
\]
If the last bracket is nonpositive, accuracy is below the disciplinary value.
If it is positive, apply the bound on \(\delta\). After multiplication by
\(1-q+qt\), the value difference is bounded below by
\[
 (1-q)g(1-t)^2+Mt(1-t)
   +Jf(1-\alpha^{\mathrm{DS}})t^2>0.
\]
Every other point on both boundaries is therefore strictly inferior.
The ray identity extends the comparison to the whole square, and
\cref{cor:whole-square-feasibility} gives feasibility and strict initial
truth-telling.
\end{proof}

There are two steps to the optimality argument. Along the agreement
boundary the comparison turns on the benefit from revision after silence
and the cost on the revealed branch. Secant bounds for \(r\) extend a
strict gain over sealing to every intermediate policy below the threshold.
Along the other boundary, disclosure of disagreements makes the concealed
pool more favourable to B's first report. The comparison there uses the
change in the decision maker's action at \(h=s\). This is why the
likelihood ratio at the silent branch, by itself, cannot determine the
optimal policy.

The sufficient conditions involve the unique continuation roots under
sealing and revealed agreement. The following family verifies them
directly in terms of signal precisions and the share of high types.

\begin{corollary}
\label{cor:analytical-optimality-interval}
Fix
\[
 (p_{L1},p_{L2},p_{H1},p_{H2})
 =\left(\frac{11}{20},\frac58,\frac{33}{40},\frac{19}{20}\right).
\]
For every \(\omega\in[3/8,2/5]\), disciplinary silence is the unique
classification-optimal reveal-or-silence policy. All policies in the square
are equilibrium-feasible. The optimal agreement-disclosure probability
decreases strictly with \(\omega\).
\end{corollary}

\begin{proof}
Feasibility and the comparative static follow from
\cref{cor:whole-square-feasibility,cor:region-nonempty-share}.
Uniformly on the stated interval, Section~OA.4 establishes
\[
 \frac{11}{50}<s<\frac14,\qquad
 \frac{53}{100}<f<\frac{11}{20},
\]
and verifies every inequality of \cref{thm:analytical-global-optimality}
strictly. Four low-degree polynomial identities give the root bounds, after
which rational inequalities complete the verification. The argument requires
neither a numerical root solution nor a search over policies.
\end{proof}

The larger share intervals in Section~\ref{subsec:explicit-feasible-region}
ensure that truthful initial reports are feasible. The smaller interval
above also satisfies the comparisons of decision values.

\subsection{Strict revision near the optimum}
\label{subsec:strict-revision-approximation}

The optimum gives B's low type a weak incentive to revise after silence.
A platform that wants her to prefer revision strictly can disclose
agreements slightly more often. At those policies, silence carries still
worse news for the first forecast. The silent branch already induces full
revision, so the cost to the platform comes entirely from persistence on
the extra agreements it reveals.

Let \(\alpha_0=\alpha^{\mathrm{DS}}\), \(r_0=\bar\lambda\), and
\(d_0=1-(1-q)\alpha_0\). For \(0<\varepsilon\leq1-\alpha_0\), consider
\begin{equation}\label{eq:strict-revision-policy}
 \pi_\varepsilon=(\alpha_0+\varepsilon,0),\qquad
 \lambda_\varepsilon
 =\frac{1-q(\alpha_0+\varepsilon)}{1-(1-q)(\alpha_0+\varepsilon)}.
\end{equation}

\begin{proposition}
\label{prop:strict-silent-revision}
Suppose the conditions of \cref{cor:whole-square-feasibility} hold. Every
policy in \cref{eq:strict-revision-policy} is equilibrium-feasible, and both
types of B strictly follow their second signal after a silent conflict.
Conditional on such a conflict, the low type's expected wage gain from revision
rather than repetition is
\begin{equation}\label{eq:strict-revision-premium}
 R_L(\varepsilon)
 =G_C(0)\frac{r_0-\lambda_\varepsilon}{\eta_L+\lambda_\varepsilon}
 \geq
 \frac{G_C(0)(2q-1)}{(\eta_L+r_0)d_0^2}\,\varepsilon>0.
\end{equation}
All four initial-report incentives retain the policy-uniform bound
\begin{equation}\label{eq:strict-revision-initial-bound}
 \Delta_{i\tau}^{\pi_\varepsilon}
 \geq(2p_{\tau1}-1)\{W_{CI}(0)-W_{II}(0)\}>0.
\end{equation}
All confirmation actions remain strict. At \(\varepsilon=0\), the low type
is indifferent after a silent conflict, but the only consistent
regular-monotone persistence rate on that branch is zero.
\end{proposition}

\begin{proof}
Increasing agreement disclosure lowers the silent likelihood ratio, with
\[
 r_0-\lambda_\varepsilon
 =\frac{(2q-1)\varepsilon}
 {d_0\{d_0-(1-q)\varepsilon\}}.
\]
Hence \(h_N=0\) and silent wages are \(W_\chi(0)\). The conflict payoff
comparison gives \cref{eq:strict-revision-premium}; bounding its denominator
by \((\eta_L+r_0)d_0^2\) yields the lower bound. The high type strictly
revises under \(\eta_H>\eta_L\). Feasibility, strict confirmation and
\cref{eq:strict-revision-initial-bound} follow from
\cref{cor:whole-square-feasibility}.

At the threshold, any positive persistence below \(h^\dagger\) raises
\(Q(h)\) above \(Q_0\), making repetition strictly worse. At or above
\(h^\dagger\), the inequality \(G_C(h)\leq0<G_I(h)\) gives the same
conclusion. Thus only zero persistence is consistent with indifference at
\(W_\chi(0)\).
\end{proof}

The strict incentive concerns a private conflict following silence.
After revealed agreement, B's low type still randomises, as does A's low
type when her own signals conflict.

\begin{proposition}
\label{prop:quantified-strict-approximation}
Under the conditions of \cref{thm:analytical-global-optimality}, every
\(0<\varepsilon\leq1-\alpha_0\) satisfies
\begin{align}
 \mathcal A^{\mathrm{DS}}-\mathcal A^{\pi_\varepsilon}
 &=Jf\varepsilon,\label{eq:strict-revision-exact-loss}\\
 \mathcal A^{\pi_\varepsilon}-\mathcal A^{\mathsf S}
 &=g-Jf\varepsilon,\notag\\
 \mathcal A^{\pi_\varepsilon}-\mathcal A^{\mathsf F}
 &=Jf(1-\alpha_0-\varepsilon).
 \label{eq:strict-revision-benchmark-gains}
\end{align}
In particular, the strict-revision policy beats both polar protocols whenever
\[
 0<\varepsilon<\min\{1-\alpha_0,\ g/(Jf)\}.
\]
For any accuracy tolerance \(\zeta>0\), the choice
\begin{equation}\label{eq:strict-revision-tolerance}
 \varepsilon_\zeta
 =\frac12\min\left\{1-\alpha_0,\frac g{Jf},\frac\zeta{Jf}\right\}
\end{equation}
has loss below \(\zeta\), beats both benchmarks, and satisfies
\cref{prop:strict-silent-revision}.

If admissible policies are required to have positive-probability silence and
strict low-type revision after a silent conflict, classification value has
supremum \(\mathcal A^{\mathrm{DS}}\), but no maximiser in that restricted
policy set.
\end{proposition}

\begin{proof}
Throughout the segment, \(h_N=0\) and persistence after revealed agreement
is \(f\). The decomposition in the proof of
\cref{thm:analytical-global-optimality} is therefore affine in \(\alpha\)
with slope \(G(f)-G(0)=-Jf\). This gives the three equalities and the
tolerance statement.

The unique optimum on the square fails the strict silent-revision requirement.
Each \(\pi_\varepsilon\) satisfies that requirement and has positive-probability
silence. Its value tends to the optimum as \(\varepsilon\downarrow0\).
The restricted problem thus has the same supremum, which no admissible
policy attains.
\end{proof}

As the policy approaches the optimum, both the cost in accuracy and B's
strict gain from revision go to zero. A platform requiring a fixed wage
margin must therefore give up some accuracy on this segment.

\begin{corollary}
\label{cor:uniform-strict-revision}
On the primitive interval of \cref{cor:analytical-optimality-interval}, every
policy with \(\alpha_0<\alpha<1\) and \(\delta=0\) makes silent-conflict
revision strict and beats both sealing and full disclosure. Uniformly over
that primitive interval, every \(0<\varepsilon\leq1/10\) is admissible and
satisfies
\begin{equation}\label{eq:uniform-strict-revision-bounds}
 R_L(\varepsilon)>\frac\varepsilon{10},\qquad
 \Delta_{i\tau}^{\pi_\varepsilon}>\frac3{400},\qquad
 \mathcal A^{\mathrm{DS}}-\mathcal A^{\pi_\varepsilon}
 <\frac{5599}{500000}\,\varepsilon.
\end{equation}
\end{corollary}

\begin{proof}
Section~OA.4 gives \(\alpha_0<783/1000\), \(g>1/100\), and
\(Jf<5599/500000\). Hence \(g>Jf(1-\alpha_0)\), implying that full
disclosure already improves on sealing throughout the interval. Every policy
strictly between the threshold and full agreement disclosure then improves
on both. The appendix derives the stated revision and initial-report margins
from the primitives.
\end{proof}

\subsection{Open-set robustness}
\label{subsec:robustness-sensitivity}

\begin{proposition}
\label{prop:open-set-robustness}
Let
\[
\xi=(\omega,p_{L1},p_{L2},p_{H1},p_{H2})
\]
collect the primitives and set
\(\xi^0=(2/5,11/20,5/8,33/40,19/20)\). There is an open neighbourhood \(U\) of \(\xi^0\), relative to the maintained parameter region, such that for every \(\xi\in U\), the recalibrated disciplinary policy
\[
\pi^{\mathrm{DS}}(\xi)
=
(\alpha^{\mathrm{DS}}(\xi),0)
\]
belongs to \(\Pi^{\mathrm{RM}}\), makes all four locked reports strictly truthful, and is the unique global maximiser of candidate classification accuracy over \(\Pi^{\mathrm{RS}}\). It is consequently the unique regular-monotone equilibrium maximiser over \(\Pi^{\mathrm{RM}}\) and strictly outperforms sealing and full disclosure.
\end{proposition}

\begin{proof}
Section~OA.4 verifies the inequalities of
\cref{thm:analytical-global-optimality} strictly at the benchmark. Since
\(Q'(h)>0\), continuation roots vary continuously with the primitives.
The inequalities therefore persist on an open neighbourhood, where the
optimality theorem applies pointwise. Its initial-report bound and gain
formulas establish the other conclusions.
\end{proof}

The proof covers simultaneous changes in all five primitives near the
benchmark. Section~OA.5 reports the levels of the optimal disclosure
probability and its gains at the benchmark itself.

\subsection{When the decision maker's losses change}
\label{subsec:blackwell}

A decision maker who faces a large loss from overlooking state one will
sometimes take action one even when she thinks state zero more likely.
Does the policy that maximises classification accuracy also serve her
interests? Each policy generates an experiment in the sense of
\citet{Blackwell1953}, since the decision maker sees the four reports and
the message. Changing her payoffs leaves the experts' incentives and hence
these experiments unchanged.

For \(k\in(0,1)\), define the posterior call value
\begin{equation}\label{eq:posterior-call}
C^\pi(k)
=
\frac12\sum_z
\left[(1-k)P_1^\pi(z)-kP_0^\pi(z)\right]_+.
\end{equation}
It is the value of a binary decision problem in which action \(1\) pays
\(1-k\) in state \(1\) and \(-k\) in state \(0\), while action \(0\)
pays zero. At \(k=1/2\),
\[
\mathcal A^\pi=\frac12+2C^\pi(1/2).
\]

\begin{proposition}
\label{prop:blackwell-incomparability}
At \((p_{L1},p_{L2},p_{H1},p_{H2})=(11/20,5/8,33/40,19/20)\)
and every \(\omega\in[3/8,2/5]\), the complete regular-monotone
equilibrium experiments generated by sealing and full disclosure are
Blackwell incomparable. Uniformly on this interval,
\[
 C^{\mathsf F}(1/10)-C^{\mathsf S}(1/10)<-\frac{19}{10000},
 \qquad
 C^{\mathsf F}(1/2)-C^{\mathsf S}(1/2)>\frac3{1000}.
\]
Moreover, sealing gives a higher value at cutoff \(1/10\) than the unique
classification optimum,
\[
 C^{\mathsf S}(1/10)-C^{\mathrm{DS}}(1/10)>\frac1{1250}.
\]
Hence the classification-optimal policy ceases to maximise expected payoff
when the decision maker uses this lower cutoff.
\end{proposition}

\begin{proof}
Set \(s=h(1)\) and \(f=h(\gamma)\), and use the pattern masses
\((A_h,B_h,C_h,D_h)=(A+\ell h,B-\ell h,C-mh,D+mh)\).
At cutoff \(1/10\), action zero is chosen only after
\((r_{A1},r_{A2},r_{B1},r_{B2})=(0,0,0,0)\) under either protocol.
Section~OA.10 verifies these decision regions throughout the primitive
interval. Always choosing action one gives expected payoff \(2/5\), so
\[
 C^{\mathsf S}(1/10)=\frac25+\frac{A_s^2-9D_s^2}{20},\qquad
 C^{\mathsf F}(1/10)=\frac25+\frac{A_sA_f-9D_sD_f}{20}.
\]
It follows that
\begin{equation}\label{eq:analytical-call-difference}
 C^{\mathsf F}(1/10)-C^{\mathsf S}(1/10)
 =-\frac{f-s}{20}(9mD_s-\ell A_s).
\end{equation}
The root bounds give \(f-s>7/25\), and the appendix establishes
\(9mD_s-\ell A_s>7/50\). Thus the difference is below
\(-49/25000<-19/10000\).

At \(1/2\), the call difference equals half the accuracy difference. Using
\(g>1/100\), \(Jf<5599/500000\), and \(\alpha_0>71/100\), we obtain
\[
 \mathcal A^{\mathsf F}-\mathcal A^{\mathsf S}
 =g-Jf(1-\alpha_0)
 >\frac1{100}-\frac{5599}{500000}\frac{29}{100}
 >\frac3{500}.
\]
The two rankings rule out Blackwell dominance in either direction.

For the selective optimum, Section~OA.10 verifies that the decision at
cutoff \(1/10\) again changes only at \(0000\). Conditional on initial
agreement and a private conflict, B's low type persists at rate \(f\)
when the agreement is disclosed and at rate zero in silence. Her average
persistence on these agreement histories is therefore \(\alpha_0 f\).
Under sealing it is \(s\). Taking the difference of the two call values gives
\begin{equation}\label{eq:selective-objective-switch}
 C^{\mathsf S}(1/10)-C^{\mathrm{DS}}(1/10)
 =\frac{(\alpha_0 f-s)(9mD_s-\ell A_s)}{20}.
\end{equation}
The first factor exceeds \((71/100)(53/100)-1/4\), and the second exceeds
\(7/50\). Their product divided by 20 exceeds \(1/1250\).
\end{proof}

The asymmetry in the decision maker's losses explains the reversal. At
cutoff \(1/10\), she ordinarily takes action one. She switches to zero
only if both experts report zero twice. When the state is zero, persistence
makes that history more frequent and saves her the cost of acting. When
the state is one, the same change makes her forgo a valuable action.
The second effect carries nine times the weight of the first, and the
inequality \(9mD_s>\ell A_s\) makes it decisive. Thus even though the
selective policy gathers more useful evidence for symmetric
classification, a decision maker facing these asymmetric losses would
prefer to seal the forecasts. This comparison suffices to establish a
change in the optimum and leaves open which rule maximises the new
objective.

\section{Deterministic confirmation from two peers}
\label{sec:two-peer-confirmation}

Suppose B initially forecasts that the event will occur. With one peer, a
rule that sends a message precisely when the peer agrees tells B the peer's
forecast in either case. Now give B two peers. If the organiser sends a
message only when both forecast the event, B can receive no message either
because the peers disagree with one another or because both disagree with
her. The organiser has some control over what B learns even when the rule
is entirely deterministic.

Let the three experts be \(A_1,A_2,B\), with the types, signal precisions
and conditional independence assumptions of Section~\ref{sec:environment}.
Each expert submits a first report, receives a second signal and submits
a final report. Neither peer receives feedback. Writing \(b\) for B's first
report, the organiser sends B

\begin{equation}\label{eq:two-peer-rule}
 M=\begin{cases}
 U,&r_{A_1,1}=r_{A_2,1}=b,\\
 N,&\text{otherwise}.
 \end{cases}
\end{equation}
before she submits her final report. Thus B receives confirmation for a
unanimous forecast in either direction. After the final reports the
decision maker sees all six forecasts and B's message. Evaluation, as
before, takes place once the state is known. We compare this rule,
\(\mathsf U\), with two alternatives for the same three forecasters. Under
\(\mathsf S_2\) each expert revises privately. Under \(\mathsf F_2\) B sees
both peers' first forecasts before revising.

A truthful peer agrees with B's first report with probability \(q\) if that
report is correct and \(1-q\) if it is wrong. Conditional independence
therefore gives the two likelihood ratios

\begin{equation}\label{eq:two-peer-likelihoods}
 \lambda_U=\gamma^2,\qquad
 \lambda_N^{(2)}=\frac{1-q^2}{1-(1-q)^2}.
\end{equation}
Confirmation now supplies two pieces of favourable evidence. Silence
counts against B's initial forecast because unanimous confirmation is more
likely when that forecast is correct. For the two peers, let
\(s=h^{\mathsf S}\) be the low type's probability of repeating her initial
report when her own signals conflict. For B, write
\(h_2=h(\gamma^2)\) for that probability after unanimous confirmation.

\begin{proposition}
\label{prop:two-peer-feasibility}
Suppose
\begin{equation}\label{eq:two-peer-feasible-region}
 \eta_H>\eta_L\geq1,\qquad
 Q_0\leq\lambda_N^{(2)}<\eta_LQ_0<1.
\end{equation}
Protocol \(\mathsf U\) has a regular-monotone sequential equilibrium in
which all three experts report their first signals strictly truthfully. Both
peers use the sealed continuation. B's low type persists with probability
\(h_2\) after a conflict and unanimous confirmation, and both types strictly
revise after a silent conflict. Confirmation actions are strict. All six
initial-report constraints satisfy
\begin{equation}\label{eq:two-peer-initial-bound}
 \Delta_{i\tau}^{\mathsf U}
 \geq(2p_{\tau1}-1)\{W_{CI}(0)-W_{II}(0)\}>0,
 \qquad i\in\{A_1,A_2,B\},\quad\tau\in\{L,H\}.
\end{equation}
\end{proposition}

\begin{proof}
Fix the state and the submitted forecasts. The likelihood factors for
the other experts cancel from each peer's posterior type odds. For B
the same cancellation holds conditional on her message. Each peer thus
faces the sealed wage schedule \(W_\chi(s)\). B faces \(W_\chi(h_2)\)
following confirmation and \(W_\chi(0)\) following silence. The strict
inequality \(\lambda_N^{(2)}<\eta_L Q_0\) gives strict revision even for
the low type on the silent branch.

An expert who changes her initial report can also change her final one.
On a mixing branch \(\lambda G_C=\eta_LG_I\), and on the silent branch
\(Q_0\leq\lambda<\eta_LQ_0\). In each case an expert who lied initially
can do no better than follow her second signal. Applying the decomposition
of Proposition~\ref{prop:analytical-initial-incentives} to these branches
gives \eqref{eq:two-peer-initial-bound}. Section~OA.8 derives the branch
probabilities conditional on the report actually submitted.

\end{proof}

To see who benefits from pooling the two kinds of disagreement, consider a
case in which the two peers initially disagree with one another. If B sees
both reports, their likelihood ratios cancel and she faces the sealed
reporting incentive. If she receives only silence, she must allow for the
possibility that both peers disagree with her. She then follows her second
signal strictly when it conflicts with her first. The next theorem gives the
value of this change to a decision maker who eventually sees the full record.

Write \(\mathcal A_2^{\mathsf U},\mathcal A_2^{\mathsf S},
\mathcal A_2^{\mathsf F}\) for the three classification probabilities. Using
the report-pattern probabilities in Section~\ref{subsec:analytical-global-optimality}, set

\begin{equation}\label{eq:two-peer-comparison-coefficients}
 \Xi=A_sD_s+B_sC_s,\qquad
 J_2=mC_s^2-\ell B_s^2,\qquad
 K_2=mA_sB_s-\ell C_sD_s.
\end{equation}

\begin{theorem}
\label{thm:two-peer-deterministic-gains}
Fix
\[
 (p_{L1},p_{L2},p_{H1},p_{H2})
 =\left(\frac{11}{20},\frac58,\frac{33}{40},\frac{19}{20}\right),
 \qquad \omega\in\left[\frac38,\frac25\right].
\]
Each of \(\mathsf U,\mathsf S_2,\mathsf F_2\) has a regular-monotone
sequential equilibrium with all six initial-report margins greater than
\(3/400\). Under \(\mathsf U\), the low type's conditional wage gain
from revision after a silent conflict is greater than \(1/250\).
Moreover,
\begin{align}
 \mathcal A_2^{\mathsf U}-\mathcal A_2^{\mathsf F}
 &=2\Xi(m-\ell)s>0,\label{eq:two-peer-full-gain}\\
 \mathcal A_2^{\mathsf U}-\mathcal A_2^{\mathsf S}
 &\geq2\{K_2+\Xi(m-\ell)\}s-J_2(h_2-s)>0.
 \label{eq:two-peer-sealed-gain}
\end{align}
\end{theorem}

\begin{proof}
Section~OA.8 checks reporting incentives under each rule. One distinction
matters after an initial lie. With full disclosure a low type who observes
unanimous opposition may benefit from disregarding her second signal. The
initial incentive bound deducts this gain.

Under full disclosure and \(\mathsf U\), B behaves alike when the two
peers' first forecasts agree. Consider the remaining histories, where the
peers initially split. If the peers subsequently agree with one another,
their evidence determines the optimal classification. If their final
reports also split, the likelihood contributions of their complete
histories cancel, and the decision maker follows B's final forecast.
On such histories the change from persistence \(s\) to revision raises
accuracy by \eqref{eq:two-peer-full-gain}, aggregated over their two
orientations. Here \(m-\ell=(1-\omega)(p_{L2}-p_{L1})>0\).

Against sealing, B revises more often after a peer split or unanimous
opposition and less often after unanimous agreement with her. The first
two changes contribute \(2\{K_2+\Xi(m-\ell)\}s\). Even retaining the
sealed decision rule on the third branch, the loss is at most
\(J_2(h_2-s)\). Section~OA.8 bounds their difference uniformly over
\(\omega\), giving \eqref{eq:two-peer-sealed-gain}.

\end{proof}

The gain in the theorem comes from the feedback given to a team that
already has three forecasts. Whether the gain would justify obtaining
a third forecast is a separate cost comparison.

\section{Direct rewards for accuracy}
\label{sec:accuracy-rewards}

An expert may value being correct even when an incorrect forecast would
leave her with a better reputation. Let her receive \(\beta\) for a correct
final forecast in addition to her reputational wage,

\begin{equation}\label{eq:accuracy-reward-payoff}
 U_i=\rho_i+\beta\ind\{r_{i2}=\theta\},\qquad \beta\geq0.
\end{equation}
We use the same reward for both types and under every feedback rule. The
organiser continues to value correct classification of the state.\footnote{A
reward may be intrinsic or paid by the organiser. In the latter case the
accuracy comparisons give gross benefits.} \citet[p.~507]{Levy2004} likewise
allows the expert to value both her reputation and a correct action.

\subsection{How an accuracy reward changes revision}

Consider an expert whose second signal contradicts her first. The message
gives likelihood ratio \(\lambda\) in favour of her first forecast, while
her conflicting private signals give odds \(1/\eta_\tau\). Repeating the
forecast therefore adds an expected accuracy payment of
\(\beta(\lambda-\eta_\tau)/(\lambda+\eta_\tau)\) to her reputational gain.
The total gain from persistence is

\begin{equation}\label{eq:accuracy-conflict-gain}
 \mathcal G_{\tau,\beta}(h;\lambda)
 =\frac{\lambda\{G_C(h)+\beta\}
           -\eta_\tau\{G_I(h)+\beta\}}{\lambda+\eta_\tau}.
\end{equation}
Denote the low-type continuation candidate by \(h_\beta(\lambda)\).

\begin{proposition}
\label{prop:accuracy-continuation}
Holding fixed truthful initial reporting, confirmation, and high-type
conflict revision, define
\[
 F_{\lambda,\beta}(h)
 =\lambda\{G_C(h)+\beta\}-\eta_L\{G_I(h)+\beta\}.
\]
This function is strictly decreasing on \([0,1]\). The low-type candidate
is zero if \(F_{\lambda,\beta}(0)\leq0\), one if
\(F_{\lambda,\beta}(1)\geq0\), and the unique interior zero otherwise.
The pure-revision cutoff is
\begin{equation}\label{eq:accuracy-pressure-threshold}
 r_\beta=\eta_L\frac{G_I(0)+\beta}{G_C(0)+\beta}.
\end{equation}
At an interior root,
\begin{equation}\label{eq:accuracy-persistence-derivative}
 \frac{\partial h_\beta(\lambda)}{\partial\beta}
 =\frac{\lambda-\eta_L}
 {\eta_LG_I'(h_\beta)-\lambda G_C'(h_\beta)}.
\end{equation}
\end{proposition}

\begin{proof}
The wage formulas remain unchanged. Using \(G_C'<0<G_I'\) in
\cref{eq:accuracy-conflict-gain} gives the first claim and the derivative.
Evaluating the same comparison at zero persistence gives
\cref{eq:accuracy-pressure-threshold}. Initial reporting and the other
continuation actions are verified below.
\end{proof}

When \(\lambda<\eta_L\), the low type's second signal is the better
forecast after conflict, and an accuracy reward reduces her incentive to
repeat. When favourable peer news is sufficiently strong, the same reward
increases persistence. To assess the value of feedback the organiser must
therefore account for different responses to silence and to agreement.

Let \(r_0=\eta_LQ_0\), \(\alpha_0=(1-r_0)/(q-(1-q)r_0)\), and
\(\pi_0=(\alpha_0,0)\). The organiser can keep the rule \(\pi_0\) adopted when reputation was the
only motive. Let \(s_\beta=h_\beta(1)\) and
\(f_\beta=h_\beta(\gamma)\) be the low type's persistence probabilities
when sealed and after revealed agreement, respectively. Write
\(J_\beta=mC_{s_\beta}-\ell B_{s_\beta}\), \(K=mA-\ell D\), and denote the
resulting accuracy under \(\pi\) by \(\mathcal A_\beta^\pi\).

\begin{theorem}
\label{thm:accuracy-robustness}
Suppose
\[
 (p_{L1},p_{L2},p_{H1},p_{H2})
 =\left(\frac{11}{20},\frac58,\frac{33}{40},\frac{19}{20}\right),
 \qquad \omega\in\left[\frac38,\frac25\right],\qquad
 0\leq\beta\leq\frac1{10}.
\]
Every reveal-or-silence policy has a regular-monotone sequential equilibrium
with all four initial-report margins greater than \(3/400\). For the fixed
policy \(\pi_0\),
\begin{align}
 \mathcal A_\beta^{\pi_0}-\mathcal A_\beta^{\mathsf F}
 &=(1-\alpha_0)J_\beta f_\beta>\frac1{500},
 \label{eq:accuracy-full-gain}\\
 \mathcal A_\beta^{\pi_0}-\mathcal A_\beta^{\mathsf S}
 &=(J_\beta+K)s_\beta-\alpha_0J_\beta f_\beta>\frac3{500}.
 \label{eq:accuracy-sealed-gain}
\end{align}
If \(\beta>0\), the low type's conditional payoff gain from revision
after a silent conflict is
\begin{equation}\label{eq:accuracy-strict-silence}
 \beta\frac{\eta_L-r_0}{\eta_L+r_0}>\frac{3\beta}{10}.
\end{equation}
Moreover, \(r_\beta\) increases with \(\beta\), and
\[
 \alpha_\beta=\frac{1-r_\beta}{q-(1-q)r_\beta}
\]
lies in \((0,\alpha_0]\). The policy
\(\pi_\beta=(\alpha_\beta,0)\) also beats both benchmarks, with
\begin{equation}\label{eq:accuracy-adjusted-policy-gain}
 \mathcal A_\beta^{\pi_\beta}-\mathcal A_\beta^{\pi_0}
 =J_\beta f_\beta(\alpha_0-\alpha_\beta)\geq0.
\end{equation}
\end{theorem}

\begin{proof}
Section~OA.9 establishes \(17/100<s_\beta<1/4\) and
\(53/100<f_\beta<29/50\) uniformly, and checks all continuation actions,
including those after an initial lie. Following the second signal remains
optimal after a lie under every policy in this range. Although equilibrium
wages change with the reward, the initial-report comparison retains the bound
\((2p_{\tau1}-1)\{W_{CI}(0)-W_{II}(0)\}>3/400\).

The classification regions from the earlier comparison remain valid on these
root bounds. Substitution of \(s_\beta,f_\beta\) gives the two exact
differences, and rational bounds establish their uniform positivity. Under
\(\pi_0\), the silent likelihood ratio is \(r_0\); evaluating
\cref{eq:accuracy-conflict-gain} at \(h=0\) gives
\cref{eq:accuracy-strict-silence}. Since \(G_C(0)>G_I(0)\),
\(r_\beta\) increases and \(\alpha_\beta\) decreases. Both policies
induce zero silent persistence. Reducing disclosure from \(\alpha_0\) to
\(\alpha_\beta\) consequently gives \cref{eq:accuracy-adjusted-policy-gain}.
\end{proof}

A small accuracy reward gives the organiser two ways to use feedback.
She can keep the old agreement probability \(\alpha_0\). Then even a
small positive reward makes the low type strictly prefer revision after a
silent conflict. Alternatively, she can send fewer agreement messages and
still induce revision, at indifference, after silence. Equation
\eqref{eq:accuracy-adjusted-policy-gain} gives the accuracy gain from this
adjustment. The comparison establishes improvements over both polar rules
without requiring a solution to the new global optimisation problem.

For a team with two peers, the organiser can likewise retain the
deterministic rule and obtain strict gains under a small reward.

\begin{corollary}
\label{cor:accuracy-two-peer}
At the same precisions and type shares, suppose
\(0\leq\beta\leq1/1000\) in the three-expert environment of
Section~\ref{sec:two-peer-confirmation}. Each of
\(\mathsf U,\mathsf S_2,\mathsf F_2\) has a regular-monotone
sequential equilibrium with all six initial-report margins greater than
\(7/1000\). Unanimous confirmation still induces a low-type
silent-conflict revision gain greater than \(1/250\), and
\[
 \mathcal A_{2,\beta}^{\mathsf U}-\mathcal A_{2,\beta}^{\mathsf F}
 >\frac{19}{10000},\qquad
 \mathcal A_{2,\beta}^{\mathsf U}-\mathcal A_{2,\beta}^{\mathsf S}
 >\frac1{625}.
\]
\end{corollary}

\begin{proof}
On this reward range, \(11/50<s_\beta<1/4\) and
\(7/10<h_\beta(\gamma^2)<71/100\), so the classification bounds in
Section~OA.8 continue to hold. Section~OA.9 verifies the altered incentives,
including the extra deviation payoff after a lie under full disclosure.
All bounds are uniform in the reward and type share.
\end{proof}

\subsection{The boundary and the accuracy-only benchmark}

At any positive reward, uniform babbling ceases to be an equilibrium.
With a constant reputational wage the expert gains by forecasting the more
likely state whenever her posterior differs from one half. Section~OA.1.2
nonetheless constructs an equilibrium in which experts report their first
signals truthfully and repeat them regardless of subsequent evidence. A revision
reduces her reputation enough to outweigh the accuracy payment. The
required inequality holds because \(W_I^{(1)}\geq7/37>1/10\). This also
applies to a team with two sealed peers.

At a larger reward the low type begins following her second signal even
under sealing. We give the threshold and compare the three protocols when
experts value only the correctness of their own final forecasts.

\begin{proposition}
\label{prop:accuracy-boundary}
At the precisions and shares of Theorem~\ref{thm:accuracy-robustness}, define
\begin{equation}\label{eq:accuracy-sealed-boundary}
 \beta_{\mathsf S}
 =\frac{G_C(0)-\eta_LG_I(0)}{\eta_L-1}.
\end{equation}
Then \(2/5<\beta_{\mathsf S}<43/100\). For
\(\beta\geq\beta_{\mathsf S}\), sealing has a truthful
regular-monotone equilibrium with zero conflict persistence. The disciplinary
threshold reaches \(\alpha=0\) at \(\beta_{\mathsf S}\).

If the reputational term is removed altogether and experts maximise only
\(\Prb(r_{i2}=\theta)\), sealing, full disclosure and \(\pi_0\)
each admit equilibria with truthful initial reports. Their classification
values satisfy
\begin{align}
 \mathcal A_{\mathrm{acc}}^{\mathsf S}
 -\mathcal A_{\mathrm{acc}}^{\pi_0}
 &=\alpha_0J_{\mathrm{acc}}>0,\notag\\
 \mathcal A_{\mathrm{acc}}^{\pi_0}
 -\mathcal A_{\mathrm{acc}}^{\mathsf F}
 &=(1-\alpha_0)J_{\mathrm{acc}}>0,
 \qquad J_{\mathrm{acc}}=mC-\ell B.
 \label{eq:accuracy-only-ranking}
\end{align}
Initial truth-telling in this accuracy-only comparison is weak, except for
B's low type under \(\pi_0\).
\end{proposition}

\begin{proof}
At \(h=0\), the sealed conflict numerator is
\(G_C(0)-\eta_LG_I(0)-(\eta_L-1)\beta\), yielding the boundary.
The wage and deviation comparisons also give truthful initial reporting
when sealed conflict persistence is zero.

With accuracy as the only motive, both types of A and B's high type follow
their second signals. B's low type follows the second signal under sealing
and after silence under \(\pi_0\), but repeats after revealed agreement.
Thus \(s=0\) and \(f=1\). Section~OA.9 verifies the reporting deviations
and classification regions, giving \cref{eq:accuracy-only-ranking}.
\end{proof}

A low type who receives confirmation can rationally retain her first
forecast despite contrary private evidence. The decision maker sees the
peer forecast in the final record anyway. What she loses in this case is
the information that a revision would have conveyed about B's second
signal. This accounts for the ranking in
\eqref{eq:accuracy-only-ranking} among the three truthful-initial-report
equilibria.

\section{Disagreement disclosure and escalation}
\label{sec:disagreement-disclosure}

An organiser may have to release some disagreements before the experts
revise. The optimum in Theorem~\ref{thm:analytical-global-optimality} keeps
every disagreement private until the decision date. Yet B can draw the same
inference from silence when some disagreements reach her earlier.
Suppose the organiser opens a fraction \(d\) of cases independently of
the first reports, delivering A's actual forecast. In the remaining cases
she follows the rule \(\pi_e=(a_e,0)\), with \(a_e<1\). B sees the peer
forecast whenever disclosure occurs. Her overall disclosure probabilities
are

\begin{equation}\label{eq:positive-disagreement-policy}
 \pi_{e,d}=\bigl(d+(1-d)a_e,d\bigr).
\end{equation}
for \(0<d<1\). Compared with \(\pi_e\), this multiplies both concealment
probabilities by \(1-d\). Silence occurs less often, but the odds that the
two experts initially agreed conditional on silence stay the same.

\begin{proposition}\label{prop:positive-disagreement-ray}
Suppose the regular-monotone assessments at \(\pi_e\) and full
disclosure are sequential equilibria. Then every policy in
\eqref{eq:positive-disagreement-policy} is equilibrium-feasible.
Every conditional silent-branch incentive is unchanged. Each initial-report
margin of B is the convex combination, with weights \(1-d,d\), of its
endpoint and full-disclosure margins; A's margins are unchanged.
For classification, put
\(g_F=\mathcal A^{\pi_e}-\mathcal A^{\mathsf F}\) and
\(g_S=\mathcal A^{\pi_e}-\mathcal A^{\mathsf S}\). Then
\begin{align}
 \mathcal A^{\pi_{e,d}}-\mathcal A^{\mathsf F}&=(1-d)g_F,
 \label{eq:disagreement-full-gain}\\
 \mathcal A^{\pi_{e,d}}-\mathcal A^{\mathsf S}&=g_S-dg_F.
 \label{eq:disagreement-sealed-gain}
\end{align}
In particular, if \(g_F,g_S>0\), both gains are positive whenever
\(0<d<\min\{1,g_S/g_F\}\).
\end{proposition}

\begin{proof}
Theorem~\ref{thm:concealment-ray} applies at scale \(1-d\). Given a
message, the continuation wages and reports agree with those on the
corresponding branch of the endpoint policies whenever that branch is
reached. This holds also for an expert who has lied. Conditional on the
submitted first report, the probability weights on the states at a given
message are rescaled by a common factor, so her optimal terminal deviation
is the same. Both truthful and optimally deviating payoffs are therefore
affine along the ray. The same argument applies to classification once
the message and the first reports are included in the public record. Its
value is \(d\mathcal A^{\mathsf F}+(1-d)\mathcal A^{\pi_e}\), giving the
two differences.

\end{proof}

This argument also solves a constrained design problem. Suppose both
agreements and disagreements must reach B with probability at least \(d\).
For example, an organiser may have agreed to open a random fraction of
all files before revision. Under the conditions of
Theorem~\ref{thm:analytical-global-optimality}, for \(0\leq d<1\), the
unique optimum subject to this common minimum and its accuracy cost are
\begin{equation}\label{eq:common-disclosure-floor}
 \pi_d^*=\bigl(d+(1-d)\alpha^{\mathrm{DS}},d\bigr),\qquad
 \mathcal A^{\mathrm{DS}}-\mathcal A^{\pi_d^*}
   =d\bigl(\mathcal A^{\mathrm{DS}}-\mathcal A^{\mathsf F}\bigr).
\end{equation}
Indeed, write any admissible policy uniquely as
\(d(1,1)+(1-d)\tilde\pi\), where \(\tilde\pi\in[0,1]^2\).
The ray theorem gives value \(d\mathcal A^{\mathsf F}+
(1-d)\mathcal A^{\tilde\pi}\), including at the full-disclosure endpoint.
All residual policies are feasible under the theorem's hypotheses. Since
\(1-d>0\), the organiser's unique choice of \(\tilde\pi\) is her original
optimum. At \(d=1\) the only admissible rule is full disclosure. A
requirement applying to disagreements alone would leave additional
policies available. The next result gives a feasible policy that both
reveals disagreements and makes revision strict.

\begin{corollary}\label{cor:strict-disagreement}
At the four precisions and type shares of
Corollary~\ref{cor:analytical-optimality-interval}, take
\(a_e=\alpha_0+\varepsilon\), with
\(0<\varepsilon\leq1/10\). Every \(0<d<1\) gives strict
silent-conflict revision and improves on both polar protocols.
For \(0<d\leq1/2\), uniformly over the primitive interval,
\[
 R_L>\frac{\varepsilon}{10},\qquad
 \Delta_{i\tau}>\frac3{400},\qquad
 \mathcal A^{\pi_{e,d}}-\mathcal A^{\mathsf F}>\frac1{2500},
 \qquad
 \mathcal A^{\pi_{e,d}}-\mathcal A^{\mathsf S}>\frac3{500}.
\]
\end{corollary}

\begin{proof}
The strict-revision result supplies the first two endpoint bounds. The
uniform initial-incentive bound also holds at full disclosure, and the ray
preserves both this bound and the conditional revision gain. Let
\(g=\mathcal A^{\mathsf{DS}}-\mathcal A^{\mathsf S}\).
Using \(g>1/100\), \(Jf<5599/500000\), and \(\alpha_0>71/100\), we obtain
\[
 \mathcal A^{\mathsf F}-\mathcal A^{\mathsf S}
 =g-Jf(1-\alpha_0)
 >\frac1{100}-\frac{5599}{500000}\frac{29}{100}
 >\frac3{500}.
\]
In addition, \(\alpha_0+\varepsilon<1\) and
\(g_F=Jf(1-\alpha_0-\varepsilon)>0\). Both endpoints therefore improve
on sealing, as does every convex combination. At zero reward, the accuracy
theorem gives \(\mathcal A^{\mathsf{DS}}-\mathcal A^{\mathsf F}>1/500\).
For \(d\leq1/2\), it follows that
\[
 (1-d)g_F>
 \frac12\left(\frac1{500}-\frac{5599}{500000}\frac1{10}\right)
 >\frac1{2500}.
\]
Mixing after revealed agreement remains. At \(d=1\), the policy becomes
full disclosure and silence disappears, so its strict gain over that
benchmark vanishes.
\end{proof}

A reward for accuracy is compatible with any opening rate below one.
Fix the primitives of Theorem~\ref{thm:accuracy-robustness},
\(0\leq\beta\leq1/10\), and take \(\pi_e=\pi_0\). We now evaluate each
policy at the same reward. The common scaling of conditional payoff
weights still applies. Subtracting the two benchmark gains in
Theorem~\ref{thm:accuracy-robustness} gives
\[
 \mathcal A_\beta^{\mathsf F}-\mathcal A_\beta^{\mathsf S}
 =(J_\beta+K)s_\beta-J_\beta f_\beta
 >\left(\frac{1877}{100000}+\frac{3737}{50000}\right)\frac{17}{100}
   -\frac{211}{10000}\frac{29}{50}
 >\frac1{300},
\]
where the bounds on the coefficients and the roots are from Section~OA.9.
Thus for every \(0<d<1\),
\begin{align}
 \mathcal A_\beta^{\pi_{0,d}}-\mathcal A_\beta^{\mathsf F}
   &>\frac{1-d}{500},\notag\\
 \mathcal A_\beta^{\pi_{0,d}}-\mathcal A_\beta^{\mathsf S}
   &>\frac1{300}+\frac{1-d}{500}.
 \label{eq:reward-disagreement-floor}
\end{align}
Here \(\pi_{0,d}=d(1,1)+(1-d)\pi_0\).
Initial truth-telling retains its margin above \(3/400\). For
\(\beta>0\) the low type strictly prefers revision after a silent
conflict, with a conditional payoff gain greater than \(3\beta/10\).
As \(d\) approaches one, she receives silence ever less often.

\section{An application to forecast review}
\label{sec:application}

\subsection{A rule for revising forecasts}

Consider a forecasting team asked whether demand next month will exceed a
specified level. The client needs a binary recommendation by a fixed date.
There is time for two rounds of forecasts before the client acts, and
realised demand will eventually allow her to assess the forecasters'
records. An expert whose record looks better may be more likely to
receive future assignments. In the model this prospect is represented
by her posterior reputation for high ability.

Some features of this procedure have counterparts in forecasting
tournaments. On many Metaculus questions the community forecast is hidden
at first, so that early forecasters must take a view before seeing the
aggregate \citep{MetaculusScores2026}. The platform announces resolution
criteria and allows for ambiguous or annulled questions
\citep{MetaculusFAQ2026}. It also uses leaderboard performance in selecting
professional forecasters \citep{MetaculusLeaderboards2026}. The comparison
suggests an institution that can control access to earlier forecasts and
in which a record of performance has career value. Agreement-contingent
access is a proposal for such a team. The platform evidence concerns
timing, scoring and selection.

For the demand forecast, fix the threshold, resolution source and reporting
dates in advance. Each forecaster files a first recommendation privately.
In a team of three, the organiser sends B a confirmation if both other
forecasters agreed with her. They see none of B's forecasts before making
their final recommendations. B may revise in light of her second signal
and the presence or absence of confirmation. At the decision date the
client receives every first and final recommendation together with B's
message. Theorem~\ref{thm:two-peer-deterministic-gains} compares this
procedure with private revision and with showing B both first forecasts,
holding fixed the three forecasters and their evidence.

Commitment matters precisely when the organiser sees a disagreement and
would like to make an exception. If forecasters expect discretionary
warnings, they will interpret silence differently. An organiser could
announce the rule before assigning a question, preserve the original
entries, and allow the client to check each interim message against the
entries when the full record is released. This would make departures
detectable. Whether the organiser can commit, and what happens after a
departure, depend on the organisation.

\subsection{When the procedure applies}

With one peer, the organiser can open a random sample of files before
revision and apply the agreement-contingent rule in the rest. Equations
\eqref{eq:common-disclosure-floor} and \eqref{eq:reward-disagreement-floor}
give the cost of the common disclosure minimum in the baseline problem
and the gains available when accuracy is rewarded. An urgent disagreement
that calls for a test, a supervisor's decision or immediate intervention
raises a different question. The present timing presumes the client can
wait for the complete file. A liability rule requiring action before that
date would change the feasible decisions as well as payoffs.

The actual form of the reward also matters. Metaculus scores binary
probability forecasts using scores based on the logarithmic rule, with each
forecast weighted by the time it stood \citep{MetaculusScores2026}.
Theorem~\ref{thm:accuracy-robustness} concerns a bonus on a correct final
binary report, measured relative to the reputational wage. It gives a range
of relative rewards over which the proposed feedback helps. Applying that
result to a probability-scoring system would require deriving the incentives
to change the reported probability at each date.

Finally, private submission prevents the forecasters from copying one
another's reports, but all three may have used the same demand model. In
that case even separately filed forecasts can reflect a common
forecast error. The two probabilities multiplying to \(q^2\) in
\eqref{eq:two-peer-likelihoods} would have to be replaced by their joint
probability. Common information could also affect what the client infers
about B's ability from the other forecasts after demand is realised.

\subsection{How much is a better forecast worth?}

Suppose either classification mistake costs the client \(L\) and a correct
decision costs zero. If selective feedback costs an additional \(\Delta C\)
per case relative to full disclosure, the client gains from adopting it
in the two-expert model precisely when
\begin{equation}\label{eq:application-cost-threshold}
 \Delta C<L\bigl(\mathcal A^{\mathrm{DS}}-\mathcal A^{\mathsf F}\bigr).
\end{equation}
At the illustrative primitives in Section~OA.5 the two classification
probabilities are about \(0.79205\) and \(0.78903\). The difference amounts
to about three additional correct decisions per thousand cases. Thus the
client can spend about \(0.003017L\) per case before the gain disappears.
A common disclosure minimum of one half halves this allowance by
\eqref{eq:common-disclosure-floor}.

Whether these amounts would cover the cost of administering the rule
depends on the loss per error and the number of cases over which fixed
costs can be spread. A comparison that adds another forecaster must also
price the extra forecast. Finally, a costly shortage and a costly surplus
need not have the same consequences. With unequal losses the client should
compare the expected loss under her own optimal decision rule. As
Section~\ref{subsec:blackwell} shows, even the ranking of the two polar
rules can then change.

\section{Conclusion}\label{sec:conclusion}

What should a forecaster learn about a colleague's opinion before submitting a revision? In the model, disclosing a little favourable news has two effects. It encourages persistence among those who receive confirmation and makes the others take contrary evidence more seriously. For a range of signal precisions and ability shares, the second effect justifies disclosing agreement up to the point where an expert who hears nothing uses her new evidence. Beyond that point there are only more confirmations. When classification mistakes carry equal costs, this trade-off determines a unique optimal disclosure probability.

An institution could obtain a similar incentive by seeking two independent peer opinions and passing on unanimous confirmation. Whether either procedure is worthwhile depends on the decision at stake. When missing state one costs nine times as much as a false alarm, the decision maker prefers sealing to the selective rule that yields the most correct decisions. And a procedure that improves on full disclosure by three decisions per thousand has to be inexpensive to run unless each error is costly.

The analysis leaves open a question about the organisation of expertise. We take the experts' sources of evidence as given and independent conditional on the state and their abilities. Forecasters who use the same data or consult one another before submitting their initial reports may agree even when the evidence is poor. An institution choosing whom to consult as well as how to disclose their reports would then have to consider the sources of agreement before deciding what an unconfirmed forecast should mean.

\clearpage
\section*{Online Appendix}
\phantomsection\label{oa:start}
\addcontentsline{toc}{section}{Online Appendix}
\setcounter{section}{0}
\setcounter{theorem}{0}
\renewcommand{\thesection}{OA.\arabic{section}}
\renewcommand{\theHsection}{OA.\arabic{section}}
\setcounter{table}{0}

The posterior wages and initial-report incentives used in the paper are
derived in Section~OA.2. Section~OA.4 proves optimality, and Sections~OA.8
and OA.9 treat a second peer and an accuracy reward. Sections~OA.3, OA.5
and OA.6 give the calculations for the illustrative profile. Section~OA.10
compares decision objectives on an interval of type shares. The equilibrium
selection arguments appear in Sections~OA.1 and OA.11.

\section{Other equilibria and perturbations}
\label{oa:equilibrium-selection}

Two quite different reporting rules survive every choice of feedback policy.
Under one, the reports convey no information. Under the other, initial reports
are truthful but the second signal never reaches the decision maker. We give
the supporting wages before considering accuracy rewards and recording errors.

\subsection{Two policy-independent equilibria}

\begin{proposition}
\label{prop:oa-babbling}
For every reveal-or-silence policy, there is a sequential equilibrium in which both types of both experts choose each report with probability $1/2$ at every information set, independently of signals and messages. Every terminal wage equals $\omega$ and the state-classification value is $1/2$.
\end{proposition}

\begin{proof}
The conditional likelihood of every complete report history is identical
under $H$ and $L$. Initial reports are independent of the state, and the
platform generates its message from these reports and the committed rule.
The message is therefore independent of both state and types. At every
positive-probability history, Bayes' rule gives wage $\omega$, so every report
is optimal. Reporting is already completely mixed, with no supporting
off-path wage choice.
\end{proof}

An informative but nonregular equilibrium is also available under every
policy. Let the reputations after a correct and incorrect initial report be
\begin{equation}\label{eq:oa-first-report-wages}
 W_C^{(1)}=
 \frac{\omega p_{H1}}
 {\omega p_{H1}+(1-\omega)p_{L1}},
 \qquad
 W_I^{(1)}=
 \frac{\omega(1-p_{H1})}
 {\omega(1-p_{H1})+(1-\omega)(1-p_{L1})}.
\end{equation}

\begin{proposition}
\label{prop:oa-full-repetition}
For every reveal-or-silence policy, there is a sequential equilibrium in which both experts report their first signals truthfully and every type subsequently repeats the locked report, independently of the second signal and the platform message.
\end{proposition}

\begin{proof}
On the equilibrium path, correct initial reports generate $CC$ and incorrect
ones generate $II$, with wages \cref{eq:oa-first-report-wages}. Consider
completely mixed terminal strategies for which the high type's revision
probability vanishes faster than the low type's at every private history.
The limiting wages following $CI$ and $IC$ are zero. If $x$ is the probability
that the initial report is correct, repetition has gain
\begin{equation}\label{eq:oa-full-repetition-terminal-gain}
 xW_C^{(1)}+(1-x)W_I^{(1)}>0.
\end{equation}
Anticipating repetition, the gain from reporting the first signal truthfully is
\begin{equation}\label{eq:oa-full-repetition-initial-gain}
 (2p_{\tau1}-1)(W_C^{(1)}-W_I^{(1)})>0.
\end{equation}
Conditional independence permits the construction for both experts after
every message. The specified trembles establish consistency.
\end{proof}

Completely mixed babbling survives refinements that restrict only off-path
beliefs in the binary game. All strategically reachable histories are already
on path, while impossible chance branches remain absent. Enlarging the
message language is a separate exercise. Full repetition shows that truthful
initial reporting alone does not make feedback effective: an expert who
always repeats does not use peer information in her final report.

\subsection{A small accuracy stake does not select}

Now consider payoffs
\begin{equation}\label{eq:oa-accuracy-stake-payoff}
 U_i=\rho_i
 +\varepsilon_1\ind\{r_{i1}=\theta\}
 +\varepsilon_2\ind\{r_{i2}=\theta\},
 \qquad \varepsilon_1,\varepsilon_2\geq0.
\end{equation}
A positive terminal stake $\varepsilon_2>0$ eliminates uniform babbling,
since reports then respond to the state posterior. Full repetition can remain.

\begin{proposition}
\label{prop:oa-accuracy-full-repetition}
If $0\leq\varepsilon_2<W_I^{(1)}$, the full-repetition equilibrium in \cref{prop:oa-full-repetition} persists for every policy and every $\varepsilon_1\geq0$.
\end{proposition}

\begin{proof}
At the same limiting off-path wages, the terminal gain from repetition is
\begin{equation}\label{eq:oa-scored-repetition-gain}
 x(W_C^{(1)}+\varepsilon_2)
 +(1-x)(W_I^{(1)}-\varepsilon_2)>0
\end{equation}
for all $x\in[0,1]$. The gain from truthful initial reporting is
\begin{equation}\label{eq:oa-scored-initial-gain}
 (2p_{\tau1}-1)
 (W_C^{(1)}-W_I^{(1)}+\varepsilon_1+\varepsilon_2)>0.
\end{equation}
\end{proof}

At the paper's benchmark,
\begin{equation}\label{eq:oa-full-repetition-profile}
 W_C^{(1)}=\frac12,
 \qquad
 W_I^{(1)}=\frac7{34}.
\end{equation}
The sufficient upper bound for repetition is therefore about one fifth
of the largest possible reputational payoff at this profile.

\subsection{Terminal recording errors}

Suppose an intended terminal report $\widehat r_2$ is recorded correctly with
probability $1-\zeta$ and reversed with probability $\zeta\in(0,1/2)$.
Recording noise is independent of types, signals, messages and policy.
The terminal accuracy payment in \cref{eq:oa-accuracy-stake-payoff} uses the
recorded report. Let $c_\tau$ and $h_\tau$ be type $\tau$'s intended repetition
probabilities after confirming and conflicting signals. Define
\begin{align}
 u_\tau
 &=\zeta+(1-2\zeta)
 \{p_{\tau2}c_\tau+(1-p_{\tau2})h_\tau\},
 \label{eq:oa-noisy-correct-first}\\
 v_\tau
 &=\zeta+(1-2\zeta)
 \{(1-p_{\tau2})(1-c_\tau)+p_{\tau2}(1-h_\tau)\}.
 \label{eq:oa-noisy-incorrect-first}
\end{align}
The observed pattern likelihoods are
\begin{equation}\label{eq:oa-noisy-likelihoods}
 \ell_{\tau,CC}=p_{\tau1}u_\tau,
 \quad
 \ell_{\tau,CI}=p_{\tau1}(1-u_\tau),
 \quad
 \ell_{\tau,IC}=(1-p_{\tau1})v_\tau,
 \quad
 \ell_{\tau,II}=(1-p_{\tau1})(1-v_\tau).
\end{equation}
All are strictly positive. Let $W_\chi$ be the corresponding Bayesian wage,
and write
\begin{equation}\label{eq:oa-noisy-wage-gaps}
 G_C=W_{CC}-W_{CI},
 \qquad
 G_I=W_{IC}-W_{II}.
\end{equation}
For posterior probability $x$ that the initial report is correct, the gain
from intended repetition, divided by $1-2\zeta>0$, equals
\begin{equation}\label{eq:oa-noisy-terminal-gain}
 x(G_C+\varepsilon_2)-(1-x)(G_I+\varepsilon_2).
\end{equation}
Every terminal action is therefore optimal at every signal history and message if
\begin{equation}\label{eq:oa-accuracy-offset-condition}
 G_C=G_I=-\varepsilon_2.
\end{equation}

For this example take $\zeta=10^{-4}$ and
\[
 (p_{L1},p_{L2},p_{H1},p_{H2},\omega)
 =(11/20,5/8,33/40,19/20,2/5).
\]
To construct the path, set, for $t\in[0,1]$,
\begin{equation}\label{eq:oa-offset-path}
 h_L(t)=c_H(t)=\frac{1-t}{2},
 \qquad
 h_H(t)=\frac{1+t}{2},
\end{equation}
and choose $c_L(t)$ to satisfy $G_C=G_I$. For fixed $t$, increasing $c_L$
reduces $G_C$ and raises $G_I$, so their difference is strictly decreasing.
Write $G_C(t)$ and $G_I(t)$ for the wage gaps along this path.
Clearing positive denominators and applying Sturm sequences gives, uniformly
on $t\in[0,1]$,
\begin{equation}\label{eq:oa-offset-endpoint-signs}
 [G_C-G_I]_{c_L=0}>0,
 \qquad
 [G_C-G_I]_{c_L=1}<0.
\end{equation}
There is consequently a unique continuous solution $c_L(t)\in(0,1)$.

At this profile, the path starts at uniform terminal
randomisation and ends at
\begin{equation}\label{eq:oa-offset-endpoint}
 c_L(1)\approx0.625106534357,
 \qquad
 (h_L,c_H,h_H)=(0,0,1).
\end{equation}
Define the endpoint stake exactly by
\(\bar\varepsilon_2=-G_C(1)=-G_I(1)\). Rational bounds give
\begin{equation}\label{eq:oa-offset-maximum-stake}
 0.4955970251548<\bar\varepsilon_2<0.4955970251549.
\end{equation}
By continuity, a fully supported accuracy-offset equilibrium exists for every
\begin{equation}\label{eq:oa-offset-range}
 0<\varepsilon_2<\bar\varepsilon_2.
\end{equation}
Intended strategies are completely mixed in the path interior, and recording
noise puts all correctness patterns on path. Also, $W_{CI}>W_{II}$ throughout,
giving truthful-versus-opposite initial-report slack
\begin{equation}\label{eq:oa-offset-first-report-slack}
 (2p_{\tau1}-1)(W_{CI}-W_{II}+\varepsilon_1)>0.
\end{equation}
Since terminal strategies are identical across messages, the platform lottery
cancels from type posteriors and deviation comparisons. The result is a
fully supported, policy-independent sequential equilibrium with strictly
truthful initial reports. It is nonmonotone: away from $t=0$, $h_H>h_L$.

An interior disciplinary policy, in turn, requires
\begin{equation}\label{eq:oa-policy-stake-bound}
 \varepsilon_2<
 \varepsilon_2^{\mathrm{policy}}(\zeta)
 =\frac{G_C^0(\zeta)-\eta_LG_I^0(\zeta)}{\eta_L-1},
\end{equation}
where $G_C^0$ and $G_I^0$ are the noisy wage gaps under pure conflict revision.
At $\zeta=10^{-4}$,
\begin{equation}\label{eq:oa-policy-stake-value}
 \varepsilon_2^{\mathrm{policy}}
 \approx0.400177385298
 <\bar\varepsilon_2.
\end{equation}
Every accuracy stake compatible with an interior disciplinary policy thus
also admits the nonmonotone equilibrium. The failure of selection therefore
need not rely on off-path wages. Recording noise still does not select regular
monotonicity.

\section{Formal assessment and analytical details}
\label{oa:formal-details}

An initial lie changes two objects for B. It changes the later reputation
attached to either final report and the probability of the message on which
she bases that report. We begin with the likelihood of the public record.
Write
\begin{equation}\label{eq:oa-signal-likelihood}
 f_{\tau t}(s\mid\vartheta)
 =\begin{cases}
 p_{\tau t},&s=\vartheta,\\
 1-p_{\tau t},&s\ne\vartheta,
 \end{cases}
\end{equation}
and let \(\mu_\pi(m\mid y,b)\) be the message kernel determined by
\((\alpha,\delta)\).

\subsection{Likelihoods, beliefs, and sequential payoffs}

Given an assessment \(\sigma\), A's type-conditional reporting likelihood is
\begin{align}
 L_{A\tau}^{\sigma}(r_A^2\mid\vartheta)
 ={}&\sum_{s_{A1},s_{A2}}
 f_{\tau1}(s_{A1}\mid\vartheta)
 f_{\tau2}(s_{A2}\mid\vartheta)
 \sigma_{A1}(r_{A1}\mid\tau,s_{A1})
 \notag\\[-1mm]
 &\quad\times
 \sigma_{A2}(r_{A2}\mid\tau,s_A^2,r_{A1}).
 \label{eq:oa-a-report-likelihood}
\end{align}
For a fixed message, the factor contributed by B is
\begin{align}
 L_{B\tau}^{\sigma}(r_B^2\mid y,m,\vartheta)
 ={}&\sum_{s_{B1},s_{B2}}
 f_{\tau1}(s_{B1}\mid\vartheta)
 f_{\tau2}(s_{B2}\mid\vartheta)
 \sigma_{B1}(r_{B1}\mid\tau,s_{B1})
 \notag\\[-1mm]
 &\quad\times
 \sigma_{B2}(r_{B2}\mid\tau,s_B^2,r_{B1},m).
 \label{eq:oa-b-report-likelihood}
\end{align}
Here $m$ is held fixed while summing over B's signals. Conditioning her
reports on observing $m$ would also weight the initial report by its
probability of generating that message. This probability enters separately
in the joint likelihood

\[
 L_{A\tau_A}^{\sigma}(r_A^2\mid\vartheta)
 \mu_\pi(m\mid y,b)
 L_{B\tau_B}^{\sigma}(r_B^2\mid y,m,\vartheta).
\]
Bayes' rule gives wages
\begin{align}
 \rho_A
 &=\frac{\omega L_{AH}^{\sigma}(r_A^2\mid\theta)}
 {\omega L_{AH}^{\sigma}(r_A^2\mid\theta)
 +(1-\omega)L_{AL}^{\sigma}(r_A^2\mid\theta)},
 \label{eq:oa-a-bayesian-wage}\\
 \rho_B
 &=\frac{\omega L_{BH}^{\sigma}(r_B^2\mid y,m,\theta)}
 {\omega L_{BH}^{\sigma}(r_B^2\mid y,m,\theta)
 +(1-\omega)L_{BL}^{\sigma}(r_B^2\mid y,m,\theta)}.
 \label{eq:oa-b-bayesian-wage}
\end{align}
The other expert's reporting factor and the policy probability cancel from
the relevant type odds, establishing the factorisation used in the main text.

For public outcome \(z=(r_A^2,r_B^2,m)\) in state \(\vartheta\), sum
\begin{align}
 &\mu_\pi(m\mid r_{A1},r_{B1})
 \prod_{i\in\{A,B\}}
 \bigl[
 \omega_{\tau_i}
 f_{\tau_i1}(s_{i1}\mid\vartheta)
 f_{\tau_i2}(s_{i2}\mid\vartheta)
 \sigma_{i1}(r_{i1}\mid\tau_i,s_{i1})
 \bigr]
 \notag\\[-1mm]
 &\hspace{20mm}\times
 \sigma_{A2}(r_{A2}\mid\tau_A,s_A^2,r_{A1})
 \sigma_{B2}(r_{B2}\mid\tau_B,s_B^2,r_{B1},m)
 \label{eq:oa-public-outcome-summand}
\end{align}
over \((\tau_A,\tau_B,s_A^2,s_B^2)\), with \(\omega_H=\omega\) and
\(\omega_L=1-\omega\). The classification and Blackwell programs use the
same summand \eqref{eq:oa-public-outcome-summand}.

\subsection{Correctness patterns and the cutoff map}

Assume truthful initial reports. On a supported message branch, both types
confirm agreeing signals, the high type follows a conflicting second signal,
and the low type repeats after conflict with probability \(h\). The high-type
pattern likelihoods are
\begin{align}
 \ell_H^{CC}(h)&=p_{H1}p_{H2},
 &\ell_H^{CI}(h)&=p_{H1}(1-p_{H2}),\label{eq:oa-high-patterns-a}\\
 \ell_H^{IC}(h)&=(1-p_{H1})p_{H2},
 &\ell_H^{II}(h)&=(1-p_{H1})(1-p_{H2}).\label{eq:oa-high-patterns-b}
\end{align}
The low-type likelihoods are
\begin{align}
 \ell_L^{CC}(h)
 &=p_{L1}\{p_{L2}+(1-p_{L2})h\},
 &\ell_L^{CI}(h)
 &=p_{L1}(1-p_{L2})(1-h),\label{eq:oa-low-patterns-a}\\
 \ell_L^{IC}(h)
 &=(1-p_{L1})p_{L2}(1-h),
 &\ell_L^{II}(h)
 &=(1-p_{L1})\{(1-p_{L2})+p_{L2}h\}.
 \label{eq:oa-low-patterns-b}
\end{align}
Inserting them into
\begin{equation}\label{eq:oa-pattern-wage}
 W_\chi(h)=
 \frac{\omega\ell_H^\chi(h)}
 {\omega\ell_H^\chi(h)+(1-\omega)\ell_L^\chi(h)}
\end{equation}
gives the four wages.

At \(h=0\), let
\begin{equation}\label{eq:oa-q0-ingredients}
 d_t=\frac{\kappa_{Ht}}{\kappa_{Lt}},
 \qquad
 O=\frac{\omega}{1-\omega}
 \frac{(1-p_{H1})(1-p_{H2})}
 {(1-p_{L1})(1-p_{L2})}.
\end{equation}
The posterior odds of \(H\), ordered as \(II,CI,IC,CC\), are
\(O,d_1O,d_2O,d_1d_2O\). Subtracting the corresponding probabilities gives
\begin{equation}\label{eq:oa-q0-closed-form}
 Q_0
 =\frac{(1+d_1O)(1+d_1d_2O)}
 {d_1(1+O)(1+d_2O)}.
\end{equation}

As \(h\) increases, the low-type masses on \(CC\) and \(II\) increase,
and those on \(CI\) and \(IC\) decrease. Thus \(W_{CC}\) and \(W_{II}\)
fall, while \(W_{CI}\) and \(W_{IC}\) rise, implying
\begin{align}
 G_C(h)&=W_{CC}(h)-W_{CI}(h)
 &&\text{falls strictly},\label{eq:oa-gap-monotonicity-a}\\
 G_I(h)&=W_{IC}(h)-W_{II}(h)
 &&\text{rises strictly}.\label{eq:oa-gap-monotonicity-b}
\end{align}
Both gaps are positive at zero. As \(h\uparrow1\), \(W_{CI}\) tends to one
and \(W_{CC}<1\). Hence \(G_C(h^\dagger)=0\) has a unique solution
\(h^\dagger\in(0,1)\), and the map
\begin{equation}\label{eq:oa-q-map}
 Q(h)=\frac{G_I(h)}{G_C(h)}:
 [0,h^\dagger)\longrightarrow[Q_0,\infty)
\end{equation}
is continuous, strictly increasing and onto.

A message with likelihood ratio \(\lambda\) gives posterior odds
\(\lambda/\eta_\tau\) after conflict and
\(\lambda\kappa_{\tau1}\kappa_{\tau2}\) after confirmation. Repetition
relative to reversal therefore has payoff difference
\begin{align}
 \text{conflict: }&
 \frac{\lambda G_C(h)-\eta_\tau G_I(h)}
 {\lambda+\eta_\tau},
 \label{eq:oa-conflict-gain}\\
 \text{confirmation: }&
 \frac{\lambda\kappa_{\tau1}\kappa_{\tau2}G_C(h)-G_I(h)}
 {1+\lambda\kappa_{\tau1}\kappa_{\tau2}}.
 \label{eq:oa-confirmation-gain}
\end{align}
The low-type branch conditions become
\begin{equation}\label{eq:oa-branch-conditions}
 \begin{array}{@{}lll@{}}
 \toprule
 \text{branch}&\text{conflict condition}&\text{confirmation condition}\\
 \midrule
 \text{revealed agreement}
 &Q(h^{\mathsf F})=\gamma/\eta_L
 &\gamma\kappa_{L1}\kappa_{L2}>Q(h^{\mathsf F})\\
 \text{revealed disagreement}
 &\gamma\eta_L>1/Q_0
 &\kappa_{L1}\kappa_{L2}/\gamma>Q_0\\
 \text{silence}
 &Q(h_N)=\lambda_N/\eta_L\ \text{if }\lambda_N>\bar\lambda
 &\lambda_N\kappa_{L1}\kappa_{L2}>Q(h_N)
 \\
 \bottomrule
 \end{array}
\end{equation}
For \(\lambda_N\leq\bar\lambda\), B follows her second signal after a
silent conflict, with \(h_N=0\); this choice is weak at equality.
High-type conflict choices are strict under \(\eta_H>\eta_L\).

\subsection{Locked-report deviations}

Let \(c(a,\vartheta)=C\) if \(a=\vartheta\), and \(I\) otherwise. Write
\begin{equation}\label{eq:oa-correctness-map}
 \chi(a_1,a_2;\vartheta)
 =\bigl(c(a_1,\vartheta),c(a_2,\vartheta)\bigr).
\end{equation}
By label symmetry, we can condition on first signal one. Under sealing,
A's truthful and deviation values are
\begin{align}
 V_{A\tau}^{T,\mathsf S}
 ={}&\sum_{\vartheta,s_2}
 f_{\tau1}(1\mid\vartheta)f_{\tau2}(s_2\mid\vartheta)
 \sum_{r_2}
 \widehat\sigma_{A2}^{\mathsf S}(r_2\mid\tau,(1,s_2),1)
 W_{\chi(1,r_2;\vartheta)}(h^{\mathsf S}),
 \label{eq:oa-a-truth-value}\\
 V_{A\tau}^{D,\mathsf S}
 ={}&\sum_{s_2}\max_{r_2}
 \sum_{\vartheta}
 f_{\tau1}(1\mid\vartheta)f_{\tau2}(s_2\mid\vartheta)
 W_{\chi(0,r_2;\vartheta)}(h^{\mathsf S}).
 \label{eq:oa-a-deviation-value}
\end{align}
The maximisation in \eqref{eq:oa-a-deviation-value} specifies the optimal
terminal response after the initial lie as well as its value.

For B, define
\begin{equation}\label{eq:oa-peer-report-likelihood}
 g_q(y\mid\vartheta)
 =\begin{cases}q,&y=\vartheta,\\1-q,&y\ne\vartheta.\end{cases}
\end{equation}
Her candidate strategy and wages give
\begin{align}
 V_{B\tau}^{T,\pi}
 ={}&\sum_{\vartheta,s_2,y,m}
 f_{\tau1}(1\mid\vartheta)f_{\tau2}(s_2\mid\vartheta)
 g_q(y\mid\vartheta)\mu_\pi(m\mid y,1)
 \notag\\[-1mm]
 &\quad\times\sum_{r_2}
 \widehat\sigma_{B2}^{\pi}(r_2\mid\tau,(1,s_2),1,m)
 \widetilde W_{\chi(1,r_2;\vartheta)}^\pi(m,y,1),
 \label{eq:oa-b-truth-value}\\
 V_{B\tau}^{D,\pi}
 ={}&\sum_{s_2,m}\max_{r_2}
 \sum_{\vartheta,y}
 f_{\tau1}(1\mid\vartheta)f_{\tau2}(s_2\mid\vartheta)
 g_q(y\mid\vartheta)\mu_\pi(m\mid y,0)
 \notag\\[-1mm]
 &\hspace{35mm}\times
 \widetilde W_{\chi(0,r_2;\vartheta)}^\pi(m,y,0).
 \label{eq:oa-b-deviation-value}
\end{align}
The deviation in \eqref{eq:oa-b-deviation-value} changes the agreement status
and thus the lottery. The four initial-report slacks are
\begin{equation}\label{eq:oa-locked-slacks}
 \Delta_{A\tau}^{\mathsf S}
 =V_{A\tau}^{T,\mathsf S}-V_{A\tau}^{D,\mathsf S},
 \qquad
 \Delta_{B\tau}^{\pi}
 =V_{B\tau}^{T,\pi}-V_{B\tau}^{D,\pi}.
\end{equation}

Under sealing, high-type terminal accuracy is \(p_{H2}\), whereas low-type
unconditional terminal accuracy is
\begin{equation}\label{eq:oa-terminal-report-accuracy}
 \Prb(r_2=\theta\mid L)
 =p_{L2}+h(p_{L1}-p_{L2}).
\end{equation}
Persistence changes how new information is used; whether it lowers terminal
accuracy depends on the relative precision of the two private signals.

\subsection{Analytical first-report incentives}
\label{oa:analytical-initial-incentives}

Maintain \(\eta_H>\eta_L\geq1\), and let \(b\) be the submitted initial
report. For a branch with likelihood ratio \(\lambda\) and persistence
\(h\), first assume
\begin{equation}\label{eq:oa-branch-binding}
\lambda G_C(h)=\eta_LG_I(h).
\end{equation}
This holds on every supported branch of \((\alpha,0)\) for
\(0\leq\alpha\leq\alpha^{\mathrm{DS}}\), including A's sealed branch.
Both \(G_C(h)>0\) and \(G_I(h)>0\) there.

After a lie, the expert knows her first signal was \(1-b\). If the second
signal is \(b\), her odds of \(\theta=b\) are \(\lambda\eta_\tau\);
if it is \(1-b\), they are \(\lambda/(\kappa_{\tau1}\kappa_{\tau2})\).
The respective gains from repeating \(b\) have signs
\begin{align}
\lambda\eta_\tau G_C(h)-G_I(h)
 &=(\eta_L\eta_\tau-1)G_I(h)\geq0,
 \label{eq:oa-lie-align-gain}\\
\frac{\lambda}{\kappa_{\tau1}\kappa_{\tau2}}G_C(h)-G_I(h)
 &=\left(\frac{\eta_L}{\kappa_{\tau1}\kappa_{\tau2}}-1\right)G_I(h)<0.
 \label{eq:oa-lie-conflict-gain}
\end{align}
The last sign follows from
\(\eta_L=\kappa_{L2}/\kappa_{L1}<\kappa_{L1}\kappa_{L2}
\leq\kappa_{\tau1}\kappa_{\tau2}\).
The second signal is therefore optimal at every post-lie information set.
Both choices are strict for \(\eta_L>1\). At \(\eta_L=1\), low-type
indifference in \cref{eq:oa-lie-align-gain} leaves the optimised payoff unchanged.

Index branches relative to \(b\). For B these are revealed agreement,
revealed disagreement and silence, omitting branches of zero mass. Define
\begin{equation}\label{eq:oa-branch-state-probabilities}
c_j=\Prb(j\mid\theta=b),\qquad
d_j=\Prb(j\mid\theta=1-b),\qquad
\lambda_j=c_j/d_j.
\end{equation}
Label symmetry of the prior and signal technology makes these probabilities
and wages the same functions of the policy whether \(b\) is truthful or
false. A has a single sealed branch with \(c_j=d_j=1\). Let
\begin{align}
T_\tau&=\sum_jc_j\{W_{CI}(h_j)+p_{\tau2}G_C(h_j)\},
\label{eq:oa-correct-first-payoff}\\
I_\tau&=\sum_jd_j\{W_{II}(h_j)+p_{\tau2}G_I(h_j)\}.
\label{eq:oa-incorrect-first-payoff}
\end{align}
Here $T_\tau$ and $I_\tau$ are the conditional wages under the hypothetical
strategy of always following the second signal after a correct and an
incorrect initial report, respectively. When the equilibrium mixes, the two actions
have equal expected payoffs given the private history and message, before
state verification. Integrating those indifferences gives the same interim
truth-telling payoff as the hypothetical strategy. The equality need not
hold separately conditional on the realised state.
After a lie, following the second signal is optimal by
\cref{eq:oa-lie-align-gain,eq:oa-lie-conflict-gain}. Thus
\begin{align}
V^T_{i\tau}&=p_{\tau1}T_\tau+(1-p_{\tau1})I_\tau,
\label{eq:oa-analytical-truth-value}\\
V^D_{i\tau}&=(1-p_{\tau1})T_\tau+p_{\tau1}I_\tau,
\label{eq:oa-analytical-lie-value}\\
\Delta_{i\tau}&=(2p_{\tau1}-1)(T_\tau-I_\tau).
\label{eq:oa-analytical-initial-slack}
\end{align}
The lie reverses the probability that the submitted report is correct and
retains the conditional branch probabilities
\cref{eq:oa-branch-state-probabilities}. This accounts for the changed lottery.

Decomposing the payoff difference gives
\begin{align}
T_\tau-I_\tau
={}&\sum_j\{c_jW_{CI}(h_j)-d_jW_{II}(h_j)\}\notag\\
&+p_{\tau2}\sum_j\{c_jG_C(h_j)-d_jG_I(h_j)\}.
\label{eq:oa-initial-slack-decomposition}
\end{align}
Under \cref{eq:oa-branch-binding}, the second sum is
\((\eta_L-1)\sum_jd_jG_I(h_j)\geq0\). Since \(W_{CI}(h)\) increases,
\(W_{II}(h)\) decreases, and \(\sum_jc_j=\sum_jd_j=1\), the first sum
is at least \(W_{CI}(0)-W_{II}(0)\).

To verify strict positivity from the primitives, set
\begin{equation}\label{eq:oa-pure-pattern-denominators}
D_\chi=\omega\ell_H^\chi(0)+(1-\omega)\ell_L^\chi(0),
\qquad \chi\in\{CC,CI,IC,II\}.
\end{equation}
Then
\begin{equation}\label{eq:oa-pure-first-report-premium}
W_{CI}(0)-W_{II}(0)
=\frac{\omega(1-\omega)(1-p_{H2})(1-p_{L2})(p_{H1}-p_{L1})}
 {D_{CI}D_{II}}>0.
\end{equation}
This positive difference gives the uniform lower bound for initial truth-telling.

For the whole square, add \(\gamma Q_0\leq1\). Every branch satisfies
\(\lambda\geq1/\gamma\geq Q_0\). A pure-revision branch has \(h=0\)
and \(\lambda\leq\eta_LQ_0\). Hence \(\lambda\eta_\tau\geq Q_0\)
and \(\lambda/(\kappa_{\tau1}\kappa_{\tau2})<Q_0\): following the second
signal remains optimal after a lie. Also,
\begin{equation}\label{eq:oa-pure-branch-premium}
c_jG_C(0)-d_jG_I(0)
=d_jG_C(0)(\lambda_j-Q_0)\geq0.
\end{equation}
The decomposition \cref{eq:oa-initial-slack-decomposition} therefore gives
the same bound on pure-revision branches. Confirmation is strict because
\(\lambda\kappa_{\tau1}\kappa_{\tau2}>Q_0\), and high-type conflict
revision is strict under \(\eta_H>\eta_L\). Every policy thus admits a
regular-monotone sequential equilibrium with strictly truthful initial reports.

\subsection{The primitive region and the share of high types}
\label{oa:explicit-primitive-region}

The ratio used in the region restrictions is
\begin{equation}\label{eq:oa-Q0-primitives}
Q_0=
\frac{(1-p_{H1})(1-p_{L1})}{p_{H1}p_{L1}}
\frac{D_{CC}D_{CI}}{D_{IC}D_{II}}.
\end{equation}
Together with the pure-revision likelihoods, this expresses the conditions
in the five primitives without an endogenous root.

For fixed signal precisions, \(D_{CC}/D_{IC}\) is a weighted average of
\(\kappa_{H1}\) and \(\kappa_{L1}\), with high-type weight
\[
\frac{\omega(1-p_{H1})p_{H2}}
 {\omega(1-p_{H1})p_{H2}+(1-\omega)(1-p_{L1})p_{L2}}.
\]
The weight increases strictly in \(\omega\). Replacing second-signal
precisions by error probabilities gives the corresponding representation of
\(D_{CI}/D_{II}\). Both ratios increase. By \eqref{eq:oa-Q0-primitives},
\(Q_0\) therefore rises strictly from \(\kappa_{L1}/\kappa_{H1}\) to
\(\kappa_{H1}/\kappa_{L1}\). The peer odds \(\gamma\) also increase strictly.

Under \(\eta_H>\eta_L>1\) and \(p_{L2}<p_{H1}\),
\(\eta_LQ_0(\omega)\) starts below one and ends above it. At its unique
crossing, \(\gamma\eta_LQ_0>1\), so
\(1/\gamma<\eta_LQ_0<1\) holds on a nonempty interval below the crossing.
The condition \(p_{L2}<p_{H1}\) is also necessary for this sufficient
region to be nonempty at a fixed technology. Otherwise monotonicity implies
\(\eta_LQ_0(\omega)>\kappa_{L2}/\kappa_{H1}\geq1\) for every \(\omega>0\).

At the benchmark precisions, \(\eta_L=15/11\) and \(\eta_H=133/33\).
The relevant endpoints satisfy
\begin{equation}\label{eq:oa-share-interval-endpoints}
\left.\gamma\eta_LQ_0\right|_{\omega=1/3}
=\frac{707805}{631283}>1,
\qquad
\left.\eta_LQ_0\right|_{\omega=1/2}
=\frac{116235}{139799}<1.
\end{equation}
Thus every \(\omega\in[1/3,1/2]\) supports the agreement-disclosure
segment. For whole-square feasibility,
\begin{equation}\label{eq:oa-whole-square-share-endpoint}
\left.\gamma Q_0\right|_{\omega=2/5}
=\frac{392931}{394279}<1.
\end{equation}
Since $\gamma Q_0$ increases with $\omega$, the inequality holds
throughout $[1/3,2/5]$.

Finally, \(q\) and \(\bar\lambda\) both increase with the high-type share.
Since \(\alpha^{\mathrm{DS}}\) decreases in each, it decreases strictly
with \(\omega\) wherever the threshold is nondegenerate. This concerns
the disclosure required for revision; it does not establish a welfare ranking.

\section{Reporting incentives at the illustrative profile}
\label{oa:selective-equilibrium}

Take the signal precisions
\begin{equation}\label{eq:oa-peer-precisions}
 p_{L1}=\frac{11}{20},
 \qquad
 p_{L2}=\frac58,
 \qquad
 p_{H1}=\frac{33}{40},
 \qquad
 p_{H2}=\frac{19}{20}.
\end{equation}
The odds $\kappa_{\tau t}=p_{\tau t}/(1-p_{\tau t})$ then give
\begin{equation}\label{eq:oa-peer-learning-rates}
 \eta_L=\frac{15}{11},
 \qquad
 \eta_H=\frac{133}{33},
 \qquad
 \eta_H-\eta_L=\frac83.
\end{equation}
At $\omega=2/5$, peer accuracy, peer odds, the wage cutoff and the silent
revision threshold are
\begin{equation}\label{eq:oa-peer-primitives}
 q=\frac{33}{50},
 \qquad
 \gamma=\frac{33}{17},
 \qquad
 Q_0=\frac{202419}{394279},
 \qquad
 \bar\lambda=\frac{3036285}{4337069}.
\end{equation}
Substitution yields
\begin{equation}\label{eq:oa-peer-continuation-region}
 \frac1{\eta_LQ_0}<\gamma
 <\frac{\kappa_{L1}\kappa_{L2}}{Q_0}.
\end{equation}

Clearing positive denominators in the sealed and revealed-agreement mixing
conditions gives
\begin{align}
 P^{\mathsf S}(h)
 &=11390625h^3-8262000h^2-8143395h+2113774,
 \label{eq:oa-peer-poly-low-sealed}\\
 P^{\mathsf F}(h)
 &=114665625h^3-129012750h^2-55925895h+49565984.
 \label{eq:oa-peer-poly-low-full}
\end{align}
Monotonicity of $Q$ selects one relevant root in each case. Exact rational
evaluation bounds them by
\begin{align}
0.2243083281662
 &<h^{\mathsf S}<0.2243083281663,
 \label{eq:oa-peer-root-low-sealed}\\
 0.5379129780400
 &<h^{\mathsf F}<0.5379129780401.
 \label{eq:oa-peer-root-low-full}
\end{align}

Disciplinary silence conceals every disagreement and reveals agreement with
probability
\begin{equation}\label{eq:oa-optimal-policy-formula}
 \alpha^*=\frac{1-\bar\lambda}{q-(1-q)\bar\lambda}.
\end{equation}
By \cref{eq:oa-peer-primitives}, this is
\begin{equation}\label{eq:oa-exact-optimal-policy-low}
 \alpha^*=\frac{2032475}{2859576}
 \approx0.710760965961387.
\end{equation}
The silent likelihood ratio satisfies
\begin{equation}\label{eq:oa-silence-identity}
 \frac{1-q\alpha^*}{1-(1-q)\alpha^*}=\bar\lambda.
\end{equation}
Thus $h_N=0$ leaves the low type indifferent after a silent conflict, while
the high type strictly follows her second signal.

Three comparisons suffice to check strict confirmation after silence,
strict revision after revealed disagreement, and strict confirmation when the
peer opposes both private signals. In the same order, the margins are
\[
 \bar\lambda\kappa_{L1}\kappa_{L2}-Q_0=\frac{359856}{394279},\qquad
 \gamma\eta_L-Q_0^{-1}=\frac{141536}{202419},\qquad
 \kappa_{L1}\kappa_{L2}/\gamma-Q_0=\frac{17117776}{31936599}.
\]
Initial truth-telling is already strict by Section~OA.2. For an indication
of the size of the incentives, substitute the root intervals above into the
truth-telling and optimal deviation payoffs in Section~OA.2.3. The wage
advantage exceeds $0.01931$ for A of type $L$, $0.14952$ for A of type $H$,
$0.02050$ for B of type $L$ and $0.15604$ for B of type $H$. Both
calculations allow the best final report after an initial lie. B's
calculation also uses the feedback lottery induced by the false report.
Section~OA.5 gives the resulting state-classification probabilities.

\section{Proof of optimality}
\label{oa:global-optimality}

\subsection{Analytical classification contributions}
\label{oa:analytical-global-optimality}

Let \(a=p_{L1}\), \(b=p_{L2}\), \(u=p_{H1}\), \(v=p_{H2}\), and
\(z=1-q\). Using the main-text notation, define
\begin{align*}
 A&=\omega uv+(1-\omega)ab,&
 B&=\omega u(1-v)+(1-\omega)a(1-b),\\
 C&=\omega(1-u)v+(1-\omega)(1-a)b,&
 D&=\omega(1-u)(1-v)+(1-\omega)(1-a)(1-b),\\
 \ell&=(1-\omega)a(1-b),&m&=(1-\omega)(1-a)b.
\end{align*}
Then \(A+B=q\), \(C+D=z\), and
\((A_h,B_h,C_h,D_h)=(A+\ell h,B-\ell h,C-mh,D+mh)\).
Write \(s=h^{\mathsf S}\), \(f=h^{\mathsf F}\), \(r(h)=\eta_LQ(h)\),
and \(\alpha_0=\alpha^{\mathrm{DS}}\).

Conditional likelihoods are products of the two experts' pattern probabilities
and the policy probability. Given the initial reports, the latter is
state-independent. Pair each history with its label reversal to cancel the
factor \(1/2\) in accuracy. Agreement and disagreement then contribute
\begin{align}
G(h)={}&\max\{A_sA_h,D_sD_h\}+\max\{A_sB_h,D_sC_h\}\notag\\
 &+\max\{B_sA_h,C_sD_h\}+\max\{B_sB_h,C_sC_h\},
 \label{eq:oa-agreement-classification}\\
H(h)={}&\max\{A_sC_h,D_sB_h\}+\max\{A_sD_h,D_sA_h\}\notag\\
 &+\max\{B_sC_h,C_sB_h\}+\max\{B_sD_h,C_sA_h\}.
 \label{eq:oa-disagreement-classification}
\end{align}
These are unconditional contributions. The complete experiment has value
\begin{equation}\label{eq:oa-classification-decomposition}
 \mathcal A^{(\alpha,\delta)}
 =\alpha G(f)+\delta H(0)+(1-\alpha)G(h_N)+(1-\delta)H(h_N).
\end{equation}
The message distinguishes branches with different continuation behaviour,
even though both initial reports are eventually observed.

In the theorem's region, \(b>a\), \(v>u\), and \(\eta_H>\eta_L\).
For \(0\leq h\leq f<1\),
\[
 A_h>D_h,\qquad C_h>B_h.
\]
The first affine difference is positive at zero and one. For the second,
\(C_h-B_h=\omega(v-u)+(1-\omega)(b-a)(1-h)>0\).
Define \(J=mC_s-\ell B_s\), \(K=mA-\ell D\), and \(L=\ell C-mB\).
Here $K>0$ because $A>D$ and $m>\ell>0$. Moreover,
\[
 L=\omega(1-\omega)
 \{(1-u)va(1-b)-u(1-v)(1-a)b\}>0,
 \qquad J-L=(m-\ell)(B_s+C_s)>0.
\]
The sign of \(L\) follows exactly from \(\eta_H>\eta_L\).

Let \(g_{10}(h)=A_sB_h-D_sC_h\) and
\(g_{01}(h)=B_sA_h-C_sD_h\). Subtraction gives
\[
 g_{01}'(h)=-J<0,\qquad
 g_{10}(h)-g_{01}(h)=(s-h)(\ell q-mz).
\]
Since \(m/\ell=\eta_L\), the condition \(\gamma\geq\eta_L\) implies
\(\ell q\geq mz\). Thus \(g_{10}(f)>0\) gives both \(g_{01}(f)>0\)
and \(g_{10}(0)>0\). Affinity makes the gaps positive on \([0,f]\).
The other agreement gaps have the signs established above, giving
\begin{equation}\label{eq:oa-affine-agreement-value}
 G(h)=A_sq+B_sA_h+C_sC_h=G(0)-Jh.
\end{equation}
For disagreement, the middle gaps are \(K(h-s)\) and \(L(h-s)\); the
remaining two retain fixed signs. Therefore
\begin{equation}\label{eq:oa-piecewise-disagreement-value}
 H(h)=
 \begin{cases}
 H(0)-Kh,&0\leq h\leq s,\\
 H(s)+L(h-s),&s\leq h\leq f.
 \end{cases}
\end{equation}

\subsection{A primitive curvature test}
\label{oa:analytical-response-curvature}

To derive convexity from the derivative at zero, define
\[
 H_* =\frac{v-b}{1-b},\quad e=\frac{v-b}{b},\quad
 x=\frac A\ell,\quad y=\frac B\ell,\quad
 t_* =\frac Cm,\quad d=\frac Dm,\quad k=\frac{1-u}{u}.
\]
Here \(H_*=h^\dagger\). Subtracting wages yields
\begin{equation}\label{eq:oa-rational-response-factorization}
 r(h)=k\frac{(h+e)(h+x)(y-h)}{(H_*-h)(t_*-h)(h+d)}.
\end{equation}
The restrictions give \(H_*<1<y<t_*\), \(x>d\), and \(d>e\).
In particular, \(L>0\) implies \(t_*>y\), \(K>0\) implies \(x>d\),
and \(D-me=(1-v)(1-q)>0\). A partial-fraction decomposition gives
\begin{equation}\label{eq:oa-positive-residue-response}
 r(h)=-k+\frac U{H_*-h}+\frac V{t_*-h}-\frac W{h+d},
\end{equation}
with
\begin{align*}
 U&=k\frac{(H_*+e)(H_*+x)(y-H_*)}{(t_*-H_*)(H_*+d)}>0,\\
 V&=k\frac{(t_*+e)(t_*+x)(t_*-y)}{(t_*-H_*)(t_*+d)}>0,\\
 W&=k\frac{(d-e)(x-d)(y+d)}{(H_*+d)(t_*+d)}>0.
\end{align*}
Hence \(r'''(h)>0\) on \(0\leq h<H_*\). The condition \(r''(0)\geq0\)
therefore establishes convexity on \([0,f]\). The derivative at zero is
the following rational function of the primitives:
\begin{align}
 r''(0)&=r(0)(E_1^2+E_2),\notag\\
 E_1&=\frac1{v-b}+\frac\ell A-\frac\ell B+\frac mC-\frac mD,\notag\\
 E_2&=\frac{1-2b}{(v-b)^2}
 -\ell^2\left(\frac1{A^2}+\frac1{B^2}\right)
 +m^2\left(\frac1{C^2}+\frac1{D^2}\right).
 \label{eq:oa-primitive-curvature-test}
\end{align}

\subsection{Global comparison of the two boundaries}
\label{oa:analytical-boundary-comparison}

Set \(V_0=G(0)+H(0)\). On \(\delta=0\), below the disciplinary threshold,
\begin{equation}\label{eq:oa-agreement-boundary-value}
 \mathcal A^{(\alpha,0)}
 =V_0-Jf\alpha-\{J(1-\alpha)+K\}h,
 \qquad h=h_N\in[0,s].
\end{equation}
At the threshold, value is \(V_*=V_0-Jf\alpha_0\). Beyond it, \(h=0\)
and value decreases strictly in \(\alpha\). Write
\(g=(J+K)s-Jf\alpha_0=V_*-\mathcal A^{\mathsf S}\) and
\(D_r=q-zr(h)>0\). Substituting the policy formula for \(0<h\leq s\)
gives the identity
\begin{align}
 \frac{(V_*-\mathcal A^{(\alpha,0)})D_r}{h}
 ={}&J\{qr(h)-z\}+K\{q-zr(h)\}\notag\\
 &-\frac{Jf(2q-1)\{r(h)-r(0)\}}
 {\{q-zr(0)\}h}.
 \label{eq:oa-agreement-gap-identity}
\end{align}
Under \(K\geq\gamma J\), the first line decreases in \(r\), so it is
at least \((2q-1)(J+K)\). Convexity gives
\((r(h)-r(0))/h\leq(1-r(0))/s\). Since
\(\alpha_0=(1-r(0))/(q-zr(0))\),
\begin{equation}\label{eq:oa-agreement-global-bound}
 V_*-\mathcal A^{(\alpha,0)}
 \geq\frac{h(2q-1)g}{s\{q-zr(h)\}}>0.
\end{equation}
The disciplinary threshold is therefore the unique optimum on this boundary.

On \(\alpha=0\), let \(F=f-s\) and \(t=(h_N-s)/F\in[0,1]\).
Convexity implies \(r(h_N)\leq1+(\gamma-1)t\). Since
\(\delta(r)=(r-1)/(qr-z)\) is increasing, \(\delta\leq t/(z+qt)\).
Using \eqref{eq:oa-affine-agreement-value} and
\eqref{eq:oa-piecewise-disagreement-value},
\begin{equation}\label{eq:oa-disagreement-boundary-value}
 \mathcal A^{(0,\delta)}
 =\mathcal A^{\mathsf S}-(J-L)Ft+\delta(Ks-LFt).
\end{equation}
If \(Ks-LFt\leq0\), this is below \(V_*\). Otherwise, the bound on
\(\delta\) gives
\begin{align}
 (V_*-\mathcal A^{(0,\delta)})(z+qt)
 \geq{}&zg+\{qg+z(J-L)F-Ks\}t+\{q(J-L)+L\}Ft^2\notag\\
 ={}&zg(1-t)^2+Mt(1-t)+Jf(1-\alpha_0)t^2>0,
 \label{eq:oa-positive-quadratic-bound}
\end{align}
where \(M=(2-q)g+z(J-L)F-Ks\). The last expression is positive throughout the unit interval for \(g>0\)
and \(M\geq0\). No monotonicity of the competing boundary is required.
Finally, ray affinity represents each interior value as a convex combination
of full disclosure and its normalised boundary value. Since full disclosure
is strictly below \(V_*\), the optimum is globally unique.

\subsection{An explicit interval of primitives}
\label{oa:analytical-optimality-interval}

Fix \((a,b,u,v)=(11/20,5/8,33/40,19/20)\) and
\(\omega\in[3/8,2/5]\). Section~OA.2 gives \(\eta_H>\eta_L>1\),
\(1/\gamma<r(0)<1\), and \(\gamma Q_0<1\) on this interval.
We verify the additional optimality conditions uniformly using rational bounds.

For a fixed rational \(h\), define
\[
 N_h=\eta_L\frac{(1-u)(1-a)}{ua}(v-b+bh)A_hB_h,
 \qquad T_h=\{v-b-(1-b)h\}C_hD_h.
\]
Then \(r(h)=N_h/T_h\), with \(T_h>0\) at each endpoint below. Let
\(\omega=3/8+t/40\), \(0\leq t\leq1\). At the sealed endpoints,
expand \(N_h-T_h\); at the agreement endpoints, expand
\((1-q)N_h-qT_h\). Each expression equals the positive scale in the table
times \(\sum_{i=0}^n b_i\binom ni t^i(1-t)^{n-i}\).
\begin{center}\scriptsize
\begin{tabular}{@{}ccc l@{}}
\toprule
root & \(h\) & positive scale & coefficients \((b_0,\ldots,b_n)\)\\
\midrule
sealed & \(11/50\) & \(1/2252800000000\) &
\((-697909725,-399156000,-111702208)\)\\
sealed & \(1/4\) & \(1/327680000000\) &
\((17469225,57987240,96990016)\)\\
agreement & \(53/100\) & \(1/346030080000000000\) &
\((-16969959799875,-17373425276400,\)\\
 & & & \(\quad -17669405049920,-17851670470656)\)\\
agreement & \(11/20\) & \(19/69206016000000000\) &
\((308079392625,298782920400,\)\\
 & & & \(\quad 290615633600,283642454016)\)\\
\bottomrule
\end{tabular}
\end{center}
The basis functions are nonnegative and sum to one. Their coefficient signs
and strict monotonicity of \(r\) therefore imply
\begin{equation}\label{eq:oa-uniform-root-bounds}
 \frac{11}{50}<s<\frac14,\qquad
 \frac{53}{100}<f<\frac{11}{20}
\end{equation}
uniformly in the type share. These conclusions follow from polynomial
identities over the interval.

For the remaining bounds, write
\begin{align*}
 K&=\omega(1-\omega)k_H+(1-\omega)^2k_L,\\
 J_{\mathrm{acc}}&=\omega(1-\omega)j_H+(1-\omega)^2j_L,
 &J&=J_{\mathrm{acc}}-(1-\omega)^2j_Ls,\\
 L&=\omega(1-\omega)l_H,
\end{align*}
where
\begin{align*}
 k_H&=uv(1-a)b-(1-u)(1-v)a(1-b),&k_L&=a(1-a)(2b-1),\\
 j_H&=(1-a)b(1-u)v-a(1-b)u(1-v),&j_L&=(1-a)^2b^2-a^2(1-b)^2,\\
 l_H&=(1-u)va(1-b)-u(1-v)(1-a)b.
\end{align*}
The derivatives of \(K\) and \(J_{\mathrm{acc}}\) in \(\omega\) are affine and negative
at both endpoints, while \(L\) increases. Combining these facts with
\cref{eq:oa-uniform-root-bounds} and the decrease of \(\alpha_0\) gives
\begin{center}
\begin{tabular}{@{}ccc@{}}
\toprule
quantity & lower bound & upper bound\\
\midrule
\(q\) & \(209/320\) & \(33/50\)\\
\(\alpha_0\) & \(71/100\) & \(783/1000\)\\
\(J\) & \(1877/100000\) & \(509/25000\)\\
\(K\) & \(3737/50000\) & \(3771/50000\)\\
\(L\) & \(531/100000\) & \(109/20000\)\\
\bottomrule
\end{tabular}
\end{center}
For curvature, the residues satisfy \(U>19/50\), \(W<61/1000\), and
\(d>119/200\), with \(H_*=13/15\). Using \(V>0\),
\[
 r''(0)>2\left\{\frac{19/50}{(13/15)^3}
             -\frac{61/1000}{(119/200)^3}\right\}>\frac12.
\]
Substitute the rational bounds and use
\[
 M=\{J+(1-q)(K+L)\}s
   +\{(1-q)(J-L)-(2-q)J\alpha_0\}f,
\]
to obtain
\begin{equation}\label{eq:oa-uniform-optimality-margins}
\begin{array}{c|c}
\text{comparison}&\text{uniform lower bound}\\ \hline
\gamma-\eta_L&1/2\\
A_sB_f-D_sC_f&7/500\\
K-\gamma J&7/200\\
g&1/100\\
M&7/10000
\end{array}
\end{equation}
The decision-gap bound uses the affine primitive ranges of
\(A,B,C,D,\ell,m\) and \cref{eq:oa-uniform-root-bounds}. All comparisons
follow by rational arithmetic from the displayed formulas. The verification
program checks their transcription. The theorem therefore holds throughout
the share interval.

Every sufficient inequality is strict at \(\omega=2/5\). The roots
\(s\) and \(f\) are uniquely defined and continuous, so the strict
conditions define an open sufficient region in all five primitives. This
does not extend optimality to the whole larger feasibility region.

\subsection{The threshold tie and strict revision}
\label{oa:strict-revision}

Let \(\alpha_0=\alpha^{\mathrm{DS}}\), \(r_0=\eta_LQ_0\),
\(d_0=1-(1-q)\alpha_0\), and \(d_\varepsilon=d_0-(1-q)\varepsilon\).
For \(0\leq\varepsilon\leq1-\alpha_0\),
\[
 \lambda_\varepsilon
 =\frac{1-q(\alpha_0+\varepsilon)}{d_\varepsilon},
 \qquad d_\varepsilon\geq q>0.
\]
Subtracting from the threshold value yields
\begin{equation}\label{eq:oa-strict-likelihood-drop}
 r_0-\lambda_\varepsilon
 =\frac{(2q-1)\varepsilon}{d_0d_\varepsilon}.
\end{equation}
Hence \(h_N=0\) throughout the segment and silent wages remain \(W_\chi(0)\).

Conditional on type \(\tau\), a truthful initial report, private conflict
and silence, the odds supporting the submitted report are
\(\lambda_\varepsilon/\eta_\tau\). The expected wage gain from revision is
\begin{equation}\label{eq:oa-strict-revision-premium}
 R_\tau(\varepsilon)
 =G_C(0)\frac{\eta_\tau Q_0-\lambda_\varepsilon}
 {\eta_\tau+\lambda_\varepsilon}.
\end{equation}
For the low type, \cref{eq:oa-strict-likelihood-drop} gives
\begin{equation}\label{eq:oa-strict-revision-rational-premium}
 R_L(\varepsilon)
 =\frac{G_C(0)(2q-1)\varepsilon}
 {d_0\bigl[(\eta_L+r_0)d_0-\{q+\eta_L(1-q)\}\varepsilon\bigr]}.
\end{equation}
The denominator is positive and bounded above by \((\eta_L+r_0)d_0^2\),
proving the linear lower bound. The premium increases strictly with
\(\varepsilon\) and tends to zero as \(\varepsilon\downarrow0\).
The high type has the uniform bound
\[
 R_H(\varepsilon)
 \geq G_C(0)\frac{(\eta_H-\eta_L)Q_0}{\eta_H+r_0}>0.
\]

To verify confirmation, set \(k_\tau=\kappa_{\tau1}\kappa_{\tau2}>1\).
On the silent branch the gain from repeating after confirmation is
\[
 G_C(0)\frac{\lambda_\varepsilon k_\tau-Q_0}
 {1+\lambda_\varepsilon k_\tau}
 \geq G_C(0)\frac{k_\tau/\gamma-Q_0}{1+k_\tau/\gamma}>0,
\]
because \(\lambda_\varepsilon\geq1/\gamma\geq Q_0\). Other branches keep
their previous wages and strict confirmation incentives. Whole-square
feasibility gives the same positive initial-report bounds for both types of
both experts, allowing the lie to change the lottery and subsequent reporting.

At \(\varepsilon=0\), \(R_L(0)=0\): the low type is indifferent at the
fixed threshold wage schedule. A different equilibrium persistence rate would
generate different Bayesian wages. For \(0<h<h^\dagger\), repetition has sign
\[
 r_0G_C(h)-\eta_LG_I(h)
 =\eta_LG_C(h)\{Q_0-Q(h)\}<0.
\]
For \(h\geq h^\dagger\), \(G_C(h)\leq0<G_I(h)\), again giving a negative
gain. No positive persistence is therefore optimal at its own wages.
The unique regular-monotone rate is zero; the argument leaves unrestricted
equilibria such as babbling available.

\subsection{Exact loss and the restricted supremum}
\label{oa:strict-approximation-value}

On \(\pi_\varepsilon=(\alpha_0+\varepsilon,0)\), silence has \(h_N=0\)
and revealed agreement has persistence \(f\). Thus
\cref{eq:oa-classification-decomposition} becomes
\[
 \mathcal A^{\pi_\varepsilon}
 =G(0)+H(0)+(\alpha_0+\varepsilon)\{G(f)-G(0)\}.
\]
Under the optimality conditions, \(G(f)-G(0)=-Jf<0\). Subtraction gives
the exact loss and both benchmark gains. The formula holds on the whole
segment, rather than only to first order.

Define the strict-action restriction by
\begin{equation}\label{eq:oa-strict-silence-policy-set}
 \Pi_{N,<}^{\mathrm{RM}}
 =\left\{\pi\in\Pi^{\mathrm{RM}}:
 \Prb_\pi(N)>0,\quad\lambda_N(\pi)<r_0\right\}.
\end{equation}
The cutoff lemma makes its inequality equivalent to strict low-type revision
after silence in the maintained class. The other continuation requirements
are unchanged. Every considered \(\pi_\varepsilon\) with \(\varepsilon>0\)
belongs to the set, and
\[
 \Prb_{\pi_\varepsilon}(N)
 =1-(\alpha_0+\varepsilon)\{q^2+(1-q)^2\}
 \geq2q(1-q)>0.
\]
The unique optimum \(\pi^{\mathrm{DS}}\) is excluded because
\(\lambda_N=r_0\). Consequently,
\begin{equation}\label{eq:oa-strict-policy-supremum}
 \sup_{\pi\in\Pi_{N,<}^{\mathrm{RM}}}\mathcal A^\pi
 =\mathcal A^{\mathrm{DS}},
 \qquad
 \operatorname*{arg\,max}_{\pi\in\Pi_{N,<}^{\mathrm{RM}}}
 \mathcal A^\pi=\varnothing.
\end{equation}
The equality follows from
\(\mathcal A^{\pi_\varepsilon}=\mathcal A^{\mathrm{DS}}-Jf\varepsilon\)
as \(\varepsilon\downarrow0\). Uniqueness of the excluded optimum implies
nonattainment. Any proposed restricted maximiser is improved upon by a
sufficiently small positive perturbation.

For tolerance \(\zeta>0\), choose
\(\varepsilon_\zeta=\tfrac12\min\{1-\alpha_0,g/(Jf),\zeta/(Jf)\}\).
The loss is at most \(\zeta/2<\zeta\), both benchmark gains are positive,
and the initial-report bounds remain fixed. The silent-revision margin is
positive for each tolerance but tends to zero with the tolerance. A fixed
positive revision margin is therefore not guaranteed at arbitrarily small loss.

\subsection{A bound for any finite downstream objective}
\label{oa:strict-general-objective}

Whole-square feasibility suffices for the behavioural and initial-incentive
claims. A value bound is also available for any finite problem \(D\).
Let \(T_D(h)\) be the unconditional optimal-payoff contribution of initial
agreement histories when A has sealed persistence and B has persistence
\(h\). Their total mass is \(q^2+(1-q)^2\), independent of \(h\).
Since initial reports are public, each affected outcome has a common
state-independent factor \(\alpha\) or \(1-\alpha\). Positive homogeneity gives
\begin{equation}\label{eq:oa-strict-general-affinity}
 V_D^{\pi_\varepsilon}-V_D^{\pi^{\mathrm{DS}}}
 =\varepsilon\{T_D(f)-T_D(0)\}.
\end{equation}
Disagreement histories are unchanged. Let
\(\operatorname{osc}(u_D)=\max_{a,\theta}u_D(a,\theta)
-\min_{a,\theta}u_D(a,\theta)\).
Both agreement contributions lie between the payoff extremes multiplied by
their common mass. Hence
\begin{equation}\label{eq:oa-strict-general-value-bound}
 \left|V_D^{\pi_\varepsilon}-V_D^{\pi^{\mathrm{DS}}}\right|
 \leq\operatorname{osc}(u_D)\{q^2+(1-q)^2\}\varepsilon.
\end{equation}
For an arbitrary objective, the value change can have either sign. The
optimality and benchmark rankings above concern symmetric classification.

\subsection{Uniform strict-approximation bounds}
\label{oa:uniform-strict-approximation}

Maintain the precisions and share interval from
\cref{oa:analytical-optimality-interval}. The rational bounds imply
\[
 \frac{71}{100}<\alpha_0<\frac{783}{1000},\qquad
 g>\frac1{100},\qquad
 0<Jf<\frac{509}{25000}\frac{11}{20}
 =\frac{5599}{500000}.
\]
and, in particular,
\[
 \mathcal A^{\mathsf F}-\mathcal A^{\mathsf S}
 =g-Jf(1-\alpha_0)
 >\frac1{100}-\frac{5599}{500000}\frac{29}{100}
 =\frac{337629}{50000000}>0.
\]
Every \(\alpha\in(\alpha_0,1)\) with \(\delta=0\) therefore improves on
both benchmarks. Since \(1-\alpha_0>217/1000\), the common range
\(0<\varepsilon\leq1/10\) is admissible.

For a root-free revision bound, use the zero-persistence wages
\[
 G_C(0)=\frac{\omega(1-\omega)p_{H1}p_{L1}(p_{H2}-p_{L2})}{AB}.
\]
The affine endpoint probabilities give
\[
 G_C(0)\geq\frac{1625}{3528}>\frac9{20},\qquad
 W_{CI}(0)-W_{II}(0)\geq\frac{125}{1624}>\frac3{40}.
\]
Since \(2p_{L1}-1=1/10\) is the smaller type coefficient, every
initial-report margin exceeds \(3/400\). Also, \(r_0<71/100\),
\(2q-1\geq49/160\), and \(d_0<3793/5000\), so the coefficient on
\(\varepsilon\) in the revision bound exceeds
\[
 \frac{(9/20)(49/160)}
 {(15/11+71/100)(3793/5000)^2}
 =\frac{3789843750}{32816402569}>\frac1{10}.
\]
These inequalities establish the uniform main-text bounds.

For example, \(\varepsilon=1/10\) gives a silent-revision premium above
\(1/100\), loss below \(5599/5000000\), gain over sealing above
\(44401/5000000\), and gain over full disclosure above
\(11639277/10000000000\). They are analytical worst-case bounds over the
primitive interval.

\subsection{Robustness of the analytical conditions}

At the benchmark, all sufficient inequalities are strict, including
\(\gamma Q_0<1\) and \cref{eq:oa-uniform-optimality-margins}.
Since \(Q'(h)>0\), both roots are continuous in the primitives. Strictness
therefore persists on an open neighbourhood in all five dimensions, where
the analytical boundary proof applies pointwise. The archived derivative-count
certificate and sensitivity rows are not used in this argument. In particular,
the non-baseline sensitivity rows do not establish global optimality.

\section{The size of the accuracy gain}
\label{oa:benchmark}

Use the four precisions in Section~OA.3 and let $\omega=2/5$. At these
primitives the optimal rule reveals roughly $71$ per cent of agreements.
The root intervals in Section~OA.3 and the history probabilities in
Section~OA.2 give the following comparisons. Values in the last column
are rounded to six decimal places.

\begin{table}[ht]
\caption{Feedback, persistence and classification at the illustrative profile}
\label{tab:main-results}
\centering\small
\begin{tabular}{@{}lccc@{}}\toprule
&\multicolumn{2}{c}{Low-type persistence after conflict}&\\
\cmidrule(lr){2-3}
Policy & After silence & After revealed agreement & Accuracy\\\midrule
Sealing & $0.2243$ & --- & $0.778347$\\
Full disclosure & --- & $0.5379$ & $0.789032$\\
Selective disclosure & $0$ & $0.5379$ & $0.792049$\\\bottomrule
\end{tabular}
\end{table}
\FloatBarrier

A dash denotes an event of probability zero. A disclosed disagreement always
induces revision. Relative to full disclosure, the selective rule raises
accuracy by about $0.302$ percentage points, or three additional correct
decisions per thousand cases. Relative to sealing the gain is $1.370$
percentage points. For a decision worth $L$ when correct and zero
otherwise, the largest additional cost the platform could recover relative
to full disclosure at these primitives is about $0.003017L$ per case.

\section{Blackwell incomparability of sealing and full disclosure}
\label{oa:blackwell-incomparability}

The following benchmark calculation independently checks the interval-wide
incomparability result proved in the main text and Section~OA.10.

For $k\in(0,1)$, define the call value
\begin{equation}\label{eq:oa-call-value}
 C^{\mathsf P}(k)
 =\frac12\sum_z
 \left[(1-k)P_1^{\mathsf P}(z)-kP_0^{\mathsf P}(z)\right]_+,
 \qquad
 \mathsf P\in\{\mathsf S,\mathsf F\}.
\end{equation}
Action 1 pays $1-k$ in state 1 and $-k$ in state 0; action 0 pays zero.
Exact rational evaluation of the complete public experiments gives
\begin{equation}\label{eq:oa-blackwell-signs}
 \begin{array}{@{}c@{\qquad}c@{}}
 \toprule
 k&C^{\mathsf F}(k)-C^{\mathsf S}(k)\\
 \midrule
 1/10&(-0.002333574526654,-0.002333574526575)\\
 1/2 &(\phantom{-}0.005342559386091,\phantom{-}0.005342559386116)\\
 \bottomrule
 \end{array}
\end{equation}
Each protocol is strictly preferred in one of these decision problems.
Since Blackwell dominance would require a weak preference in every problem,
neither complete experiment dominates the other.

\section{Reproducing the calculations}
\label{oa:reproducibility}

The ancillary file \path{Verification_Code.zip} contains the Python programs. Extract it into an empty directory.
From that directory run \texttt{python3 run\_checks.py}. The program uses
the standard library and, for the symbolic polynomial checks, SymPy.
The instructions accompanying the code give the versions used and the
location of the output.

The programs in \path{working/code/} compute the benchmark mixing roots,
initial-report margins and decision values in Sections~OA.3, OA.5 and OA.6.
They also evaluate the recording-error example in Section~OA.1. For that
example the endpoint is defined by the wage-gap equation. The path argument
uses inequalities on the entire parameter interval. The sample table in
\path{moderate_accuracy_feasibility.py} sets
$\varepsilon_1=\varepsilon_2$, while the initial incentive bound also
covers $\varepsilon_1=0$.

The \path{verification/} directory contains the polynomial and interval
checks for optimality, the reward and two-peer extensions, and the objective
comparison in Section~OA.10. It also contains a separate reconstruction of
the full public experiment and the calculation of gains per thousand
decisions. The program \path{verify_neologisms.py} enumerates the terminal
announcements for the benchmark babbling and repetition equilibria and for
both experts at the benchmark under informative play. Section~OA.11 gives the analytical argument for
both experts under informative play. To include two earlier benchmark and
sensitivity calculations, run \texttt{python3 run\_checks.py
--include-archived}.

\section{Deterministic confirmation with two peers}
\label{oa:two-peer-confirmation}

Add a second peer with the same signal technology and reputational payoff.
Peers \(A_1,A_2\) receive no feedback. B receives confirmation exactly when
both initial peer reports equal her initial report \(b\). All six reports
are eventually public. The sealing and full-disclosure benchmarks use this
same three-expert environment.

\subsection{Information, wages, and continuation choices}

Let \(K\) be the number of peer reports equal to \(b\), and set \(z=1-q\).
Under truthful initial reporting,
\[
 K\mid\theta=b\sim\operatorname{Bin}(2,q),\qquad
 K\mid\theta\ne b\sim\operatorname{Bin}(2,z).
\]
Confirmation, \(K=2\), has likelihood ratio \(\gamma^2\). The pooled
silent event \(K<2\) has ratio \(\lambda_N^{(2)}=(1-q^2)/(1-z^2)\).
Full disclosure gives ratios \(\gamma^2,1,\gamma^{-2}\) at \(K=2,1,0\).
These probabilities arise from the reports; the platform's rule is deterministic.

Given the state and initial reports, the message is fixed. Joint likelihoods
factor into each peer's reporting likelihood and B's reporting likelihood
conditional on her message. A peer's posterior ability therefore depends
only on her own history and the state. For B it depends on her history,
message and state. The exact peer configuration within a pooled message adds
no type-dependent term. Wages remain \(W_\chi(h)\).

Let \(s=h(1)\) and \(h_2=h(\gamma^2)\). Under
\(\eta_H>\eta_L\geq1\) and
\(Q_0\leq\lambda_N^{(2)}<\eta_LQ_0<1\), each peer has persistence
\(s\), while B has \(h_2\) after confirmation and zero after silence.
High types strictly revise after conflict. On mixing branches,
\(\lambda G_C=\eta_LG_I\) and
\(\eta_L\kappa_{\tau1}\kappa_{\tau2}>1\) ensure strict confirmation.
For silence, \(\lambda_N^{(2)}\geq Q_0\) gives the same result.

Following an initial lie, type \(\tau\)'s odds supporting the submitted
report are \(\lambda\eta_\tau\) when the second signal agrees with it,
and \(\lambda/(\kappa_{\tau1}\kappa_{\tau2})\) otherwise. On both
mixing and silent branches, the second signal is an optimal terminal choice.
Apply the decomposition from Section~OA.2, with \(c_j,d_j\) conditional
on a correct and incorrect submitted report. The second sum is nonnegative:
mixing gives \(c_jG_C=\eta_Ld_jG_I\), and silence gives
\(c_jG_C\geq d_jG_I\). The first sum is at least
\(W_{CI}(0)-W_{II}(0)>0\). This establishes B's uniform initial bound;
the sealed-peer argument is the same.

The deviation calculation conditions the confirmation probability on the
report actually submitted, including a lie. Every compatible report history
has positive probability because high types follow full-support signals.
Messages inconsistent with the deterministic rule are impossible. Independent
reporting trembles give the Bayesian assessment on supported histories, and
optimal post-lie continuation completes the sequential equilibrium.

\subsection{Primitive and root bounds}

Fix \((p_{L1},p_{L2},p_{H1},p_{H2})=(11/20,5/8,33/40,19/20)\) and
\(\omega\in[3/8,2/5]\). The previous bounds and monotonicity of \(Q_0\) give
\begin{equation}\label{eq:oa-two-peer-primitive-bounds}
 \frac{209}{320}\leq q\leq\frac{33}{50},\quad
 \frac{49}{100}<Q_0<\frac{13}{25},\quad
 \frac{671}{1000}<\eta_LQ_0<\frac{71}{100},\quad
 \frac{319}{500}<\lambda_N^{(2)}<\frac{163}{250}.
\end{equation}
Here \(\eta_L=15/11\) and \(\eta_H=133/33\).
Since \(\lambda_N^{(2)}\) decreases with \(q\), its endpoint values
give the last bounds and the strict confirmation-protocol inequalities.

The sealed root satisfies \(11/50<s<1/4\). The additional bound is
\begin{equation}\label{eq:oa-two-peer-root-bound}
 \frac7{10}<h_2<\frac{71}{100}.
\end{equation}
To verify it algebraically, write \(r(h)=\eta_LQ(h)=N_h/T_h\), where
\begin{align*}
 N_h&=\eta_L\frac{(1-p_{H1})(1-p_{L1})}{p_{H1}p_{L1}}
 \{p_{H2}-p_{L2}+p_{L2}h\}(A+\ell h)(B-\ell h),\\
 T_h&=\{p_{H2}-p_{L2}-(1-p_{L2})h\}(C-mh)(D+mh).
\end{align*}
Denominators are positive at both proposed bounds. Let
\(x=40\omega-15\in[0,1]\) and
\(\mathcal B_i(x)=\binom4i x^i(1-x)^{4-i}\). Expansion yields
\begin{align*}
 z^2N_{7/10}-q^2T_{7/10}
 &=\frac{\sum_{i=0}^4 c_i^-\mathcal B_i(x)}{692060160000000000},\\
 z^2N_{71/100}-q^2T_{71/100}
 &=\frac{\sum_{i=0}^4 c_i^+\mathcal B_i(x)}{138412032000000000000},
\end{align*}
with
\[
\begin{array}{c|r|r}
 i&c_i^-&c_i^+\\\hline
 0&-4866826089375&4320868683841875\\
 1&-9396052699500&3422735338725000\\
 2&-13829159045600&2542682347316000\\
 3&-18163387272960&1681242474551040\\
 4&-22395964686336&838951748960256
\end{array}
\]
The basis functions are nonnegative and sum to one, making the first
polynomial negative and the second positive on the interval. Strict
monotonicity of \(r\) then proves \cref{eq:oa-two-peer-root-bound} without
a numerical root solution.

\subsection{Full-disclosure incentives}

The primitive bounds imply
\[
 \gamma^2Q_0<\frac{14157}{7225}<2<\kappa_{L1}\kappa_{L2}=
 \frac{55}{27},\qquad
 \gamma^{-2}\eta_L<Q_0<\gamma^{-2}\eta_H.
\]
Confirmation is therefore strict under full disclosure, even after unanimous
opposition. Low-type conflict persistence is \(h_2,s,0\) at \(K=2,1,0\);
the high type follows her second signal. Following a lie, the low type at
\(K=0\) chooses \(1-b\) even when the second signal equals \(b\).
On other branches she follows the second signal, as the high type does on
every branch.

We must deduct this extra low-type deviation gain from the second-signal
comparison. Define
\begin{align}
 P_0&=W_{CI}(0)-W_{II}(0),\qquad
 D_{CI}=W_{CI}(h_2)-W_{CI}(0),\notag\\
 \mathcal T&=(\eta_L-1)\{z^2G_I(h_2)+2qzG_I(s)\}
             +z^2G_C(0)-q^2G_I(0),\notag\\
 E_L&=p_{L1}(1-p_{L2})q^2G_I(0)
       -(1-p_{L1})p_{L2}z^2G_C(0)>0.
 \label{eq:oa-two-peer-deviation-terms}
\end{align}
If both truthful and false initial reports are followed by the second signal,
the payoff difference is
\begin{align*}
 (2p_{\tau1}-1)\bigl[ P_0
 &+q^2\{W_{CI}(h_2)-W_{CI}(0)\}
 +2qz\{W_{CI}(s)-W_{CI}(0)\}\\
 &+z^2\{W_{II}(0)-W_{II}(h_2)\}
 +2qz\{W_{II}(0)-W_{II}(s)\}
 +p_{\tau2}\mathcal T\bigr].
\end{align*}
Truth-telling has this equilibrium continuation value because mixing occurs
at indifference. Optimal continuation after a lie adds \(E_L\) for the low
type and zero for the high type. In particular,
\begin{align}
 \Delta_{BL}^{\mathsf F_2}
 &\geq(2p_{L1}-1)\{P_0+q^2D_{CI}+p_{L2}\mathcal T\}-E_L,
 \label{eq:oa-two-peer-full-initial-low}\\
 \Delta_{BH}^{\mathsf F_2}
 &\geq(2p_{H1}-1)\{P_0+q^2D_{CI}+p_{H2}\mathcal T\}.
 \label{eq:oa-two-peer-full-initial-high}
\end{align}
All omitted terms are nonnegative by wage monotonicity.

Endpoint fractions give uniform wage bounds. Wages increase in \(\omega\).
For \(h\in[h_-,h_+]\), bound \(W_{CC},W_{II}\) at
\((3/8,h_+)\) and \((2/5,h_-)\), and \(W_{CI},W_{IC}\) at
\((3/8,h_-)\) and \((2/5,h_+)\). Apply these bounds to \(h=0\),
\(s\in[11/50,1/4]\), and \(h_2\in[7/10,71/100]\).
Substitution into \cref{eq:oa-two-peer-deviation-terms} gives
\[
 P_0\geq\frac{125}{1624},\qquad D_{CI}>\frac16,\qquad
 \mathcal T>\frac1{100},\qquad E_L<\frac{39}{5000}.
\]
Hence
\[
 \Delta_{BL}^{\mathsf F_2}
 >\frac1{10}\left\{\frac{125}{1624}
       +\frac16\left(\frac{209}{320}\right)^2+\frac1{160}\right\}
       -\frac{39}{5000}
 =\frac{47591767}{6236160000}>\frac3{400}.
\]
The high-type margin exceeds \((13/20)(125/1624)>3/400\). The earlier
bound, also above \(3/400\), applies to both peers and to B under the other
protocols. Thus all six initial constraints hold under each protocol,
including optimised continuation after a lie.

Under \(\mathsf U\), the low-type silent revision gain is
\[
 G_C(0)\frac{\eta_LQ_0-\lambda_N^{(2)}}{\eta_L+\lambda_N^{(2)}}
 >\frac{(9/20)(671/1000-163/250)}{15/11+163/250}
 =\frac{1881}{443440}>\frac1{250}.
\]
The bound uses $G_C(0)>9/20$ from Section~OA.4.

\subsection{Classification contributions and benchmark comparisons}

For these deterministic protocols, the three initial reports determine the
message. Once all six reports are retained, the message can be omitted from
the classification calculation. This simplification does not apply to the
original disclosure lottery.

Write \(a=A_s,b_s=B_s,c=C_s,d=D_s\), and define
\[
 \Psi(P,Q;h)=\max\{P A_h,Q D_h\}+\max\{P B_h,Q C_h\}.
\]
Here \(b_s\) denotes a pattern probability; \(b\) elsewhere is B's initial
report. The unconditional contributions of unanimous agreement, unanimous
opposition and a split are
\begin{align}
 U(h)&=\Psi(a^2,d^2;h)+2\Psi(ab_s,dc;h)+\Psi(b_s^2,c^2;h),\notag\\
 V(h)&=\Psi(c^2,b_s^2;h)+2\Psi(cd,ab_s;h)+\Psi(d^2,a^2;h),\notag\\
 T(h)&=2\{\Psi(ac,b_sd;h)+\Psi(ad,ad;h)
                  +\Psi(b_sc,b_sc;h)+\Psi(b_sd,ac;h)\}.
 \label{eq:oa-two-peer-classification-contributions}
\end{align}
State-label symmetry cancels the prior factor \(1/2\). The additional factor
two in \(T\) counts the two split orientations. Therefore
\begin{align*}
 \mathcal A_2^{\mathsf S}&=U(s)+V(s)+T(s),\\
 \mathcal A_2^{\mathsf F}&=U(h_2)+V(0)+T(s),\\
 \mathcal A_2^{\mathsf U}&=U(h_2)+V(0)+T(0).
\end{align*}

The root bounds give strict rational enclosures
\[
\begin{array}{c|cc}
 &\text{lower bound}&\text{upper bound}\\\hline
 a&5359/10000&552/1000\\
 b_s&108/1000&1172/10000\\
 c&1913/10000&201/1000\\
 d&1418/10000&1527/10000
\end{array}
\]
Substitute these and the affine formulas for \(A_h,B_h,C_h,D_h\) into
the two terms of each \(\Psi\). Writing P or Q for the first or second
maximising term gives
\[
\begin{array}{c|c|c}
 \text{contribution}&\text{choices in row order}&
 \text{lower bound on all signed margins}\\\hline
 U(s)&PP,\ PP,\ PQ&1/20000\\
 V(0),\ V(s)&PP,\ PQ,\ QQ&1/1000\\
 T(0),\ T(s)&PP,\ PQ,\ PQ,\ QQ&1/1000
\end{array}
\]
Margins for the equal-weight rows of \(T\) include those weights. Each
comparison is affine in \(h\), so its endpoint signs hold throughout
\([0,s]\) for \(V\) and \(T\). Letting \(\Xi=ad+b_sc\), we obtain
\begin{align*}
 V(h)&=c^2q+2\{cd A_h+ab_s C_h\}+a^2z,\\
 T(h)&=2ac+2\Xi(A_h+C_h),\qquad 0\leq h\leq s.
\end{align*}
The slopes are $-2K_2$ and $-2\Xi(m-\ell)$, respectively. At \(s\), the first
decision pattern also gives
\[
 U(s)=(a^2+2ab_s)q+b_s^2A_s+c^2C_s.
\]
Keeping that rule at \(h_2\) yields
\(U(h_2)-U(s)\geq-J_2(h_2-s)\), even if the rule ceases to be optimal.
Thus no fixed decision region at \(h_2\) is needed. These identities give
the exact full-disclosure gain and the sealing lower bound.

The same enclosures imply
\[
 \Xi>\frac{12}{125},\qquad K_2>\frac{29}{5000},\qquad
 0<J_2<\frac{57}{10000},\qquad m-\ell\geq\frac9{200}.
\]
Using \(s>11/50\) and \(h_2-s<49/100\),
\begin{align}
 \mathcal A_2^{\mathsf U}-\mathcal A_2^{\mathsf F}
 &>\frac{297}{156250}>\frac{19}{10000},\notag\\
 \mathcal A_2^{\mathsf U}-\mathcal A_2^{\mathsf S}
 &>\frac{8299}{5000000}>\frac1{625}.
 \label{eq:oa-two-peer-uniform-gains}
\end{align}
Both inequalities hold for the same three experts at every share on the
specified interval. The two differences therefore measure the value of
changing the feedback rule with the available information held fixed.

\section{Direct accuracy rewards}
\label{oa:accuracy-rewards}

Consider payoff \(\rho_i+\beta\ind\{r_{i2}=\theta\}\), corresponding to
\(\varepsilon_1=0\), \(\varepsilon_2=\beta\) in Section~OA.1.
We now apply it to the informative response. The reward is fixed across
policies, and the evaluator's information and Bayesian wage rule are unchanged.

\subsection{Continuation and initial reporting}

As in Section~OA.4, write \(a=p_{L1},b=p_{L2},u=p_{H1},v=p_{H2}\).
The effective terminal payoff gaps are
\[
 \widetilde G_C(h)=G_C(h)+\beta,\qquad
 \widetilde G_I(h)=G_I(h)+\beta.
\]
The low-type conflict numerator is
\(F_{\lambda,\beta}(h)=\lambda\widetilde G_C(h)-\eta_L\widetilde G_I(h)\).
Wage monotonicity makes it strictly decreasing on \([0,1]\). At an interior
root, \(\widetilde G_C>0\) because \(\widetilde G_I>0\).
Implicit differentiation gives the main-text derivative. The dependence of
the numerator on \(\beta\) is \((\lambda-\eta_L)\beta\).

For the two-expert result, fix precisions \((11/20,5/8,33/40,19/20)\),
\(\omega\in[3/8,2/5]\), and \(0\leq\beta\leq1/10\). The earlier
primitive bounds imply
\begin{align*}
 &\eta_L=\frac{15}{11},\quad \eta_H=\frac{133}{33},\quad
 \frac1\gamma<r_0<\frac{71}{100}<1,\\
 &\gamma Q_0\leq1,\quad
 \frac{49}{100}<Q_0<\frac{13}{25},\quad G_C(0)>\frac9{20}.
\end{align*}
In particular, \(G_I(0)=Q_0G_C(0)>11/50\). Define
\(Q_{\beta,0}=(G_I(0)+\beta)/(G_C(0)+\beta)\). Since \(Q_0<1\), it
increases in \(\beta\) and decreases in \(G_C(0)\) at fixed \(Q_0\). Thus
\begin{equation}\label{eq:oa-accuracy-effective-bounds}
 Q_{\beta,0}
 <\frac{(13/25)(9/20)+1/10}{9/20+1/10}
 =\frac{167}{275}<\frac58<\frac{\eta_L}{\gamma},\qquad
 r_\beta=\eta_LQ_{\beta,0}<1.
\end{equation}
We also have \(r_\beta\geq r_0>1/\gamma\).

The bounds below give \(0<s_\beta< f_\beta<29/50<h^\dagger\).
Every supported message has \(\lambda\in[1/\gamma,\gamma]\), so its root
is zero or belongs to \((0,f_\beta]\). At an interior root,
\(\eta_H>\eta_L\) gives strict high-type revision. At zero persistence,
the same follows from \(\lambda\leq\eta_LQ_{\beta,0}\).
Mixing-branch confirmation is strict by
\(\lambda\widetilde G_C=\eta_L\widetilde G_I\). On pure branches,
\(\lambda\geq1/\gamma\) and \cref{eq:oa-accuracy-effective-bounds} give
\(\lambda\kappa_{\tau1}\kappa_{\tau2}>Q_{\beta,0}\).

After an initial lie, the odds supporting the submitted report are
\(\lambda\eta_\tau\) if the second signal agrees with it, and
\(\lambda/(\kappa_{\tau1}\kappa_{\tau2})\) otherwise. On mixing branches,
substituting \(\lambda\widetilde G_C=\eta_L\widetilde G_I\) makes
following the second signal optimal, as in Section~OA.2. On pure branches,
\[
 \lambda\eta_\tau>Q_{\beta,0},\qquad
 \frac{\lambda}{\kappa_{\tau1}\kappa_{\tau2}}
 \leq\frac{\eta_LQ_{\beta,0}}{\kappa_{\tau1}\kappa_{\tau2}}
 <Q_{\beta,0}.
\]
So the optimal choice on these branches also follows the second signal.
Both calculations condition the lottery on the report the expert submits.

Let \(c_j,d_j\) be the conditional branch probabilities from Section~OA.2.
Under truth-telling, following the second signal gives the equilibrium total
payoff because departures occur only at mixing indifference. Its accuracy
is \(p_{\tau2}\) after either a correct or incorrect initial report.
The direct reward therefore cancels from the initial comparison, giving
\begin{equation}\label{eq:oa-accuracy-initial-decomposition}
 \Delta_{i\tau,\beta}
 =(2p_{\tau1}-1)\left[
 \sum_j\{c_jW_{CI}(h_j)-d_jW_{II}(h_j)\}
 +p_{\tau2}\sum_j\{c_jG_C(h_j)-d_jG_I(h_j)\}\right].
\end{equation}
The first sum is at least \(P_0=W_{CI}(0)-W_{II}(0)\).
At an interior root, a term of the second sum is
\[
 c_jG_C-d_jG_I
 =d_j\{(\eta_L-1)G_I+\beta(\eta_L-\lambda_j)\}>0,
\]
because
\[
 (\eta_L-1)G_I(0)-\beta(\gamma-\eta_L)
 >\frac4{11}\frac{11}{50}-\frac1{10}\frac7{11}>0.
\]
At a zero root, the term is \(d_jG_C(0)(\lambda_j-Q_0)\geq0\), since
\(\gamma Q_0\leq1\). Hence every policy satisfies
\begin{equation}\label{eq:oa-accuracy-initial-bound}
 \Delta_{i\tau,\beta}\geq(2p_{\tau1}-1)P_0
 \geq(2p_{\tau1}-1)\frac{125}{1624}>\frac3{400}.
\end{equation}
This holds for each expert and type. The high type's full-support signals make every
mechanically possible public history occur with positive probability.
Reporting trembles as in Section~OA.2 provide consistency; the optimised
private deviations complete sequential rationality.

\subsection{Uniform root and classification bounds}

For the root bounds, set \(z=1-q\) and define, for \(k=0,1,2\),
\begin{align}
 \mathcal P_k(h,\beta)
 ={}&\omega(1-\omega)
 \bigl[q^k ua\{v-b-(1-b)h\}C_hD_h\notag\\
 &\hspace{19mm}-\eta_Lz^k(1-u)(1-a)
       \{v-b+bh\}A_hB_h\bigr]\notag\\
 &+\beta(q^k-\eta_Lz^k)A_hB_hC_hD_h.
 \label{eq:oa-accuracy-root-polynomial}
\end{align}
Wage subtraction gives
\[
 \mathcal P_k(h,\beta)
 =z^kA_hB_hC_hD_h F_{\gamma^k,\beta}(h).
\]
The multiplier is positive at every endpoint used below. At fixed \(h\)
and \(\beta\), substitute \(\omega=3/8+x/40\). The resulting polynomial
has degree at most \(n=k+4\). Write
\[
 \mathcal P_k=\sum_{j=0}^n a_jx^j
 =\sum_{i=0}^n b_i\binom ni x^i(1-x)^{n-i},\qquad
 b_i=\sum_{j=0}^i\frac{\binom ij}{\binom nj}a_j.
\]
Substitution into \cref{eq:oa-accuracy-root-polynomial} bounds every
coefficient in each row as follows:
\begin{center}
\begin{tabular}{@{}ccccc@{}}
\toprule
\(k\)&\(h\)&\(\beta\)&\(n\)&coefficient bound\\
\midrule
0&\(17/100\)&\(1/10\)&4&\(b_i>3/10^6\)\\
1&\(29/50\)&\(1/10\)&5&\(b_i<-6/10^6\)\\
0&\(11/50\)&\(1/1000\)&4&\(b_i>4/10^6\)\\
2&\(71/100\)&\(1/1000\)&6&\(b_i<-4/10^7\)\\
0&0&\(2/5\)&4&\(b_i>2/10^7\)\\
0&0&\(43/100\)&4&\(b_i<-3/10^6\)\\
\bottomrule
\end{tabular}
\end{center}
These inequalities follow by finite expansion of the polynomial and binomial
formula. Nonnegative Bernstein basis functions sum to one, so the signs hold
over the whole share interval.

The sealed root decreases with \(\beta\), while the \(\gamma\) and
\(\gamma^2\) roots increase because \(\gamma>\eta_L>1\).
The first two rows and the zero-reward bounds from Section~OA.4 imply
\begin{equation}\label{eq:oa-accuracy-uniform-roots}
 \frac{17}{100}<s_\beta<\frac14,\qquad
 \frac{53}{100}<f_\beta<\frac{29}{50},\qquad
 0\leq\beta\leq\frac1{10}.
\end{equation}
The next two rows give the deterministic-extension bounds
\begin{equation}\label{eq:oa-accuracy-two-peer-roots}
 \frac{11}{50}<s_\beta<\frac14,\qquad
 \frac7{10}<h_\beta(\gamma^2)<\frac{71}{100},\qquad
 0\leq\beta\leq\frac1{1000}.
\end{equation}
The last two rows bound the reward at which the sealed expert always revises.

For two experts, define \(G_\beta(h),H_\beta(h)\) by the Section~OA.4
formulas with \(s\) replaced by \(s_\beta\). That argument requires
\(\gamma\geq\eta_L\) and
\(A_{s_\beta}B_{f_\beta}-D_{s_\beta}C_{f_\beta}>0\).
The affine primitive ranges and \cref{eq:oa-accuracy-uniform-roots} give
\begin{equation}\label{eq:oa-accuracy-classification-bounds}
 A_{s_\beta}B_{f_\beta}-D_{s_\beta}C_{f_\beta}>\frac1{100},\qquad
 \frac{1877}{100000}<J_\beta<\frac{211}{10000}.
\end{equation}
For the latter, use \(J_\beta=J_{\mathrm{acc}}-(1-\omega)^2j_Ls_\beta\), the formula
and monotonicity of \(J_{\mathrm{acc}}\), and the enlarged sealed-root interval.
The previous remaining bounds are unchanged:
\[
 K>\frac{3737}{50000},\qquad
 \frac{71}{100}<\alpha_0<\frac{783}{1000}.
\]
It follows that
\[
 G_\beta(h)=G_\beta(0)-J_\beta h\quad(0\leq h\leq f_\beta),
 \qquad H_\beta(h)=H_\beta(0)-Kh\quad(0\leq h\leq s_\beta).
\]
Under \(\pi_0\), silence has ratio \(r_0\leq r_\beta\) and zero
persistence. Substitution into the complete-history decomposition gives the
exact main-text gains. Uniform lower bounds follow from
\begin{align*}
 (1-\alpha_0)J_\beta f_\beta
 &>\left(1-\frac{783}{1000}\right)
    \frac{1877}{100000}\frac{53}{100}>\frac1{500},\\
 (J_\beta+K)s_\beta-\alpha_0J_\beta f_\beta
 &>\left(\frac{1877}{100000}+\frac{3737}{50000}\right)
     \frac{17}{100}
   -\frac{783}{1000}\frac{211}{10000}\frac{29}{50}>\frac3{500}.
\end{align*}
At the fixed policy's silent branch, the low type's reputational conflict
term is zero. Her revision gain is therefore
\(\beta(\eta_L-r_0)/(\eta_L+r_0)\), and
\[
 \frac{\eta_L-r_0}{\eta_L+r_0}
 >\frac{15/11-71/100}{15/11+71/100}>\frac3{10}.
\]
Since \(Q_0<1\), \(r_\beta\) has positive derivative. The map
\(r\mapsto(1-r)/(q-zr)\) has derivative \((z-q)/(q-zr)^2<0\), proving
the adjusted-threshold comparison.

\subsection{Accuracy rewards with two peers}

Restrict the reward to \(0\leq\beta\leq1/1000\).
The intervals in \cref{eq:oa-accuracy-two-peer-roots} coincide with those
used in Section~OA.8. Its decision regions, exact full-disclosure gain and
sealing lower bound therefore remain valid at the reward-dependent roots.
The uniform accuracy bounds are unchanged.

For incentives, the zero-persistence bounds are
\[
 Q_{\beta,0}<\frac{13}{25}+\frac{1/1000}{9/20}
 <\frac{55/27}{(33/17)^2},\qquad
 Q_{\beta,0}<\frac{319}{500}<\lambda_N^{(2)}.
\]
The first gives strict confirmation after unanimous opposition under full
disclosure. The second gives the lower likelihood bound for silence under the
unanimous-confirmation rule, including after a false initial report. High-type conflict choices stay strict. On mixing branches
with \(\lambda\leq\gamma^2<4\),
\[
 (\eta_L-1)G_I(0)-\beta(4-\eta_L)
 >\frac4{11}\frac{11}{50}-\frac1{1000}\frac{29}{11}>0.
\]
The initial-report argument therefore applies to both peers and B under
\(\mathsf U\), with margins above \(3/400\).

Under \(\mathsf F_2\), the low type's optimal post-lie choice reverses the
submitted report when both peers oppose it and follows the second signal
otherwise. The high type follows the second signal everywhere. The relevant
zero-persistence inequalities are
\[
 \gamma^{-2}\eta_L<Q_0\leq Q_{\beta,0}
 <\gamma^{-2}\eta_H.
\]
The extra deviation payoff is now
\begin{equation}\label{eq:oa-accuracy-two-peer-extra-lie}
 E_{L,\beta}
 =a(1-b)q^2\{G_I(0)+\beta\}
  -(1-a)b z^2\{G_C(0)+\beta\}.
\end{equation}
Let \(t_\beta=h_\beta(\gamma^2)\) and
\(D_{CI,\beta}=W_{CI}(t_\beta)-W_{CI}(0)\). The unadjusted wage-gap sum
in the initial decomposition is
\begin{align}
 \mathcal T_\beta
 ={}&(\eta_L-1)\{z^2G_I(t_\beta)+2qzG_I(s_\beta)\}
       +z^2G_C(0)-q^2G_I(0)\notag\\*
 &+\beta\{\eta_L(z^2+2qz)-(q^2+2qz)\}.
 \label{eq:oa-accuracy-two-peer-wage-sum}
\end{align}
The term proportional to the reward reflects the change in mixing required
for terminal indifference.

Applying the Section~OA.8 endpoint wage bounds gives
\[
 D_{CI,\beta}>\frac16,\qquad
 \mathcal T_\beta>\frac1{100}-\beta\geq\frac9{1000},\qquad
 E_{L,\beta}<\frac{39}{5000}+\frac9{100}\beta
 \leq\frac{789}{100000}.
\]
For the second bound, the coefficient on \(\beta\) in
\cref{eq:oa-accuracy-two-peer-wage-sum} exceeds \(-1\). For the third,
the added coefficient in \cref{eq:oa-accuracy-two-peer-extra-lie} is at most
\(a(1-b)q^2<9/100\). Thus the low-type full-disclosure margin is
\begin{align*}
 \Delta_{BL,\beta}^{\mathsf F_2}
 &\geq\frac1{10}
   \{P_0+q^2D_{CI,\beta}+b\mathcal T_\beta\}-E_{L,\beta}\\
 &>\frac1{10}\left\{\frac{125}{1624}
       +\frac16\left(\frac{209}{320}\right)^2+\frac9{1600}\right\}
       -\frac{789}{100000}>\frac7{1000}.
\end{align*}
The high type has no extra deviation payoff and a larger positive bound.
All six constraints hold under each protocol. Under \(\mathsf U\), silent
revision gives the zero-reward gain at \(h=0\), already above \(1/250\),
plus \(\beta(\eta_L-\lambda_N^{(2)})/(\eta_L+\lambda_N^{(2)})\geq0\).
Factorisation and consistency follow as in Section~OA.8.

\subsection{The sealed boundary, accuracy alone, and selection}

The sealed root becomes zero at
\[
 \beta=\beta_{\mathsf S}
 =\frac{G_C(0)-\eta_LG_I(0)}{\eta_L-1}.
\]
The last two polynomial rows prove \(2/5<\beta_{\mathsf S}<43/100\)
uniformly. At larger rewards, \(r_\beta\geq1\), and zero is the unique
low-type sealed conflict candidate. Confirmation and high-type revision stay
strict. Since \(1/\eta_L\leq Q_{\beta,0}<1\), the second signal is also
optimal after a lie. Each initial margin is
\[
 (2p_{\tau1}-1)\{P_0+p_{\tau2}[G_C(0)-G_I(0)]\}>0.
\]
This establishes a truthful sealed equilibrium. The disciplinary formula
reaches zero disclosure at the boundary, where agreement disclosure is no
longer required for silent revision. It gives no global ranking at larger
finite rewards.

For accuracy alone, remove \(\rho_i\) and normalise the positive accuracy
coefficient to one. Experts choose the more likely state. The primitives give
\[
 1<\eta_L<\gamma<\eta_H,\qquad
 \gamma<\kappa_{L1}\kappa_{L2},\qquad
 \eta_Lr_0<1.
\]
Thus A follows her second signal. Under full disclosure, B's low type retains
the first report after conflict and revealed agreement, and follows the
second signal after conflict and revealed disagreement. Confirmation is
strict. Under \(\pi_0\), she also follows the second signal after silence.
B's high type follows the second signal throughout.

A's initial report is payoff-irrelevant under all three rules. B's is likewise
irrelevant under sealing and full disclosure, since it does not affect her
information. Under \(\pi_0\), the high type follows her second signal even
after a lie and is initially indifferent. For the low type, a lie changes
the meaning of silence. Since \(\eta_Lr_0<1\), she retains her actual first
signal after a silent private conflict and follows the second signal after
revealed agreement with the submitted report. Relative to always following
the second signal, the gains after truth-telling and a lie are
\begin{align*}
 T_{\mathrm{acc}}-b
 &=\alpha_0\{a(1-b)q-(1-a)b z\},\\
 D_{\mathrm{acc}}-b
 &=a-b+\alpha_0\{(1-a)b q-a(1-b)z\}.
\end{align*}
Their difference equals
\begin{equation}\label{eq:oa-accuracy-only-initial-margin}
 T_{\mathrm{acc}}-D_{\mathrm{acc}}
 =(b-a)(1-\alpha_0)>0.
\end{equation}
Truthful initial reports are therefore equilibrium choices under all three
rules, with the stated weak and strict incentives. Bayesian beliefs and
optimal terminal choices also apply after a lie.

For classification set \(s=0\). Agreement contributes
\[
 G_{\mathrm{acc}}(h)
 =\max\{AA_h,DD_h\}+\max\{AB_h,DC_h\}
  +\max\{BA_h,CD_h\}+\max\{BB_h,CC_h\}.
\]
On \(0\leq h\leq1\), the first, first, first and second terms maximise
the respective comparisons. Each comparison is affine in \(h\). Primitive
bounds give a signed margin above \(1/2000\) at both endpoints, using at one
the exact masses
\begin{align*}
 A_1&=\omega uv+(1-\omega)a,& B_1&=\omega u(1-v),\\
 C_1&=\omega(1-u)v,& D_1&=\omega(1-u)(1-v)+(1-\omega)(1-a).
\end{align*}
Consequently,
\[
 G_{\mathrm{acc}}(h)=G_{\mathrm{acc}}(0)-J_{\mathrm{acc}}h,
 \qquad J_{\mathrm{acc}}=mC-\ell B>\frac{11}{500}>0.
\]
Disagreement has zero persistence under every protocol. On agreement,
sealing uses \(h=0\), full disclosure \(h=1\), and \(\pi_0\) uses
\(h=1\) with probability \(\alpha_0\). Keeping the message in the public
experiment gives the two main-text differences. Initial reporting need not
be unique, and the comparison does not optimise over all feedback rules.

Finally, \(W_I^{(1)}=\omega(1-u)/(1-q)\) increases with \(\omega\) and
is at least \(7/37\). The full-repetition construction thus survives for
\(0\leq\beta\leq1/10\), including with two peers, since other experts'
likelihood factors cancel conditional on the state and initial reports.
Positive rewards eliminate uniform terminal babbling without selecting the
informative equilibrium.

\section{Analytical objective dependence}
\label{oa:analytical-blackwell}

Fix the four precisions in the main proposition and let
\(\omega\in[3/8,2/5]\). The earlier policy-uniform incentive theorem
establishes feasibility of the maintained equilibrium on this interval.
The roots satisfy \(11/50<s<1/4\) and \(53/100<f<11/20\).
Write \(M_h=(A_h,B_h,C_h,D_h)\). For a public history
\(z=(a,b,c,d)\), define \(i_\theta(a,b)=2\ind\{a\ne\theta\}
+\ind\{b\ne\theta\}\), where the entries of \(M_h\) are indexed from
zero. Conditional independence then gives
\[
 P_\theta^{\mathsf S}(z)
 =M_s[i_\theta(a,b)]M_s[i_\theta(c,d)],\qquad
 P_\theta^{\mathsf F}(z)
 =M_s[i_\theta(a,b)]M_{h(a,c)}[i_\theta(c,d)],
\]
where \(h(a,c)=f\) if \(a=c\), and zero otherwise. The correctness
masses integrate out abilities and private signals. Both expressions retain
all four public reports.

At cutoff \(1/10\), action one is optimal if and only if
\(9P_1(z)-P_0(z)>0\). We establish its sign using rational bounds valid
throughout the parameter interval. Each primitive mass is affine in
\(\omega\), so its extrema occur at \(\omega=3/8,2/5\).
Combining these extrema with the root bounds gives, for example,
\(A_s\in[A_{\min}+\ell_{\min}(11/50),
A_{\max}+\ell_{\max}/4]\) and
\(B_s\in[B_{\min}-\ell_{\max}/4,
B_{\max}-\ell_{\min}(11/50)]\).
Applying the same interval operations to the remaining masses and substituting
into the two history-probability formulas yields
\begin{center}
\begin{tabular}{@{}lll@{}}
\toprule
protocol & history & bound on \(9P_1-P_0\)\\
\midrule
sealing & \(0000\) & less than \(-1/20\)\\
sealing & all other histories & greater than \(1/50\)\\
full disclosure & \(0000\) & less than \(-1/50\)\\
full disclosure & other than \(0000,1000\) & greater than \(1/50\)\\
\bottomrule
\end{tabular}
\end{center}

For the remaining full-disclosure history, \(1000\), the relevant
expression is \(9B_sD-C_sA\). Since it is affine in \(s\), it is enough
to examine \(s=11/50,1/4\). Substituting
\(\omega=3/8+x/40\) gives two quadratics in \(x\), whose coefficients
in the Bernstein basis \(((1-x)^2,2x(1-x),x^2)\) are, respectively,
\[
 \left(\frac{30921}{2560000},\frac{497123}{64000000},
       \frac{71577}{20000000}\right),\qquad
 \left(\frac{56199}{5120000},\frac{173723}{25600000},
       \frac{21681}{8000000}\right).
\]
Every coefficient exceeds \(1/400\). The basis functions are nonnegative
and sum to one, so the same bound holds at both endpoints in \(s\).
Affinity extends it to the whole root interval. This establishes the decision
regions without approximating the roots.

The primitive interval bounds also imply
\[
 9mD_s-\ell A_s>\frac7{50}.
\]
To verify the lower bound, substitute
\(9m_{\min}(D_{\min}+m_{\min}(11/50))
-\ell_{\max}(A_{\max}+\ell_{\max}/4)\).
The endpoint primitives are given by the affine functions in OA.4. Hence the
factorisation and both strict rankings in the main proposition hold over the
entire interval.

For the selective policy, a revealed agreement yields the same conditional
decision as under full disclosure. On disagreement the recipient's persistence
is zero both under full disclosure and under silence at the selective optimum.
It remains to check the histories in which the initial reports agree but no
feedback is sent. Given these reports, the probability of concealment is
independent of the state. Removing this common factor, the state likelihoods
for a history \(z=(a,b,c,d)\) are
\[
 L_\theta(z)=M_s[i_\theta(a,b)]M_0[i_\theta(c,d)],\qquad a=c.
\]
Use the componentwise bounds for \(M_s\) above and the two endpoint values
of each component of \(M_0\). For \(0000\), an upper bound on
\(9L_1-L_0\) is \(-63092093/512000000<-3/25\).
Among the other seven histories the smallest of the resulting lower
bounds is \(45306603/512000000>2/25\), attained by \(1010\).
These bounds use the order \(CC,CI,IC,II\) in \(M_h\) and
\(11/50\leq s\leq1/4\). Thus the only history at which the decision
maker chooses zero is again \(0000\), whether agreement is revealed or
concealed. Summing the two contributions gives
\[
 C^{\mathrm{DS}}(1/10)=\frac25+
 \frac{A_s[\alpha_0 A_f+(1-\alpha_0)A]
       -9D_s[\alpha_0 D_f+(1-\alpha_0)D]}{20}.
\]
The corresponding expression under sealing replaces the terms in brackets by
\(A_s\) and \(D_s\). Since \(A_h=A+\ell h\) and \(D_h=D+mh\), the
difference factors as in Section~5.4 of the paper.

\section{A terminal-message test of credible announcements}
\label{oa:neologisms}

Farrell (1993, pp.~520--522) asks whether an unused message has a
credible literal meaning. Here the expert can announce at the terminal date
that her private history belongs to a nonempty set $S$. She receives the
posterior wage obtained when the evaluator accepts the announcement and
subsequently observes the state and the complete public record. The peer
submits her terminal report before learning of the announcement. All other
reports retain their equilibrium interpretation.

\subsection{Private histories, beliefs and the test}

Write $t=(\tau,s_1,s_2)$ and $\pi_H=\omega$, $\pi_L=1-\omega$.
At an observed initial report $b$ and any realised feedback known to the
expert, retain every private history compatible with that information.
Under babbling all eight histories remain possible, whereas truthful initial
reports restrict the set to $s_1=b$. Let
\[
 L_t(q)=\pi_\tau f_{\tau1}(s_1\mid q) f_{\tau2}(s_2\mid q),
 \qquad q\in\{0,1\},
\]
using the signal probabilities from Section~OA.2.
If the evaluator accepts $S$, the wage after verification of state $q$ is
\[
 W_S(q)=\frac{\sum_{t\in S:\tau=H} L_t(q)}{\sum_{t\in S} L_t(q)}.
\]
For the equilibria studied here the likelihood of the other expert's
public record cancels in this ratio, conditional on $q$, the initial report
and any feedback relevant to the announcer.

The expert accounts for feedback when computing the probability of the
state. Let $k_q$ be the probability of her observed branch conditional on
$q$ and $b$. For A, set $k_0=k_1=1$. For B in truthful play use the
probabilities of the message implied by the policy and the peer's truthful
report. Under babbling feedback carries no information and this factor also
cancels. With the equal state prior,
\[
 x_t=\frac{k_1L_t(1)}{k_0L_t(0)+k_1L_t(1)}.
\]
If $U_t$ is the equilibrium continuation payoff, the gain from the
announcement is $g_t(S)=(1-x_t)W_S(0)+x_tW_S(1)-U_t$. The
self-signalling requirement is
\begin{equation}\label{eq:oa-self-signalling}
  S=\{t:g_t(S)>0\}.
\end{equation}
A member must strictly gain. An outsider may be indifferent.

\subsection{Babbling and repetition give different answers}

\begin{proposition}\label{prop:oa-terminal-neologisms}
At $(p_{L1},p_{L2},p_{H1},p_{H2},\omega)
=(11/20,5/8,33/40,19/20,2/5)$, under every reveal-or-silence policy:
(i) uniform babbling admits no self-signalling terminal announcement for
either expert; (ii) full repetition admits such an announcement by A.
These statements concern the terminal extension specified above.
\end{proposition}

\begin{proof}
Under babbling, $U_t=2/5$ for each of the eight types. Their ordering by
$x_t$ is
\[
\begin{array}{c|rrrrrrrr}
 t&H00&H10&L00&L10&L01&L11&H01&H11\\
 \hline
 x_t&7/634&33/166&27/82&11/26&15/26&55/82&133/166&627/634
\end{array}
\]
Here $H10$, for example, denotes $(H,1,0)$. Since $g_t(S)$ is affine in
$x_t$, the types that strictly gain form an upper tail, a lower tail, the
whole set, or the empty set. For the whole set,
$W_S(0)=W_S(1)=2/5$, so no type strictly gains. The empty set cannot be an
announcement. We therefore need only exclude the seven proper upper tails
and the corresponding lower tails. Number the ordered types from 1 to 8 and
let $S_k$ contain positions $k+1,\ldots,8$. Substitution gives
\begin{center}
\begin{tabular}{@{}clll@{}}
\toprule
$k$ & witness type & membership & $g_t(S_k)$\\
\midrule
1 & $H10$ & inside & $-249773139/1135601435$\\
2 & $L00$ & inside & $-78063/311395$\\
3 & $L10$ & inside & $-150582/796055$\\
4 & $L01$ & inside & $-2778/36995$\\
5 & $L01$ & outside & $129342/2957045$\\
6 & $L00$ & outside & $3/5$\\
7 & $L00$ & outside & $3/5$\\
\bottomrule
\end{tabular}
\end{center}
Each row violates the self-signalling condition. Complementing both signal
labels exchanges the states and maps an upper tail into a lower tail with
the same gains. This excludes all nonempty sets. Under babbling, feedback
and other reports are independent of the state and abilities, so the argument
applies to either expert under every policy.

Under full repetition, consider A with initial report $b=0$. The announcement
$S=\{L00,H00\}$ states that her second signal confirms her first. It gives
wages $W_S(0)=38/63$ and $W_S(1)=14/419$. Her original wages are $1/2$ if the
first report is correct and $7/34$ otherwise. Hence
\[
\begin{array}{c|rrrr}
 t&L00&H00&L01&H01\\
 \hline
 g_t(S)&\frac{228397}{18398709}&\frac{4748003}{47417811}&
 -\frac{25063}{448749}&-\frac{1460953}{12415389}
\end{array}
\]
Both members strictly gain and both outsiders strictly lose. Label symmetry
gives the same conclusion for $b=1$. Since A receives no feedback and her
wages conditional on the state do not depend on the peer's reports, the
announcement is available under every policy.
\end{proof}

\subsection{An informative equilibrium}

\begin{proposition}\label{prop:oa-informative-announcements}
Suppose $\eta_H>\eta_L\geq1$ and the equilibrium is regular-monotone.
At every on-path initial report, and for B at every supported feedback
message, neither expert has a self-signalling terminal announcement under
purely reputational payoffs.
\end{proposition}

\begin{proof}
Fix the expert's initial report $b$, and for B fix the message received.
Write $C$ for a confirming second signal and $D$ for a conflicting one.
There are four private histories, $HC,LC,LD,HD$. Let $x$ be the expert's
probability that $b$ is wrong. With $\lambda$ the branch likelihood ratio
favouring $b$ ($\lambda=1$ for A), the odds $x/(1-x)$ are, in this order,
\[
 \frac{1}{\lambda\kappa_{H1}\kappa_{H2}},\quad
 \frac{1}{\lambda\kappa_{L1}\kappa_{L2}},\quad
 \frac{\eta_L}{\lambda},\quad
 \frac{\eta_H}{\lambda}.
\]
They increase strictly. Each binary report pays a wage conditional on
whether $b$ is right or wrong. The same is true of each announcement.
Indeed, conditional on the state, $b$ and B's message, the likelihood of
A's two reports factors out of the posterior over B's private histories.
This also holds after silence, when the evaluator eventually sees the
concealed initial report. For A, conditional on the state and her initial
report, the likelihood of B's reports and of feedback factors out. Thus
the two wages for any given action apply to all four histories on the branch.

Expected payoffs from a fixed action are affine in $x$. The expert's
equilibrium payoff is the maximum of the two binary-report payoffs.
A new announcement is strictly preferred on the intersection of two
half-lines, one for each incumbent report. Consequently the types that
strictly gain must form a consecutive block of $HC,LC,LD,HD$.

Consider the possible blocks. A set of low types alone earns zero. A
set of high types alone earns one and attracts low types as well. Every
incumbent wage lies strictly between zero and one because each type uses
both reports with positive probability conditional on either state.
For a block of three, one of the two high types is absent. Consider the
binary report used by the high type in the block. Every type making that
report belongs to the block. In each state the announcement adds only low
types to its pool, keeping the high mass fixed. Its wage is therefore at
most that of the binary report in each state. The high member cannot
strictly gain.

The two remaining mixed pairs are $\{HC,LC\}$ and $\{LD,HD\}$. Write
$h\in[0,1)$ for the probability that $LD$ repeats. In the first pair the
announcement pays exactly the repetition wage if $h=0$. If $h>0$, it
removes a positive mass of $LD$ from repetition in both states. Its wage
then rises strictly in both states. The excluded type $LD$ can earn her
equilibrium payoff by repeating and would strictly prefer the announcement.
For the second pair the announcement adds the repeating fraction of $LD$
to the revision pool in each state. Its wage is at most the revision wage,
so $HD$ cannot strictly gain.

Finally, an announcement by the whole block gives the evaluator the same
information as suppressing the expert's binary terminal report. Averaging
over the four histories and the state, conditional on the public branch,
the expected posterior probability of high ability is the same with or
without that report, by iterated expectations. All four types cannot
strictly gain. These cases exhaust the consecutive blocks.
\end{proof}

The argument concerns a unilateral announcement about the expert's private
history at the terminal date. An announcement naming histories from several
public branches must satisfy the test on each branch where it names any
types, since the evaluator observes the branch. Initial reports, which
determine the feedback lottery, still have the original binary language.
Both uniform babbling at the benchmark and the informative response there
pass this terminal test, although it rules out full repetition.

\end{document}